\documentclass[]{aastex631}

\usepackage{booktabs}
\usepackage{color, bm}
\renewcommand{\edit}[1]{{\textcolor{black}{#1}}}
\newcommand{\reedit}[1]{{\textcolor{black}{#1}}}
\newcommand{\cedit}[1]{{\textcolor{black}{#1}}}

\shorttitle{Microlensing Discovery and Characterization in LSST}
\shortauthors{Abrams et al.}
\graphicspath{{./}{figures/}}

\begin{document}

\title{Microlensing Discovery and Characterization Efficiency in the Vera C. Rubin Legacy Survey of Space and Time}

\author[0000-0002-0287-3783]{Natasha S.~Abrams}
\affiliation{University of California, Berkeley, Astronomy Department, Berkeley, CA 94720, USA}

\author[0000-0003-0961-5231]{Markus P.G.~Hundertmark}
\affiliation{Zentrum f{\"u}r Astronomie der Universit{\"a}t Heidelberg, Astronomisches Rechen-Institut, M{\"o}nchhofstr. 12-14, 69120 Heidelberg, Germany}

\author[0000-0002-1910-7065]{Somayeh Khakpash}
\affiliation{Rutgers University, Department of Physics \& Astronomy, 136 Frelinghuysen Rd, Piscataway, NJ 08854, USA}

\author[0000-0001-6279-0552]{Rachel A.~Street}
\affiliation{Las Cumbres Observatory (LCOGT), 6740 Cortona Drive, Suite 102, Goleta, CA 93117, USA}

\author[0000-0001-5916-0031]{R. Lynne Jones}
\affiliation{Aerotek and Rubin Observatory, Tucson, AZ, USA}

\author[0000-0001-9611-0009]{Jessica R. Lu}
\affiliation{University of California, Berkeley, Astronomy Department, Berkeley, CA 94720, USA}

\author[0000-0002-6578-5078]{Etienne Bachelet}
\affiliation{IPAC, Caltech, Pasadena, CA 91125, USA}

\author[0000-0001-8411-351X]{Yiannis Tsapras}
\affiliation{Zentrum f{\"u}r Astronomie der Universit{\"a}t Heidelberg, Astronomisches Rechen-Institut, M{\"o}nchhofstr. 12-14, 69120 Heidelberg, Germany}

\author[0000-0001-8716-6561]{Marc Moniez}
\affiliation{Universit\'e Paris-Saclay, CNRS/IN2P3, IJCLab, France}

\author{Tristan Blaineau}
\affiliation{Universit\'e Paris-Saclay, CNRS/IN2P3, IJCLab, France}

\author[0000-0003-0972-1376]{Rosanne Di Stefano}
\affiliation{Harvard-Smithsonian Center for Astrophysics, Cambrdige, MA, 02138, USA}

\author[0000-0003-2206-2651]{Martin Makler}
\affiliation{International Center for Advanced Studies \& Instituto de Ciencias Físicas, ECyT-UNSAM \& CONICET, San Martín, Buenos Aires, 1650, Argentina}
\affiliation{Centro Brasileiro de Pesquisas Físicas, Rio de Janeiro, RJ, 22290--180, Brazil}

\author{Anibal Varela}
\affiliation{International Center for Advanced Studies \& Instituto de Ciencias Físicas, ECyT-UNSAM \& CONICET, San Martín, Buenos Aires, 1650, Argentina}

\author[0000-0003-2935-7196]{Markus Rabus}
\affiliation{Departamento de Matem{\'a}tica y F{\'i}sica Aplicadas, Facultad de Ingenier{\'i}a, Universidad Cat{\'o}lica de la Sant{\'i}sima Concepci{\'o}n, Alonso de Rivera 2850, Concepci{\'o}n, Chile}



\begin{abstract}
The Vera C. Rubin Legacy Survey of Space and Time will discover thousands of microlensing events across the Milky Way Galaxy, allowing for the study of populations of exoplanets, stars, and compact objects. We evaluate numerous survey strategies simulated in the Rubin Operation Simulations (OpSims) to assess the discovery and characterization efficiencies of microlensing events. We have implemented three metrics in the Rubin Metric Analysis Framework: a discovery metric and two characterization metrics, where one estimates how well the lightcurve is covered and the other quantifies how precisely event parameters can be determined. We also assess the characterizability of microlensing parallax, critical for detection of free-floating black hole lenses. We find that, given Rubin's baseline cadence, the discovery and characterization efficiency will be higher for longer duration and larger parallax events. Microlensing discovery efficiency is dominated by \cedit{the} observing footprint, where more time spent looking at regions of high stellar density including the Galactic bulge, Galactic plane, and Magellanic clouds, leads to higher discovery and characterization rates. However, if the observations are stretched over too wide an area, including low-priority areas of the Galactic plane with fewer stars and higher extinction, event characterization suffers by $> 10\%$. \edit{This} could impact exoplanet, binary star, and compact object events alike. We find that some rolling strategies (where Rubin focuses on a fraction of the sky in alternating years) in the Galactic bulge can lead to a 15-20\% decrease in microlensing parallax characterization, so rolling strategies should be chosen carefully to minimize losses. 

\end{abstract}



\section{Introduction} 
\label{sec:intro}

Microlensing occurs when light coming from a distant star (source) is deflected by a foreground object (lens) located along the observer-source line of sight. As a result, multiple images of the source are formed, and \edit{as} the images are usually unable to be resolved, the images appear blended. \cedit{The net observational effect is that} the source that then appears to be photometrically magnified \citep{pacynski1986halo}. Since the effect depends on the gravitational influence of the lens and not its luminosity, microlensing is a powerful tool to find and weigh dim objects like cool low-mass stars, planets \citep[e.g.][]{Gaudi:2012, Tsapras:2018}, neutron star candidates, and stellar-mass black hole candidates \citep[e.g.][]{Lu:2016, Lam:2022, Sahu:2022, Mroz:2022, Lam:2023-OB110462} that are otherwise hard to observe.

The microlensing discovery rate increases with the stellar density; therefore, it is highest when observing crowded parts of the sky like the Galactic bulge, Galactic plane, and Large and Small Magellanic Clouds (LMC and SMC). Previous and ongoing dedicated microlensing surveys (e.g. OGLE \citep{ogleIV:Udalski:2015}, MOA \citep{moa:Sumi:2003}, KMTNet \citep{kmtnet:kim:2018}, MACHO \citep{macho:alcock:2000}, and EROS \citep{Moniez:2017}) have focused on these areas of high stellar density. All-sky surveys offer the opportunity to explore microlensing throughout the Galaxy. Observing throughout the Galaxy gives us the opportunity to probe Galactic structure \citep[e.g.][]{Moniez:2010, Moniez:2017} and constrain how the mass function of the lenses, such as black holes, changes throughout the galaxy. Observations of the Magellanic Clouds also offer an opportunity to explore compact halo objects and extragalactic stellar remnants,
\reedit{that} can then be compared with those of the Milky Way Galaxy.\footnote{See the cadence note \textit{Microlensing towards the Magellanic Clouds: searching for long events} by Blaineau et al. for a more detailed explanation. \url{https://docushare.lsst.org/docushare/dsweb/Get/Document-37634/LMC_SMC.pdf}}

The Vera C. Rubin Observatory Legacy Survey of Space and Time (LSST) will survey 18,000 deg$^2$ including parts of the Galactic plane along with LMC and SMC as part of its Wide Fast Deep (WFD) survey. Surveying at least every 2-3 days is imperative to \cedit{achieving} a high discovery rate of microlensing events \citep{Gould:2013, street2018diverse, sajadian2019lsst}. The Vera C. Rubin Observatory (Rubin) is going through a community-driven cadence optimization process described in detail in \cite{bianco2022}. \edit{In a community-driven cadence optimization process, it is critical to understand how cadence affects one's science case. When optimizing, the Survey Cadence Optimization Committee could make many decisions that could render a particular science case impossible, including choosing to exclude a region where most of a particular science case can be done (i.e. excluding the Galactic bulge and plane for microlensing) or choosing to revisit fields on a cadence longer than the timescale of a science case. As this process must balance many science cases with conflicting operational requirements, the community driven optimization is used to ensure that the survey strategy does not make science cases impossible. In order to evaluate these tensions and to strike a balance between them, statistical metrics have been developed to represent the observational requirements of each science case. The evaluation of these metrics for a range of science goals is described in the collection of papers lead by \cite{bianco2022}. Here, we present an analysis of how potential cadences will affect microlensing.} 

\edit{We determine how cadence will affect microlensing by using a} series of hundreds of cadence simulations called Operation Simulations (OpSims) \cedit{which} were created to mock scheduled observations, using the Rubin scheduler \citep{Naghib:2019} with LSST simulations framework \citep{Connolly:2014}. \reedit{They simulate observations including their times, durations, airmasses, moon position, and seeing from mock weather, among other statistics.} There are families of simulations \reedit{that} focus on optimizing particular qualities such as the region of sky covered in the WFD (or ``footprint"), the frequency of observations (or ``cadence"), filter balance \reedit{(fraction of observations in each filter)}, and rolling \reedit{cadence}. 
A rolling cadence is when we divide the sky into multiple sections and in some years Rubin will observe some sections with an increased number of observations and in other years it will focus its observations to the other sections.\footnote{This animation of the \texttt{baseline\_v2.0\_10yrs} simulation (which has a rolling cadence in years 2-9) by Lynne Jones illustrates how the observations build up over time with a rolling cadence. \url{https://epyc.astro.washington.edu/~lynnej/opsim_downloads/baseline_v2.0_10yrs__N_Visits.mp4}} These are evaluated using the Metric Analysis Framework \citep[MAF,][]{Jones:2014} \reedit{that} contains both metrics from the Rubin project development team and contributed by the community for particular science cases. 

In this work, we have written and tested a multi-faceted \texttt{MicrolensingMetric} (see Section \ref{sec: microlensing metric}) on the set of OpSims from v2.0 to v3.0 \reedit{that} investigates the detection and characterization of microlensing events. 
We have also tested the effect of changes to footprint and a rolling cadence on the characterization of microlensing events with a parallax signal outside the context of the MAF.
The rest of the paper is organized as follows. In Section \ref{microlensing model} we introduce standard microlensing terminology used throughout. In Section \ref{sec:methodology} we introduce the metric used to assess the microlensing yields, the sample of microlensing events assessed, and our methodology for determining microlensing parallax characterization. In Section \ref{sec: results} we explore the results of the \texttt{MicrolensingMetric} for relevant OpSims and parallax characterization for select OpSims. Finally, in Section \ref{sec: discussion and conclusion} we discuss implications and summarize conclusions.

\subsection{Microlensing Parameters}
\label{microlensing model}

We will introduce the standard microlensing parameters that \reedit{are} used to model microlensing events \citep{pacynski1986halo}. The characteristic length scale of a microlensing event is known as the angular Einstein radius, which is given by
\begin{equation}
    \theta_{\rm E} = \sqrt{\frac{4GM}{c^2}\left(\frac{1}{D_{\rm L}} - \frac{1}{D_{\rm S}}\right)},
    \label{eq:einstein_radius}
\end{equation}
where $M$ is the lens mass, $D_{\rm L}$ is the distance to the lens from the observer, and $D_{\rm S}$ is the distance to the source from the observer.
$\theta_{\rm E}$ and the relative proper motion of the source and lens ($\mu_{\rm rel}$) can be used to define the characteristic timescale, the Einstein crossing time:
\begin{equation}
\label{eq:tE}
    t_{\rm E} = \frac{\theta_{\rm E}}{\mu_{\rm rel}}.
\end{equation}
So events with more massive lenses tend to have longer $t_{\rm E}$. Neglecting the effects of parallax, the projected separation between the lens and source in units of Einstein radii as a function of time is the impact parameter:
\begin{equation}
    u(t) = \sqrt{u_0^2 + \left(\frac{t - t_0}{t_{\rm E}}\right)^2},
\end{equation}
where $u_0$ is the closest projected separation and $t_0$ is the time of closest approach. Microlensing events are detected as a temporary photometric magnification. The amplification of a point source-point lens microlensing event is given by:
\begin{equation}
    A(t) = \frac{u^2 + 2}{u\sqrt{u^2 +4}}.
    \label{eq:pspl}
\end{equation}
The amplification is maximized when $u = u_0$.  We can use this signal in inferring the lens mass, but a measurement of the lens mass requires either a galactic model or additional observable constraints on $D_L$, $D_S$, and $\mu_{\rm rel}$ \cedit{(and/or the lens flux)}. 
When the lens is luminous or there are neighboring stars, the light from the source may be blended with them. \cedit{The total light is} characterized by the source flux ($F_{\rm S}$) and the blend flux:
\begin{equation}
    F_{\rm B} = F_{\rm L} + F_{\rm N},
\end{equation}
where $F_{\rm L}$ is the lens flux and $F_{\rm N}$ is the neighbor flux. So the flux as a function of time is then:
\begin{equation}
    F(t) = F_{\rm S} A(t) + F_{\rm B}.
\end{equation}
The blend source flux fraction or blend fraction is $b_{\rm sff} = F_{\rm S}/(F_{\rm B} + F_{\rm S})$.

The motion of the Earth around the Sun introduces a higher order effect in microlensing events known as the microlensing parallax \citep{Gould:1992-parallax}, $\bm{\pi_{\rm E}}$, the magnitude of which is defined as 
\begin{equation}
    \pi_{\rm E} = \frac{\pi_{\rm rel}}{\theta_{\rm E}},
    \label{eq: piE}
\end{equation}
where $\pi_{\rm rel}$ is the relative lens-source parallax ($\pi_{\rm rel} = (1 \textrm{AU}) (\frac{1}{D_{\rm L}} - \frac{1}{D_{\rm S}})$ ). 
The signal strength depends on the difference between the lens parallax and the source parallax, \edit{the direction of the lens-source proper motion, and} on the time of the year. The changing position of an observer following the Earth's orbit changes the optical axis of the lensing configuration. Microlensing events much shorter than a year tend to exhibit a negligible parallax signal. $\bm{\pi_{\rm E}}$ is a vector that can be broken down into East and North components, $\bm{\pi_{\rm E}} = [\pi_{E,E}, \pi_{E, N}]$. \reedit{It is} pointed in the direction of $\mu_{\rm rel}$, the relative lens-source proper motion. Along with $\mu_{\rm rel}$, which can \edit{have a prior put on \reedit{it} by a} galactic model \edit{\citep[e.g.][]{Beaulieu:2006, Bennett:2014}}, and $t_{\rm E}$, $\pi_{\rm E}$ can be used to determine the mass of the lens:
\begin{equation}
M = \frac{\mu_{\rm rel} t_{\rm E}}{\kappa \pi_{\rm E}}
\label{eq:mass_measurement}
\end{equation}
where $\kappa \equiv \frac{4G}{(\textrm{1 AU})c^2} \approx 8.144 ~\mathrm{mas}/M_{\odot}$.

\begin{figure}
    \centering
    \includegraphics[scale=0.5]{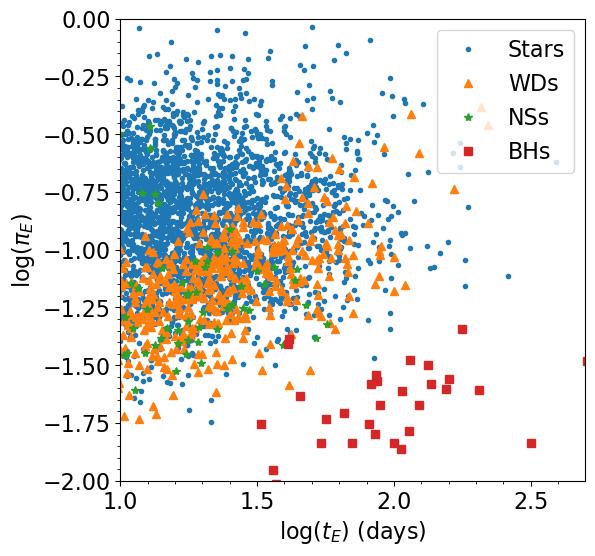}
    \caption{Microlensing parallax (Eq. \ref{eq: piE}) vs Einstein crossing time (Eq. \ref{eq:tE}) for microlensing events simulated using \texttt{PopSyCLE} \citep{Lam:2020}. Blue circles are stars, orange triangles are white dwarfs, green stars are neutron stars, and red squares are black holes. The ranges of $t_{\rm E}$ and $\pi_{\rm E}$ are cropped to match Figures \ref{fig:parallax_char_baseline_v3.0}-\ref{fig:parallax_char_retro_baseline}. \edit{As shown in Eq.~\ref{eq:mass_measurement}}, more massive lenses tend to be towards the bottom right of this plot.}
    \label{fig:popsycle_piE_tE}
\end{figure}

\edit{$\pi_{\rm E}$ and $t_{\rm E}$ thus gives you a statistical constraint on the mass of the lens}. For a physical understanding of where microlensing events lie in the $\pi_{\rm E}-t_{\rm E}$ plane see Figure \ref{fig:popsycle_piE_tE}. Here we plot simulated events from Population Synthesis for Compact-object Lensing Events \citep[\texttt{PopSyCLE},][]{Lam:2020}. Events with less massive, often stellar, lenses tend to have shorter $t_{\rm E}$ and larger $\pi_{\rm E}$; whereas, events with more massive, often compact object, lenses tend to have longer $t_{\rm E}$ and smaller $\pi_{\rm E}$.

\section{Methodology}
\label{sec:methodology}
In this section we describe \edit{OpSims and the MAF framework} and a suite of metrics \reedit{that} are used to assess how the relative number of detected and characterized microlensing events is affected by the survey strategy. This is not expected to produce a realistic yield for Rubin LSST. Instead, it is expected to inform which strategies are beneficial for microlensing\cedit{, which make it impossible to do microlensing science,} and \cedit{which may negatively affect the microlensing yield without making the science case inviable}. There are three metrics within the MAF framework \reedit{that} are simulated without microlensing parallax and one metric outside the MAF framework simulated with microlensing parallax (see Table \ref{tab:metric summary}). The results for the three metrics in the \texttt{MicrolensingMetric} are described in Sections \ref{sec: baseline metrics}-\ref{sec:rolling cadence} and the parallax characterization metric results are in Section \ref{sec: parallax characterization}.

\begin{table}[h]
    \centering
    \begin{tabular}{|c|c|c|c|}
    \hline
        Metric & Simulated Sample & \reedit{Simulated Mags} & Metric Description \\
        \hline
         \texttt{MicrolensingMetric} - Discovery Metric & $t_{\rm E}$-only (Sec.~\ref{sec: sample}) & \reedit{1 mean star} & 2 points \edit{on rise} with $\geq 3 \sigma$ difference \\
         \hline
         \texttt{MicrolensingMetric} - Npts Metric & $t_{\rm E}$-only (Sec.~\ref{sec: sample}) & \reedit{1 mean star}  & 10 points \cedit{(w/ SNR > 3$\sigma$)} within $t_{\rm 0} \pm t_{\rm E}$\\
         \hline
         \texttt{MicrolensingMetric} - Fisher Metric & $t_{\rm E}$-only (Sec.~\ref{sec: sample}) & \reedit{1 mean star}  & Analytic Fisher with $\frac{\sigma_{t_{\rm E}}}{t_{\rm E}} < 0.1$ \\
         \hline
         Parallax Characterization Metric & $t_{\rm E} + \pi_{\rm E}$ (Sec.~\ref{sec: parallax characterization methods}) & \reedit{Rand TRILEGAL}  & Numerical Fisher w/ $\pi_{\rm E} > 2\sigma_{\pi_{\rm E}}$ \& $t_{\rm E} > 2\sigma_{t_{\rm E}}$ \\
         \hline
    \end{tabular}
    \caption{Description, name, and input sample of metrics used to analyze microlensing efficiency. The Sample column refers to the sample of microlensing events that were evaluated with the metric. \reedit{The Simulated Mags column refers to the source magnitudes of the simulated microlensing events. For those in the \texttt{MicrolensingMetric}, there was a single average star (with different mags in each filter) from the TRILEGAL map used (see Section \ref{sec: sample}). For the parallax characterization, random stars from TRILEGAL were drawn and their magnitudes were used for source magnitudes of the microlensing event (see Section \ref{sec: parallax characterization methods}).} The three metrics in the \texttt{MicrolensingMetric} are not simulated with parallax and are all-sky. The parallax characterization metric used a sample with parallax for two small patches of the sky.}
    \label{tab:metric summary}
\end{table}

\subsection{OpSims and MAF Framework}
\label{sec:opsims}
The Rubin OpSim team ran the \texttt{MicrolensingMetric} on 360+ OpSims. \cedit{The OpSims} evaluated in this paper are summarized in Table \ref{tab:opsim_summary}. See \cite{Jones_baseline_2018-2020} and Rubin technical note \href{https://pstn-055.lsst.io/}{PSTN-055} for more detailed descriptions. We also include summary statistics of OpSims discussed in this paper for reference in Table \ref{tab:general_metrics}. 

These OpSims have ``families" \reedit{that vary} a particular aspect of the cadence so that they can be tested in isolation and the many aspects of the cadence can be optimized. For example, OpSims in the \reedit{\texttt{vary\_gp\_} family vary the fraction of the Galactic plane included in the WFD footprint.} The first order OpSims are known as the \texttt{baseline} OpSims. They are marked with version numbers, and the most recent baseline OpSim is considered the nominal current suggested plan for LSST. The baseline OpSims are also the jumping-off point for the other OpSims. So OpSims marked with a \texttt{v2.1} are changing an aspect relative to the \texttt{baseline\_v2.1}\reedit{\texttt{\_10yrs}} OpSim.

\edit{There are a few main quantities that OpSims tend to vary:
\begin{enumerate}
    \item Footprint. This mostly refers to which areas will be included in the WFD and which will be included but at a lower cadence. This also refers to the location of the small Deep Drilling Fields (DDFs) which will be surveyed at a higher cadence. 
    \item Cadence. This is the frequency at which observations are taken. This can be both internight cadence, for example if triplet observations should be taken in a single night, and intranight cadence, for example ``rolling."
    \item Filter Balance. This refers to what fraction of exposures are taken in which filter.
    \item Exposure Length. This refers to the durations of the exposures \reedit{that} may affect both the science in each individual exposure and overall mission efficiency. 
    \item Additional Surveys. Rubin may have time to do small additional surveys \reedit{that} are not part of the DDFs or WFD, such as observing Near Earth Objects in a twilight survey. These OpSims explore those possibilities.
\end{enumerate}
There are also a number of OpSims that explore the technical aspects such as how the performance of the telescope or how different simulated weather will change the survey strategy, which we do not evaluate here.} In this paper we also do not evaluate filter balance \cedit{or} exposure length due to using a single average magnitude in \texttt{MicrolensingMetric} (see Section \ref{sec: sample}).

\edit{There are hundreds of metrics that are evaluated for each OpSim in the MAF \citep{Jones:2014} including general metrics, for example one which measures the 5$\sigma$ limiting magnitude in each filter, and specific ones for each science case. These metrics must be lightweight and are generally designed to test one result so that they can be run on all the OpSims and interpreted collectively \citep{COSEP}. In this paper we analyze the effect \cedit{on} microlensing, so we use the metrics described in Section \ref{sec: microlensing metric}. We also analyze the ability to characterize events with microlensing parallax outside of the context of the MAF due to computation speed (Section \ref{sec: parallax characterization methods}) on a few key OpSims (Section \ref{sec: parallax characterization}).}

\subsection{\texttt{MicrolensingMetric}}
\label{sec: microlensing metric}

The \texttt{MicrolensingMetric} is integrated with the \texttt{rubin\_sim}\footnote{\url{https://github.com/lsst/rubin_sim}} package and provides a set of metrics for evaluating the efficacy of cadences in detecting, alerting, and characterizing microlensing events. The metric relies on a simulated population of microlensing events described in detail in Section \ref{sec: sample}. We can calculate the discovery efficiency \cedit{using the Discovery Metric in default mode} by \reedit{computing} the fraction of events with at least 2 points on the rise of the lightcurve with at least a $3\sigma$ difference between the highest and lowest magnitude points.
\cedit{Events that meet this criteria are considered ``discovered"/``detected" (the terms are used interchangeably).}
We can also specify a number of days before the peak time that the event must be ``triggered" by, since ensuring the observation of the peak in follow-up is important for lightcurve characterization. (Note that while this functionality is in the metric and is important for event follow-up, including exoplanet and black hole candidates, an exploration of alerting efficiency is beyond the scope of this paper). \cedit{The Discovery Metric can also compute the fraction of events with at least 2 points on the rise and fall of the lightcurve with at least a $3\sigma$ difference in magnitude. This serves a similar function to the Npts and Fisher Metrics described below, so we do not explore the results of that metric in this paper.} \cedit{The default mode of the Discovery Metric} is based on a common first step in detecting microlensing events \citep[e.g.][]{Mroz:2017}. These typically use 3 points above baseline, but we found minimal difference in relative OpSim performance between requiring 2 or 3 points. 

\begin{figure}
    \centering
    \includegraphics[scale=0.9]{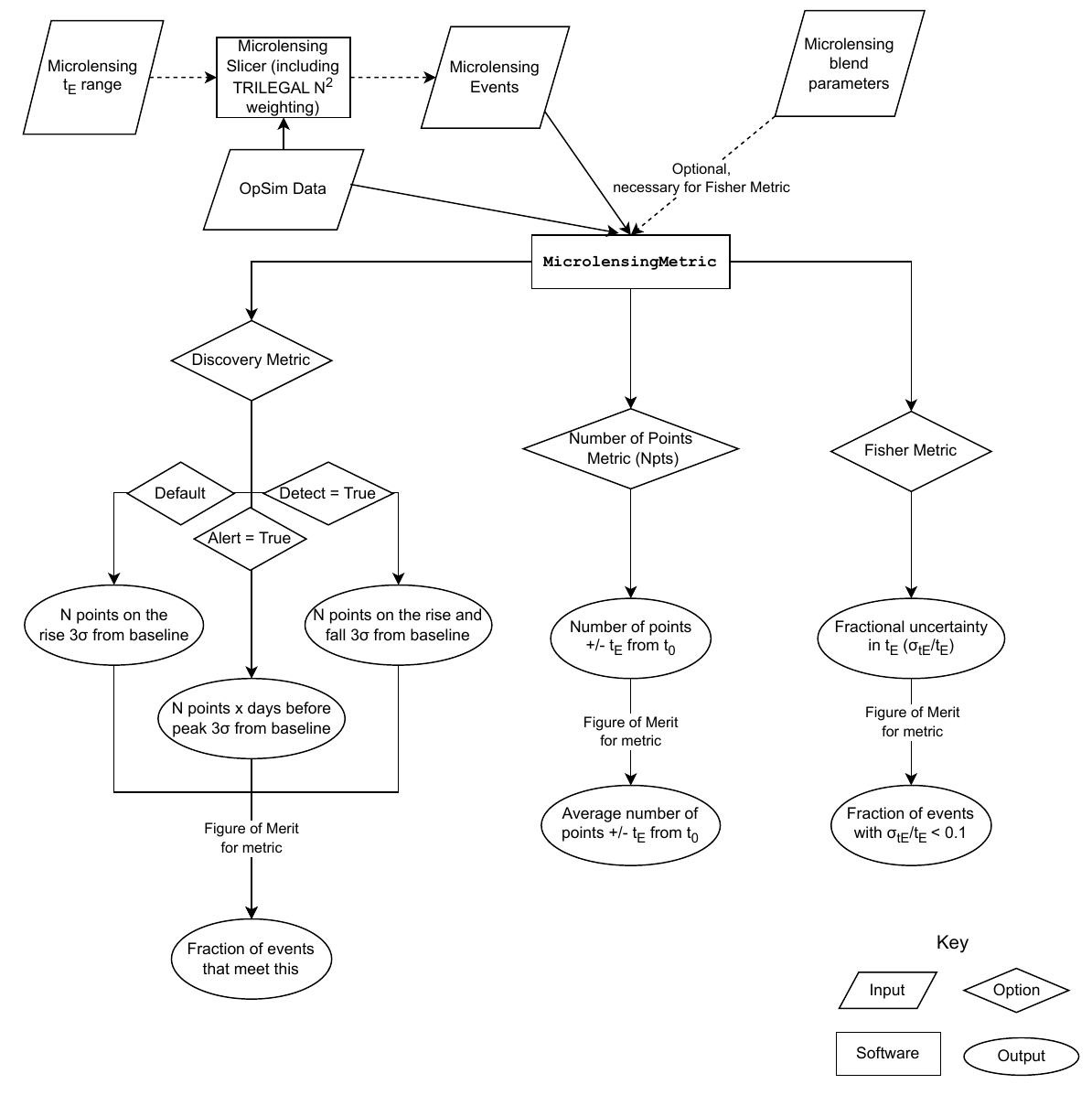}
    \caption{Flowchart of the \texttt{MicrolensingMetric} functionality. The user can either input a $t_{\rm E}$ range and generate microlensing events with a function known as a ``Slicer", or the user can input the events. The user can then choose to use the Discovery Metric, Npts Metric, or Fisher Metric. The Discovery Metric, by default, finds the fraction of events with N points on the rising side with at least a 3$\sigma$ difference from baseline. The user can select detect=True to find at least N points on both the rising and falling side or alert=True to require the N points on the rising side to be a certain number of days before the peak. \cedit{In this paper, we only explore the default mode of the Discovery Metric.} Npts Metric returns the number of points within $t_{\rm 0} \pm t_{\rm E}$ \cedit{with SNR $> 3\sigma$} and Fisher Metric returns the fraction of events with a fractional $t_{\rm E}$ uncertainty $< 0.1$.}
    \label{fig:metric schema}
\end{figure}

There are also two metrics for the characterization of the lightcurves purely from Rubin observations. There is a basic metric that quantifies the number of points (Npts) within $t_{\rm 0} \pm t_{\rm E}$ \reedit{that} estimates the coverage of the lightcurve. \edit{We require that the signal to noise ratios (SNR) of the points are above a 3$\sigma$ threshold. The SNR is calculated using the \texttt{rubin\_sim} function \texttt{m52snr()}, which requires the magnitude and 5$\sigma$ depth and takes into account a number of effects including weather and instrumental effects.} Npts Metric was used as a proxy for characterization and figure of merit is the fraction of events with at least 10 points \cedit{(w/ SNR $> 3\sigma$)} within $t_{\rm 0} \pm t_{\rm E}$. \edit{Requiring 10 points as a proxy is rooted in inspection of microlensing lightcurves \citep[e.g. from inspecting Zwicky Transient Facility lightcurves, it was determined that 10 points were necessary to establish a detection][]{Medford:2023}}. There is also a metric that calculates the Fisher matrix for each event and returns the fractional uncertainty in $t_{\rm E}$ (see Section \ref{sec: fisher matrix}). See Figure \ref{fig:metric schema} for a summary of the \texttt{MicrolensingMetric} functionality.

\reedit{When discussing the results of these metrics, we may roughly refer to regions in the sky that are affected differently by OpSims. For reference, the Galaxy is $|b|$ $< 10^{\circ}$, the Galactic bulge is $|l|$ $<20^{\circ}$, and the Galactic plane is the l values outside that (though some of this is not observable by Rubin).}

\subsubsection{Fisher Matrix}
\label{sec: fisher matrix}
To characterize events, we want to ensure an adequate photometric cadence during the information-bearing part of the lightcurve. This is essential for determining microlensing parameters, such as $t_{\rm E}$, \reedit{that} allow us to infer physical properties of the lens and source populations. We adapt the Fisher matrix approach, which is widely used in many fields such as cosmology \citep[e.g.][]{jungman1996,albrecht2006}. To quantify how precisely parameters can be recovered, we use a fiducial model, namely \cedit{assuming} that our simulated parameters correspond to the actual parameter estimates. According to the Cramer-Rao inequality, the Fisher matrix allows us to calculate a lower bound on the uncertainty. 

In the MAF, since we have simulated the microlensing event population, we can use their known parameters to evaluate the Fisher matrix
\begin{equation}
I_{i,j} = \sum_{k=1}^{N_{\rm{data}}} \frac{1}{\sigma^2_F(t_k)}\left(\frac{\partial F_{\mathrm{model}}(t_k)}{\partial p_i}\right)\left(\frac{\partial F_{\mathrm{model}}(t_k)}{\partial p_j}\right)
\label{eq:fisher_matrix}
\end{equation}
where $p_{i}$ and $p_{j}$ denote the event parameters: $t_{\rm E}, t_{0}, u_{0}$, and the blend and baseline flux parameters for each passband; $t_k$ is each time of observation; $N_{\rm data}$ is the length of the entire dataset (including all passbands); $F$ is the flux; and $\sigma_F$ is the error in the flux. Assuming Gaussian errors on each observable, the Fisher information matrix is approximately the inverse of the covariance matrix. One element of this matrix will be the uncertainty in $t_{\rm E}$, so we can calculate the fractional uncertainty $\sigma_{t_{\rm E}}/t_{\rm E}$. We treat an event as well characterized if $\sigma_{t_{\rm E}}/t_{\rm E} <  10\%$. This threshold indicates if we can constrain the lens mass or if its error budget is dominated by the unknown $t_{\rm{E}}$. An advantage of using a Fisher information matrix is that it can account for the contribution of both the lightcurve coverage and the uncertainty of the observations,  $\sigma^2_F(t_k)$, to the uncertainties of the model parameters.

We can evaluate the Fisher matrix by taking analytic derivatives with respect to each of the parameters ($t_{\rm E}$, $t_0$, $u_0$, $F_{\rm S}$, and $F_{\rm B}$) using \texttt{SymPy} \citep{SymPy}.
Speeding up the calculation of the \texttt{MicrolensingMetric} is key for evaluating many OpSims, and Eq.~\ref{eq:fisher_matrix} is the best suited approximation since it only relies on first derivatives. To optimize the evaluation of the analytic Fisher metric, we use the common subexpression elimination part of \texttt{SymPy} jointly finding suitable substitutions for the parameters \edit{to reduce the number of coefficients}. \edit{We then use the analytic equations that \texttt{SymPy} had optimized to carry out the calculation.} 

\subsection{Sample of Microlensing Events}
\label{sec: sample}
In order to cover the phase space of microlensing events in a heuristic way, we simulate across the whole sky in HEALPIX\footnote{\reedit{A division of the spherical sky into surface area pixels (see \url{https://healpix.sourceforge.io/}).}} and weight the probability of having a microlensing event at a particular RA and Dec with the number of stars squared in the TRILEGAL stellar map simulated for LSST \citep{DalTio:2022-LSST-TRILEGAL}. 
The number of events should scale with the square of the visible density which traces the square of all compact objects. \edit{While the purpose of this paper is to compare between OpSims, if one were to determine the number of detectable or characterizable microlensing events, one could use a compact object tracer and source star tracer  \citep[e.g.][]{Poleski:2016} or carry out a full mock-microlensing survey using a Galactic model \citep[e.g.][]{Lam:2020}. A full simulation is beyond the scope of this paper, but will be subject of future work (see Section \ref{sec: end-to-end pipeline}).}

For each metric, we split the population into $t_{\rm E}$ bins and generated 10,000 events for each $t_{\rm E}$ bin. The events were simulated with uniform distributions of $t_{\rm E}$ from minimum to maximum $t_{\rm E}$ in the bin and $t_0$ from the minimum to maximum observation date in the given OpSim. 
For the discovery metric we make 9 $t_{\rm E}$ bins: 1-5 days, 5-10 days, 10-20 days, 20-30 days, 30-60 days, 100-200 days, 200-500 days, and 500-1000 days. For the Npts and Fisher metrics we analyze a subset of these bins for computational efficiency (10- 20 days, 20-30 days, 30-60 days, and 200-500 days).
We break the events up into $t_{\rm E}$ bins so that we can analyze different populations of objects separately. 
Since $t_{\rm E} \propto \frac{\sqrt{M_{\rm lens}}}{\mu_{\rm rel}\sqrt{\pi_{\rm rel}}}$, various $t_{\rm E}$ \reedit{values} are related to different populations of objects, from low mass stars and free floating planets with short $t_{\rm E}$, to black holes with long $t_{\rm E}$. \cedit{Since} longer $t_{\rm E}$ events are more likely to be \cedit{detected and characterized} than shorter $t_{\rm E}$ events (see Figure \ref{fig:baseline_metrics}), and a rolling cadence (which is where Rubin focuses on a fraction of the sky in alternating years) could leave large parts of long $t_{\rm E}$ events unobserved (see Section \ref{sec:rolling cadence}). So if we evaluated events with $1 < t_E < 1000\,{\rm d}$ together, then we would not see the strength of the effect of changing the cadence.
The events were also simulated with uniform distribution of $u_0$ from 0 to 1. While Rubin will likely detect events with lower magnifications than this, as OGLE and other surveys do, this is an approximation to compare between OpSims. 

To simulate a typical star as the source of our microlensing events, we found the mean stellar magnitude in each filter of the TRILEGAL map and used that as the source magnitude for all the events. These values are: $u$: 25.2, $g$: 25.0, $r$: 24.5, $i$: 23.4, $z$: 22.8, $y$: 22.5. Note these are close to the detection limit of Rubin. \reedit{Using a brighter star does not change the overall effects of cadence in most cases. However, in the case of exposure time, using a faint star can lead to an effect where it is detected in longer exposures, but not in shorter ones leading to longer exposures being preferred. Whereas, if it had been brighter, shorter exposures mean \cedit{there would be} more time for more data points, so shorter exposures are preferred. Given this limitation in this paper, we do not include an assessment of exposure time.} In general, brighter stars are less strongly affected by the cadence as they are over the detection threshold a higher fraction of the time. In addition, using \cedit{a single set of assumed magnitudes} leads to the limitation that we cannot effectively assess filter balance since it would only be affected by the single magnitudes used.

For the Fisher matrix calculation (see Section \ref{sec: fisher matrix}), it is important to take blending into account. A high blend fraction can \reedit{increase the uncertainty} of parameters, since for blended events the apparent baseline becomes brighter and the blend fraction, $t_{\rm E}$, and $u_{0}$ are degenerate \citep[e.g.][]{Yee:2012}. We estimated that in the locations of high stellar density where most of these events occur that the blend fraction is $\sim 50\%$, \cedit{i.e.} the flux from neighboring stars + the lens ($F_{\rm B}$) is approximately equal to the source flux ($F_{\rm S}$) \citep[see Figure 3 of][]{Tsapras:2016}. \edit{This blend fraction of $\sim 50\%$ is used in our Fisher matrix calculation.}

\subsection{Parallax characterization}
\label{sec: parallax characterization methods}
The methods for evaluating the effect of the cadence on microlensing in Section \ref{sec: microlensing metric} did not include the parallax effect. Characterizing the microlensing parallax is important for inferring the mass and nature of the lenses (see  Eq.~\ref{eq:mass_measurement}; see Figure 13 of \cite{Lam:2020}). Hence, on a subset of OpSims we simulated 100,000 events in a representative bulge and disk field including parallax to determine how the characterization of lightcurves including parallax are affected by cadence. We are particularly interested in how \reedit{a rolling cadence} affects the characterization of microlensing parallax, as this is a periodic effect in long enough events.

We determined how well we could characterize each event by taking numerical derivatives of each of the parameters ($t_{0}, t_{\rm E}, u_0, \pi_{E, E}, \pi_{E, N}$, blend parameter, and source magnitude) and applied Eq. \ref{eq:fisher_matrix} to determine the Fisher information matrix. Numerical derivatives were computed by simulating models where the parameters differed by a tolerance of (0.01 $\times$ value of the parameter) and the slope was calculated. The errors on the magnitude of each observation were determined using the \texttt{calc\_mag\_error\_m5()} in \texttt{rubin\_sim}. The lightcurves were modeled using Bayesian Analysis of Gravitational Lensing Events (\texttt{BAGLE})\footnote{\url{https://github.com/MovingUniverseLab/BAGLE_Microlensing/tree/main}}, an open-source microlensing event modeling and fitting code.

We simulated two patches of 100,000 events, one at RA = 263.89$^{\circ}$, Dec = -27.16$^{\circ}$ \reedit{(l = 0.33$^{\circ}$, b = 2.82$^{\circ}$)}, and the other at RA = 288.34$^{\circ}$, Dec = 9.66$^{\circ}$ \reedit{(l = 44.02$^{\circ}$, b = -0.42$^{\circ}$)}. The first is a representative bulge field and the second is a representative disk field in one of the pencil beam fields described in \cite{Street:2023}. We used the observations in a square field of view 3.5$^{\circ}$ across to mimic a Rubin field of view. The events were simulated with uniform distributions of $u_0$ from -1 to 1, log($t_{\rm E}$) from \reedit{0.70 to 2.78 (i.e. 5 to 600 days)}, $t_0$ from the minimum to maximum observation date in the given OpSim, log($\pi_{\rm E}$) from \reedit{-2 to 0 (i.e. 0.01 to 1)}, and $\phi$ from 0 to $2\pi$. Where $\phi$ determines the direction of the parallax by
\begin{eqnarray}
    \pi_{\rm E, N} &=& \pi_{\rm E}\sin(\phi) \\
    \pi_{\rm E, E} &=& \pi_{\rm E}\cos(\phi).
\end{eqnarray}

\begin{figure}
    \centering
    \hspace*{-1.6cm}
    \includegraphics[scale=0.4]{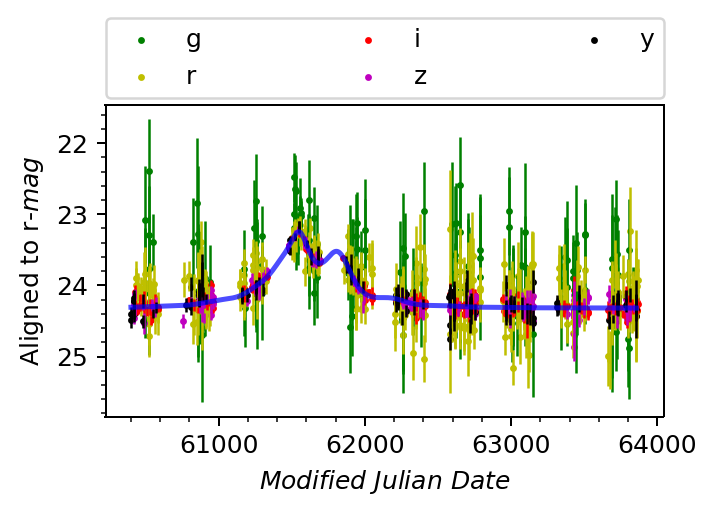}
    \caption{Example plot of a characterizable event with $t_{\rm E} = 161$ days, $\pi_{\rm E} = 0.21$, \reedit{$u_0 = 1.5$, and $\phi = 0.79$. It was simulated in a Galactic disk patch (RA = 288.34$^{\circ}$, Dec = 9.66$^{\circ}$) with \texttt{baseline\_v3.0\_10yrs}.}
    Details of the simulation are described in Section \ref{sec: parallax characterization methods}. The parallax is characterized with a relative uncertainty of $\sigma_{\pi_{\rm E}}/\pi_{\rm E} = 0.06$ and the Einstein time is characterized with $\sigma_{t_{\rm E}}/t_{\rm E} = 0.06$. As is standard for achromatic microlensing events, all lightcurves are aligned and rescaled to the best data set. In this case, the r band serves as reference baseline.
    }
    \label{fig:example_characterized}
\end{figure}

\reedit{Note that, in reality, $\phi$ is not uniform and is dependent on Galactic location due to the spatial and velocity distributions of the Galaxy. A variation of characterization depending on $\phi$ changes on the $\sim 15\%$ level, but this is constant across all evaluated OpSims, so does not affect our relative comparisons that are done on individual Galactic patches.}

\reedit{Instead of a single average source magnitude, as was used for the metrics in the \texttt{MicrolensingMetric}, here we used a distribution of magnitudes. This is because brighter stars enable events to be more easily characterized, so we are better able to explore the entire parameter space where $t_{\rm E}$ and $\pi_{\rm E}$ may be characterized.
We selected $10^6$ random stars in each representative patch with an r-mag brighter than 24.03 (the single image $5\sigma$ r-mag depth\footnote{\url{https://www.lsst.org/scientists/keynumbers}}) in the TRILEGAL LSST simulation \citep{DalTio:2022-LSST-TRILEGAL}. We then randomly matched a star to a microlensing event and kept the star's color.}

The criteria for characterization of a lightcurve is for both $t_{\rm E} > 2\sigma_{t_{\rm E}}$ and $\pi_{\rm E} > 2\sigma_{\pi_{\rm E}}$. \reedit{The criteria on the characterization of each parameter here is less strict than the one in MAF ($\sigma_{t_{\rm E}}/t_{\rm E} < 0.1 \rightarrow t_{\rm E} > 10\sigma_{t_{\rm E}}$) since we are characterizing two parameters instead of just one.} An example characterized event can be seen in Figure \ref{fig:example_characterized}. These results are  outside the context of the \texttt{MicrolensingMetric} and explored in Section \ref{sec: parallax characterization}.

\section{Results}
\label{sec: results}
 We will describe the results of the \texttt{MicrolensingMetric} for each of the families of OpSims. See Table \ref{tab:opsim_summary} for descriptions of the OpSims discussed here with descriptions of their relevance to microlensing and Galactic science. In this simulation, the Discovery Metric, Npts Metric, and Fisher Metric were all run. The Discovery Metric was configured such that 2 points $3\sigma$ above baseline were required on the rising side of the lightcurve. When we refer to the discovery efficiency, this refers to the fraction of simulated events that meet the Discovery Metric criteria, and when we refer to characterization efficiency, this refers to the fraction of simulated events that meet the Npts or Fisher Metric criteria. The retro baseline is described in detail in \cite{Jones_baseline_2018-2020} and the v2.0-v3.0 are described in detail in a Rubin technical note \href{https://pstn-055.lsst.io/}{PSTN-055} (where \texttt{baseline\_v3.0\_10yrs} is the same as \texttt{draft2\_rw0.9\_uz\_v2.99\_10yrs}).

In general, the larger the footprint dedicated to the Galactic bulge and plane, the more microlensing events we can see and characterize. There is also a trend where longer duration microlensing events are less affected by observing strategy since they last long enough that most strategies eventually accumulate enough observations\edit{; however}, the exact cadence of observations is still important, especially for characterization of $t_{\rm E}$ and $\pi_{\rm E}$. We quantify how well the metrics perform by comparing their performance to the \texttt{baseline\_v2.0\_10yrs} OpSim. The results of the sample without parallax analyzed \edit{with} the \texttt{MicrolensingMetric} are in Sections \ref{sec: baseline metrics}-\ref{sec:rolling cadence}, and the parallax characterization metric results are in Section \ref{sec: parallax characterization}. \edit{A few OpSim families that include additional surveys unrelated to microlensing or do not have a strong effect on microlensing were analyzed with the \texttt{MicrolensingMetric} are included in Appendix \ref{appendix: additional opsims}.} 
We do not include \cedit{OpSims} that explicitly vary the filter balance or exposure length due to using a single set of average  stellar magnitudes (see Section \ref{sec: sample}). All results from the \texttt{MicrolensingMetric} are summarized in Table \ref{tab:detect_metric} (\cedit{Discovery} Metric) and Table \ref{tab:char_metrics} (Npts and Fisher Metrics).

\subsection{Baseline Family}
\label{sec: baseline metrics}
\reedit{The \texttt{baseline} strategies show the general evolution over time of the survey and the rest of the OpSims are variations on these baselines. In most of this paper, we compare OpSims to \texttt{baseline\_v2.0\_10yrs}, which includes the Galactic bulge and parts of the Galactic plane at a WFD cadence. It does not perform rolling \cedit{(see Section \ref{sec:rolling cadence} for a detailed definition)} in the Galaxy and does perform rolling in the rest of the sky in years 2-9 of the survey.} Most estimates of the number of microlensing events Rubin is expected to discover are based on the \texttt{baseline\_2018a} OpSim \citep[i.e.][]{sajadian2019lsst}, \reedit{which does not include the Galactic bulge or denser parts of Galactic plane}. Along with \cedit{updating }the survey strategy, the software to simulate survey strategies has changed, so instead of using \texttt{baseline\_2018a}, we use \texttt{retro\_baseline\_v2.0\_10yrs}, which uses a similar observation strategy as \texttt{baseline\_2018a}, but implemented with updated software and updated throughput and weather inputs. 

\begin{figure}
    \centering
    \hspace*{-1.6cm}
    \includegraphics[scale=0.3]{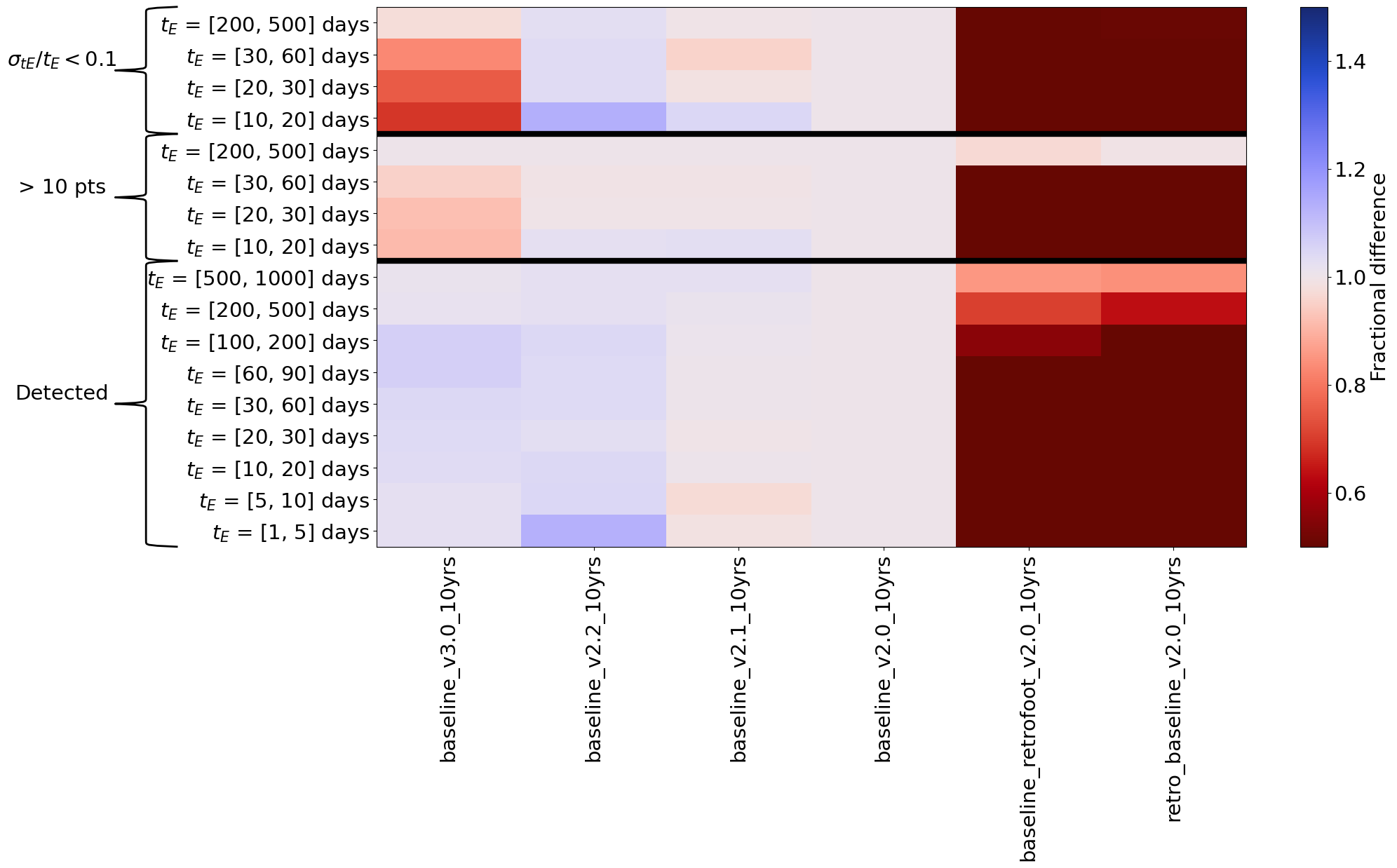}
    \caption{Comparison between \texttt{MicrolensingMetric} results for several distinct iterations of the Rubin baseline strategy. The colors show the fractional improvement relative to \texttt{baseline\_v2.0\_10yrs}, where blue means the metric has performed better relative to the \texttt{baseline} metric and red means it has performed worse. On the x-axis are each of the OpSims, where \texttt{10yrs} means it was simulated for a 10 year LSST survey. On the y-axis are each of the metrics for a $t_{\rm E}$ range compared to their values for \texttt{baseline\_v2.0\_10yrs}. Shorter events (smaller $t_{\rm E}$) correspond \reedit{to} less massive objects such as brown dwarfs and low-mass stars, whereas longer events (larger $t_{\rm E}$) \cedit{can} correspond to black holes. \reedit{The y-axis groups each set of metrics and each line is a different $t_{\rm E}$ range. The black horizontal lines indicate the separate metrics. ``$\sigma_{tE}/t_{\rm E} < 0.1$" refers to the Fisher Metric; ``$> 10$ pts" refers to the Npts Metric, and ``Detected" refers to the Discovery Metric. For example, ``$\sigma_{tE}/t_{\rm E} < 0.1$ $t_{\rm E}$=[30, 60] days"} refers to the fraction of the 10,000 events with $30~\mathrm{days} \leq t_{\rm E} \leq 60~\mathrm{days}$ simulated as described in Section \ref{sec: sample} with $\frac{\sigma_{t_{\rm E}}}{t_{\rm E}} < 0.1$ as calculated by the Fisher Information Matrix.  All of the current baselines show a $\gtrsim$ 50\% improvement over the retro baseline.}
    \label{fig:baseline_metrics}
\end{figure}

More recent versions of the baseline (\texttt{v2.0} and higher) of the OpSims lead to a $\gtrsim 50\%$ improvement in both discovery and characterization over the \texttt{retro} baseline due to the inclusion of the Galactic bulge and parts of the Galactic plane. \texttt{baseline\_retrofoot\_v2.0\_10yrs} adopts the old footprint, but uses the v2.0 baseline strategy; whereas \texttt{retro\_baseline\_v2.0\_10yrs} is a version of the retro footprint and strategy. See Figure \ref{fig:baseline_metrics} for a comparison of \cedit{the metrics for} the baseline OpSims. The \texttt{v2.0} performs slightly better than \texttt{v2.1}, since \texttt{v2.1} includes the Virgo cluster which is not a traditional microlensing target and takes time away from other areas. \texttt{v2.2} included optimizations to the code and a change in DDF strategy \reedit{that} should not significantly affect microlensing. \texttt{baseline\_v3.0\_10yrs} spends less time on the galactic bulge and spreads out observations across the plane \citep[see][for detailed discussion]{Street:2023}. Covering this larger area leads to $\sim 10-20\%$ fewer events being characterized (Fisher and Npts metric), but $\sim 5-10\%$ more being discovered (Discovery metric), since there are fewer events in the Galactic plane than the bulge due to the decrease in stellar density. \cedit{We also note that the $t_{\rm E}$ distribution changes as a function of position. Due to changes in relative proper motion, $t_{\rm E}$ is on average longer in the Galactic plane than the bulge \citep{Mroz:2020-galplane, Medford:2023}. We include Table \ref{tab:location breakdown} with results broken down by general location for \texttt{baseline\_v2.0\_10yrs} for reference.} A strategy similar to this could allow Rubin to better probe galactic structure, but may require increased follow-up to characterize the discovered microlensing events.

\subsection{Galactic bulge and plane coverage and footprint}
\label{sec: vary gp}
\begin{figure}
    \centering
    \hspace*{-1.6cm}
    \includegraphics[scale = 0.3]{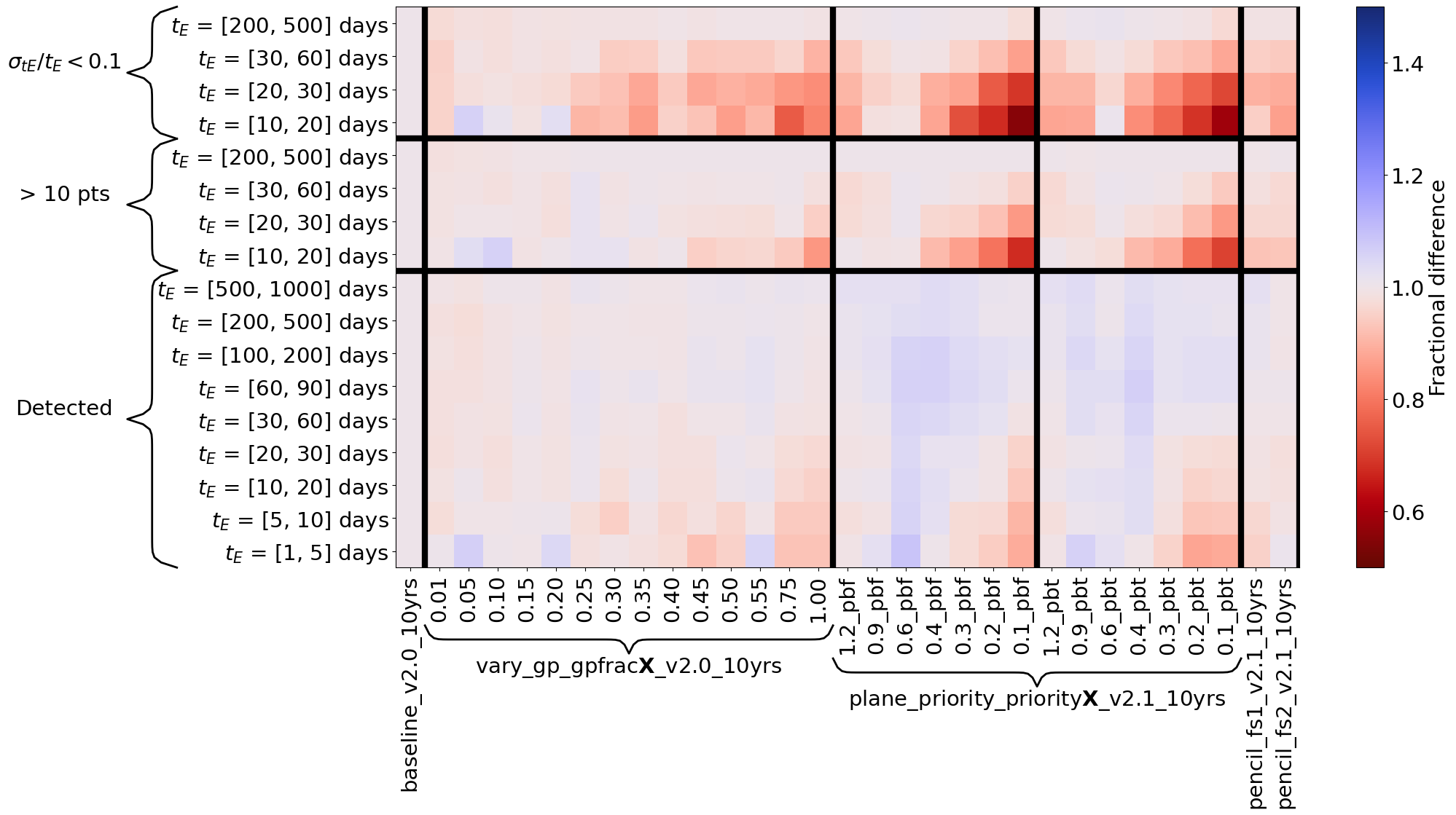}
    \caption{Same as Figure \ref{fig:baseline_metrics} but for the \texttt{vary\_gp}, \texttt{plane\_priority}, \texttt{pencil}, and \texttt{galplane\_priority}, families. \reedit{The colors show the fractional improvement relative to \texttt{baseline\_v2.0\_10yrs}.} Black vertical lines indicate separate OpSim families (summarized in Table \ref{tab:opsim_summary}) \reedit{and black horizontal lines indicate separate metrics}.  Covering regions of lower stellar density and/or high extinction (areas of priority less than 0.4 as defined in \cite{Street:2023}) leads to significant decrease in microlensing characterization efficiency. The \texttt{pbf} strategies do not include pencil beams selected in \cite{Street:2023} of scientific interest; whereas, the \texttt{pbt} strategies include them. While the number of detected and characterized microlensing events does not significantly differ between them, since the pencil beams were specifically chosen to optimize our ability to probe the Galaxy, the strategy is preferable. The size of the pencil beams does not appear to significantly affect microlensing efficiency. 
    }
    \label{fig:vary_gp and plane_priority metrics}
\end{figure}

\begin{figure}
    \centering
    \hspace*{-0.8cm}
    \includegraphics[scale = 0.5]{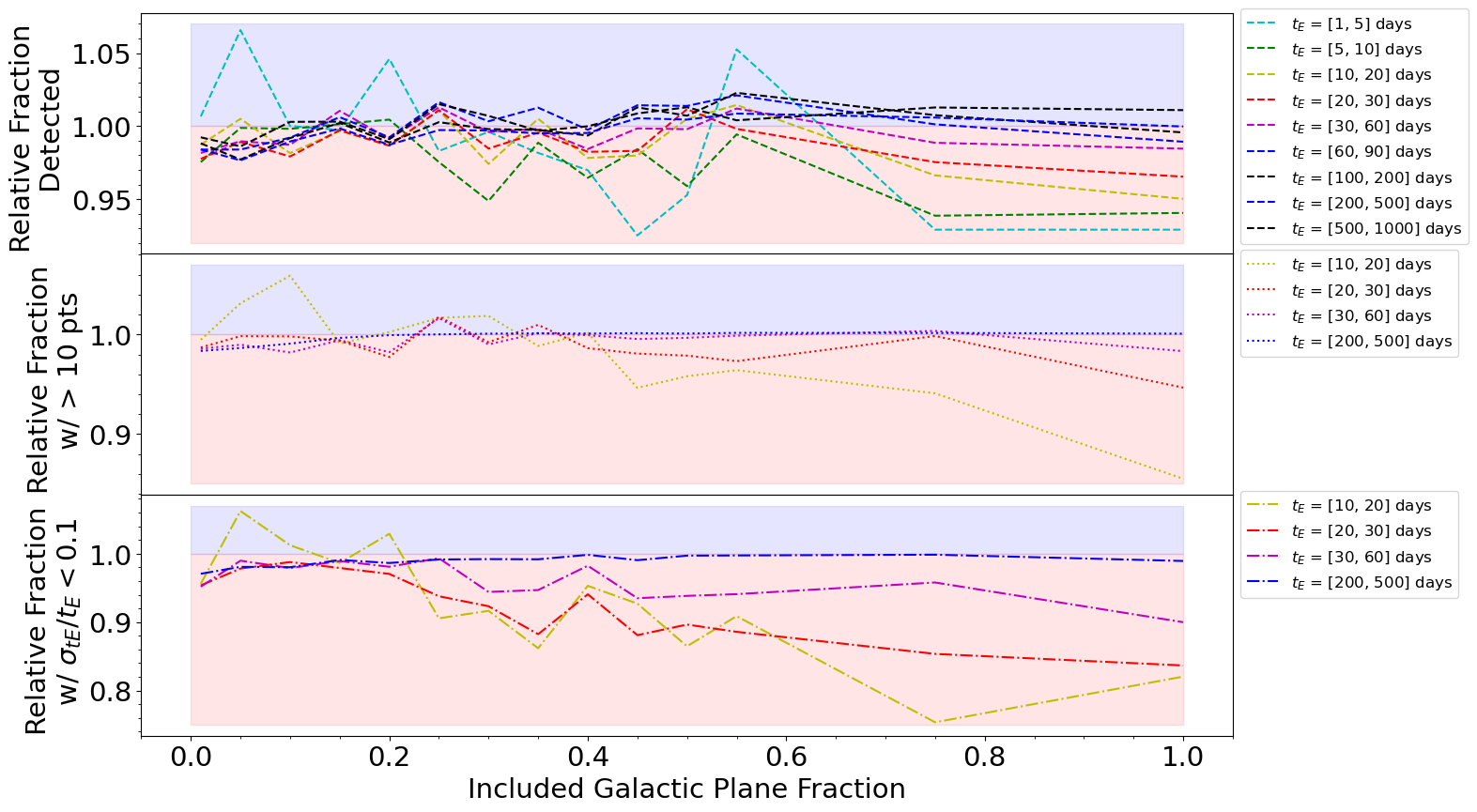}
    \caption{\reedit{Plot of fraction of events detected (top), fraction of events with at least 10 points (middle), and fraction of events with at $\sigma_{\rm tE}/t_{\rm E} < 0.1$ (bottom), relative to \texttt{baseline\_v2.0\_10yrs} as a function of the fraction included of the Galactic plane. These correspond to the results from the \texttt{vary\_gp} family. The colors of the lines correspond to the timescale of the events and the linestyle corresponds to the type of metric (dashed for Detect, dotted for Npts, and dot-dash for Fisher). If the line is in the blue shaded region, this indicates the OpSim performed better than \texttt{baseline\_v2.0\_10yrs}, and if it is in the red shaded region, it performed worse. In general, shorter $t_E$ events are more affected by the change in cadence including random fluctuations. There is a significant decrease of microlensing characterization above \texttt{gpfrac $\geq 0.55$} since most microlensing events are in the Galactic bulge.}
    }
    \label{fig:vary_gp line plot}
\end{figure}

\edit{There are a number of families of OpSims \reedit{that} explore the Galactic plane coverage and footprint. The \texttt{vary\_gp} family} varies the visits to fields in the Galactic plane as a percentage of the WFD survey from \reedit{\texttt{gpfrac} =} 1-100\%. We see a significant decrease of microlensing characterization in strategies with \texttt{gpfrac} $\geq 0.55$. We see 
that technically if we cover the Galactic plane more, that we characterize fewer microlensing events overall, since many of the microlensing events are concentrated towards the Galactic bulge (see Figure \ref{fig:vary_gp and plane_priority metrics} \reedit{and Figure \ref{fig:vary_gp line plot}}). However, it is scientifically interesting to be able to probe microlensing events throughout the Galactic plane.

\begin{figure}
    \centering
    \hspace*{-0.8cm}
    \includegraphics[scale = 0.5]{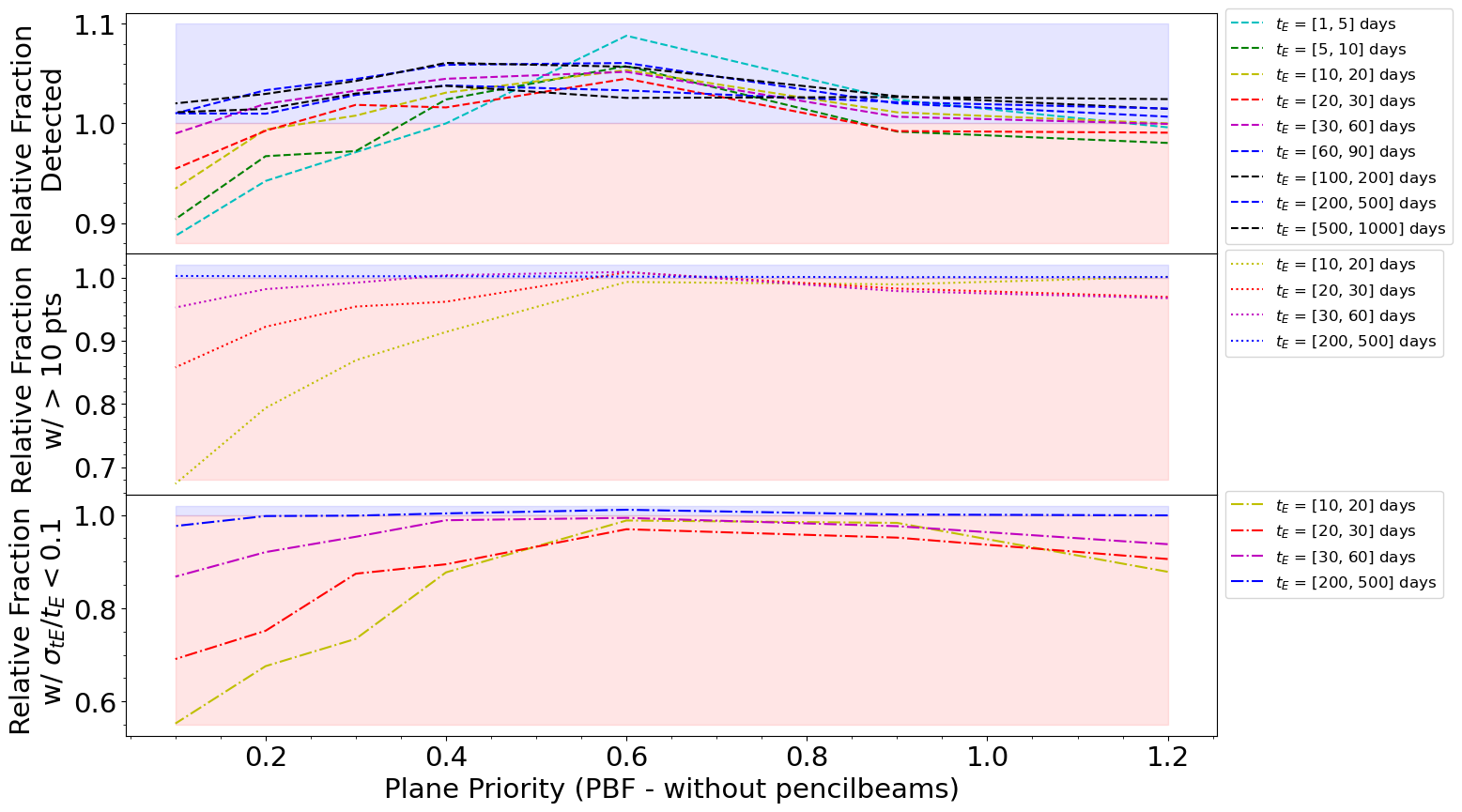}
    \caption{\reedit{Same as Figure \ref{fig:vary_gp line plot}, but for each metric as a function of plane priority without pencil beams (\texttt{pbf} = pencil beams false). If the line is in the blue shaded region, this indicates the OpSim performed better than \texttt{baseline\_v2.0\_10yrs}, and if it is in the red shaded region, it performed worse. As plane priority decreases, this means areas of the Galaxy of lower priority are covered at a WFD-level cadence. The priorities are scientifically motivated regions of interest as defined in \cite{Street:2023}. There is a peak around 0.6, but there is a significant drop when \cedit{areas} of priority 0.4 and lower are covered. This is because a finite number of visits are distributed over too large an area.}
    }
    \label{fig:pp_pbf line plot}
\end{figure}

\begin{figure}
    \centering
    \hspace*{-0.8cm}
    \includegraphics[scale = 0.5]{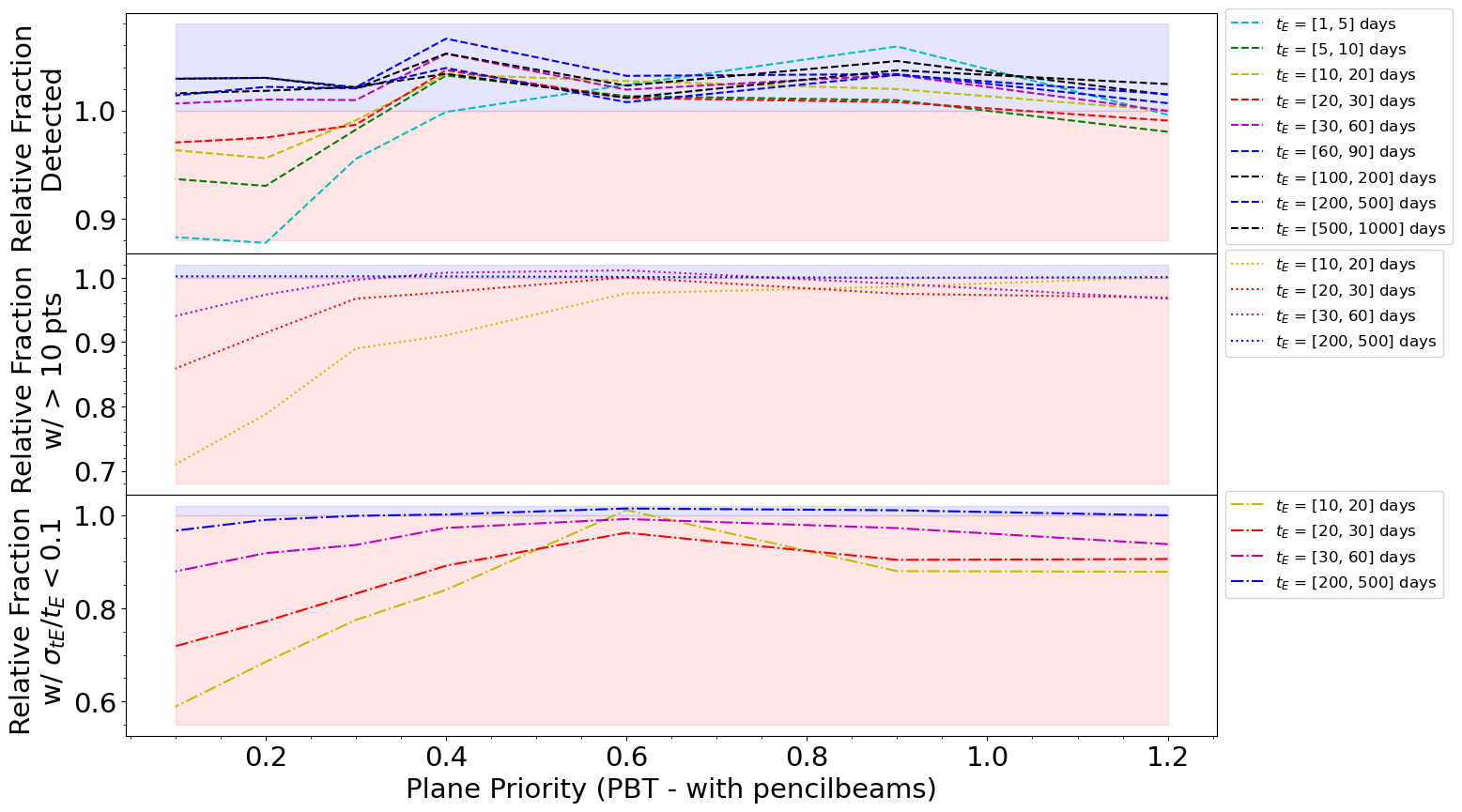}
    \caption{\reedit{Same as Figure \ref{fig:pp_pbf line plot}, but for each metric as a function of plane priority with pencil beams (\texttt{pbt} = pencil beams true). If the line is in the blue shaded region, this indicates the OpSim performed better than \texttt{baseline\_v2.0\_10yrs}, and if it is in the red shaded region, it performed worse. Comparatively, the results are similar to those without pencil beams with a significant drop when areas with priority 0.4 and below are included. Though a similar number of events are detected and characterized, the pencil beam fields were specifically chosen in \cite{Street:2023} to probe Galactic structure (and other scientific goals) throughout the Galactic plane.}
    }
    \label{fig:pp_pbt line plot}
\end{figure}

\edit{Besides time spent on the Galactic plane, we can also probe the optimal Galactic bulge and plane footprint.} \reedit{We used} a map of the Galactic plane with scientifically motivated priorities assigned to each region \citep{Street:2023} ranging between 0.1-1.2. Generally regions of lower stellar density and/or high extinction correspond to lower priority regions. The \edit{\texttt{plane\_priority\_priorityX\_} OpSim family} adds regions of progressively less priority to the WFD footprint, so \texttt{plane\_priority\_priority0.4} includes regions assigned priority $\geq 0.4$. 
We find that there is a drop in characterization efficiency for the long duration events in the plane priority map when it covers regions of priority of ~0.4 or lower (see Figure \ref{fig:vary_gp and plane_priority metrics}), as a finite number of visits are distributed over too large a spacing in time for characterization \reedit{(see Figures \ref{fig:vary_gp and plane_priority metrics}, \ref{fig:pp_pbf line plot}, and \ref{fig:pp_pbt line plot})}. This matches what is found by the general Galactic plane metrics \citep{Street:2023}. 

\edit{In addition to areas ranked by their priority, \cite{Street:2023} defines a series of designated ``pencil beam"} fields selected for their scientific interest. \edit{The \texttt{plane\_priority\_priorityX.X\_pbf\_} do not include those pencil beams \reedit{(pencil beams false)} and \texttt{plane\_priority\_priorityX.X\_pbt\_} include them \reedit{(pencil beams true)}.} We technically find similar results to the OpSims without pencil beams since we detect and characterize a similar number of events (see Figures \ref{fig:vary_gp and plane_priority metrics}, \reedit{\ref{fig:pp_pbf line plot}, and \ref{fig:pp_pbt line plot}}). However, the pencil beam fields were picked specifically to optimize our ability to probe Galactic structure (along with other goals) throughout the Plane. Decades of microlensing surveys have looked at the Galactic bulge \edit{with some surveys delving into the Galactic plane 
(e.g. Gaia \citep{Wyrzykowski:2023}, ZTF \citep{Rodriguez:2022, Medford:2023, Zhai:2023}, and OGLE \citep{Mroz:2020-galplane})}, 
but Rubin will enable us to look much deeper \edit{across a larger area of} the Galactic plane, so looking in strategic spots is helpful. The \edit{\texttt{pencil\_} OpSim family varies} the size of the pencil beam fields. As can be seen in Figure \ref{fig:vary_gp and plane_priority metrics}, the size of the pencil beams does not appear to affect the microlensing results.

\subsection{Image quality}

\edit{The \texttt{good\_seeing\_} OpSim family} adds the requirement of at least 3 good seeing images per year per pointing. \reedit{By adding a requirement, here we mean a requirement is added in the scheduler. So when the scheduler decides what part of the sky to look at, it will attempt to ensure at least 3 good seeing images per year per pointing are obtained. This is balanced against other requirements such as maintaining a particular cadence.} As the good seeing metric is prioritized, detection and characterization metrics worsen on the 10\% level \reedit{relative to \texttt{baseline\_v2.1\_10yrs}} for characterization since it appears as though the footprint decreases and we end up with fewer events (see Figure \ref{fig:v2.1 and later}). However, better template images for DIA could improve alerts and photometric accuracy but there is insufficient data to assess a suitable trade-off.

\subsection{\cedit{Triplet Observations}}
\begin{figure}
    \centering
    \hspace*{-1.6cm}
    \includegraphics[scale=0.3]{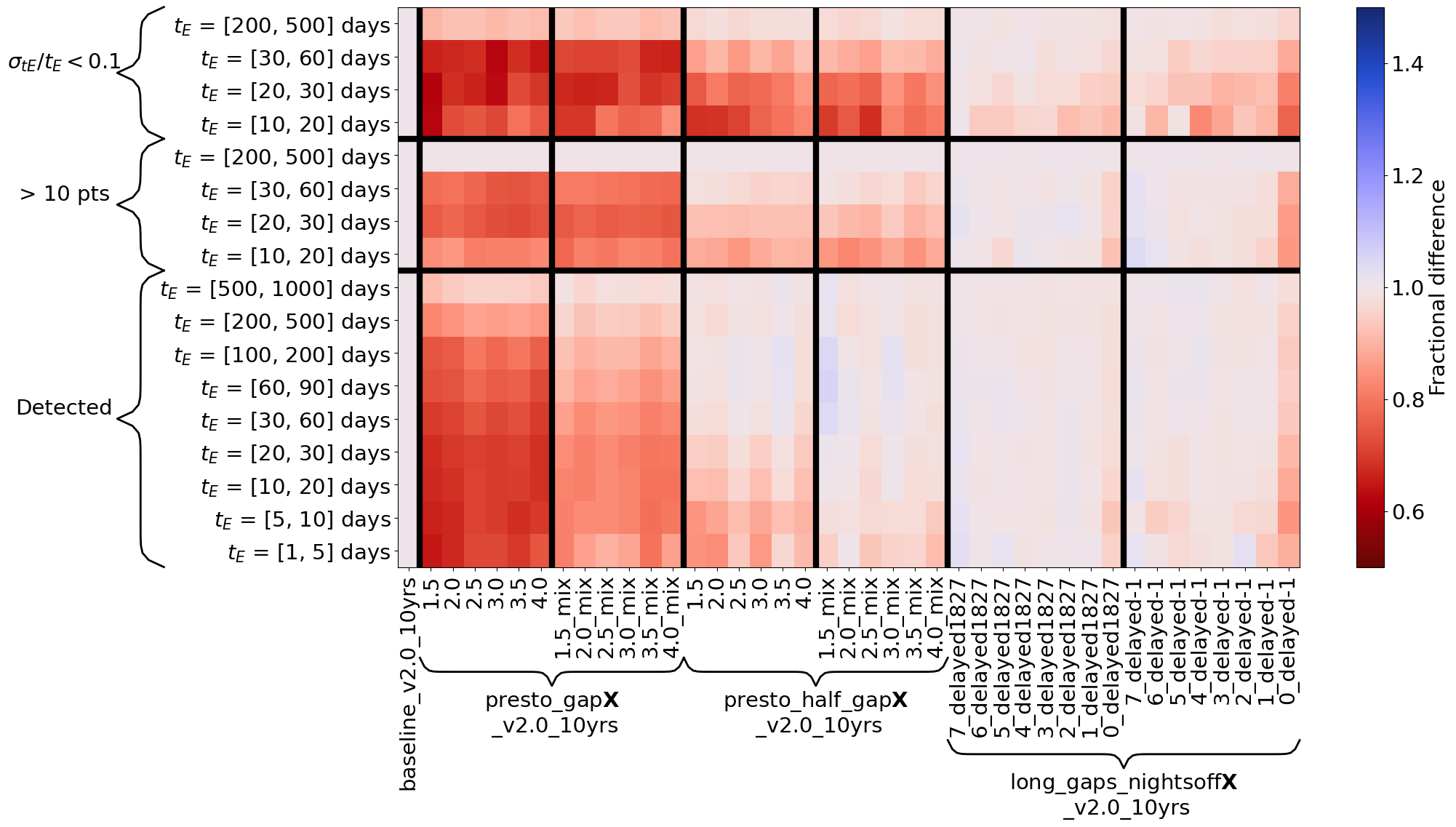}
    \caption{Same as Figure \ref{fig:baseline_metrics} but for the presto family which explores triplets of observations within one night. \reedit{The colors show the fractional improvement relative to \texttt{baseline\_v2.0\_10yrs}.} \edit{Black vertical lines indicate separate OpSim families (summarized in Table \ref{tab:opsim_summary}) \reedit{and black horizontal lines indicate separate metrics}.} Since most microlensing events do not vary significantly over the course of a night, when observations are taken over a less varied time period, this decreases detection and characterization efficiency.}
    \label{fig:presto_metrics}
\end{figure}

The \reedit{\texttt{presto\_gapX}, \texttt{presto\_half\_gapX}, and \texttt{long\_gaps\_nightsoffX}} set of families explore ``triplets" of observations described in detail in \cite{Bianco:2019}. This means there will be a third visit on the same night, (in the case of the \texttt{presto} family, see Figure \ref{fig:presto_metrics}), \reedit{a third visit on the same night after \texttt{X} hours (in the case of the \texttt{presto\_gapX} family), every other night (in the case of the \texttt{presto\_half\_gapX} family)}, or a third visit \reedit{every \texttt{X} nights} (in the case of the \texttt{long\_gaps\_nightsoffX} family).
Microlensing events decrease in discovery and characterization by 10-30\% in the presto family. In general, microlensing events do not change sufficiently in a single night to warrant a third visit that night, and taking time away from looking at more varied points in time greatly decreases the efficacy of microlensing detection and characterization. In some strategies, there is an improvement in discovery to events with $t_{\rm E}$ 1-10 days, which do change at this timescale, but at the large expense of the majority of events ($t_{\rm E} > 10$ days) and to their characterization. There could also be an improvement to events with short-duration features such as microlensing events with a binary lens, though events with such features are a small fraction of events.

\subsection{Microsurveys}
Microsurveys are ``micro" observing surveys that take up to a few percent of the LSST observing time (explored in detail in a Rubin technical note \href{https://pstn-053.lsst.io/}{PSTN-053}). The two microsurveys of relevance for microlensing are \texttt{roman\_v2.0\_10yrs} and \texttt{smc\_movie\_v2.0\_10yrs}. Since the rest of the surveys do not focus on microlensing targets, they only negatively impact microlensing on the 5-10\% level since it takes time away from microlensing targets. See Figure \ref{fig:v2.0 and earlier} for a summary.

The \texttt{roman\_v2.0\_10yrs} microsurvey is designed to look at the footprint of the Nancy Grace Roman Galactic Bulge Time Domain Survey \citep[GBTDS,][]{roman-Spergel:2015, roman-Penny:2019}. Observing the Roman field both during Roman’s $\sim 60-72$ day survey ``seasons" and also filling in the multi-month gaps between its observations would be impactful. During Roman’s observing windows, concentrating more of the Galactic bulge observations on the Roman field could allow for simultaneous observations \reedit{that} could be used to \cedit{measure} satellite parallax \reedit{that} can be used to constrain the mass of the lens \citep{Yee:2014}. The number of increased visits to the Roman field should not be at a level that visits to the rest of the Galactic bulge are significantly reduced, but perhaps $\sim\texttt{rw}0.5$ rolling \cedit{(i.e. the scheduler allocating 50\% more observations to the Roman field than if no rolling were performed)} since that did not seem to significantly negatively impact detection and characterization of events $> 30$ days (see Section \ref{sec:rolling cadence}). The nominal GBTDS is planned to have seasons of $60-72$ days with multiple months-long gaps. While, some of these gaps are at times \cedit{when} Rubin cannot observe the Galactic bulge, there are some where Rubin could fill in Roman gaps. Filling in the photometry is particularly beneficial for characterizing long duration events that span multiple Roman seasons \citep{Lam2023-Roman}. The impact of a lack of space based astrometry during those times, though, is still to be determined. 
Work is in progress to further quantify the synergy between Roman and Rubin; see \cite{Street:2023Roman} for more details.

The \texttt{smc\_movie\_v2.0\_10yrs} includes two nights of high intensity observations of the SMC. Though the \texttt{smc\_movie\_v2.0\_10yrs} survey decreases the characterization fraction of short duration events by 5-10\%, the SMC is a target of scientific interest for microlensing for compact halo objects and to probe galactic structure in a nearby dwarf galaxy. 

\subsection{Rolling Cadence}
\label{sec:rolling cadence}
In OpSims with a rolling cadence, the sky is broken up into \reedit{2-6 regions (indicated by the number following \texttt{ns})} and observations are focused \cedit{on} \reedit{one region each year}, alternating between years. \reedit{These regions are not necessarily continuous.}
\reedit{The strength of the rolling indicates the aggressiveness of reshuffling visits into active or inactive rolling seasons, with higher numbers corresponding to pushing more visits into active rolling seasons. For example, \texttt{rw0.8} or \texttt{strength0.8} means 80\% rolling, so if there were 100 visits in a non-rolling season, the scheduler would try to have 180 visits in an active rolling season and 20 visits in an inactive rolling season. However, since there are other scheduling requirements (i.e. coverage of the WFD footprint with observations with good seeing), it tends to perform rolling less than the programmed strength.} There is already \reedit{a rolling cadence} incorporated in the baseline, though the \edit{\texttt{rolling}, \texttt{roll\_}, \texttt{six\_rolling} OpSim families} explore different configurations of strips and \reedit{the effect of having} a rolling cadence in the Galactic bulge. As can be seen in Figures \ref{fig:v2.0 and earlier} and \ref{fig:v2.1 and later}, the rolling cadence generally does increase discovery of short events ($< 10$ days) by $\sim 5\%$, especially \reedit{OpSims with a rolling cadence in the Galactic bulge}. However, events with $t_{\rm E} > 10$ days have a decrease in detection and characterization. In strategies in which the Galactic bulge and plane are explicitly included in \reedit{areas with a rolling cadence}, detection efficiency decreases by 5-15\% and characterization efficiency decreases by 10-40\%. Most microlensing events have $t_{\rm E} > 10$ days and it was expected that Rubin would alert on shorter duration events, not that it would be able to completely characterize them. 
Beyond \reedit{cadences that explicitly include the Galactic bulge and plane in a rolling cadence}, even if a region is not explicitly part of the rolling cadence, due to slew times and survey efficiency, those regions may also be affected. There is a decrease in characterization efficiency by 10-20\% in \texttt{six\_rolling\_ns2\_rw0.9\_v2.0\_10yrs} (see Figure \ref{fig:v2.0 and earlier}) even though no region with significant microlensing population is part of the \reedit{footprint with a rolling cadence}.
Note that we have not included microlensing parallax here, see Section \ref{sec: parallax characterization}. 

\subsection{Parallax Characterization}
\label{sec: parallax characterization}
\begin{figure}[ht]
\centering
\begin{tabular}{cc}
\hspace*{-0.5cm}
\includegraphics[scale=0.5]{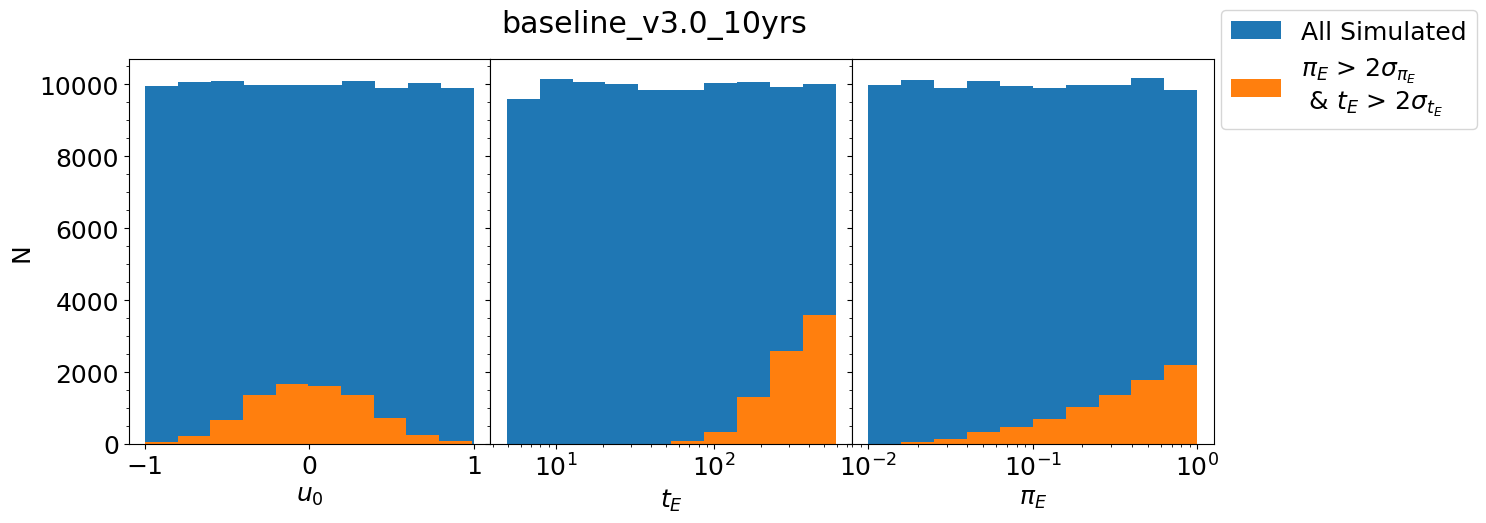}
\end{tabular}
\caption{Distribution of $u_0$, $t_{\rm E}$, and $\pi_{\rm E}$ for simulated and characterized events. In blue are all of the simulated events and in orange are those that are characterized ($\pi_{\rm E} > 2\sigma_{\pi_{\rm E}}$ and $t_{\rm E} > 2\sigma_{t_{\rm E}}$) by the \texttt{baseline\_v3.0\_10yrs} OpSim in the bulge field RA = 263.89$^{\circ}$, Dec = $-27.16^{\circ}$.} \label{fig:parallax char parameters}
\end{figure}

In this section we explore the characterizability of events with a microlensing parallax signal as described in Section \ref{sec: parallax characterization methods} \reedit{($t_{\rm E} > 2\sigma_{t_{\rm E}}$ and $\pi_{\rm E} > 2\sigma_{\pi_{\rm E}}$)}. In Figure \ref{fig:parallax char parameters}, we show a histogram of simulated (blue) and characterized (orange) event parameters. This is for the \texttt{baseline\_v3.0\_10yrs} in a field around RA = 263.89$^{\circ}$, Dec = $-27.16^{\circ}$. We see those with smaller $|u_0|$ are characterized more easily due to their higher magnification. Longer $t_{\rm E}$ are characterized more frequently since events with longer $t_{\rm E}$ are more likely to be covered eventually. Large $\pi_{\rm E}$ are characterized more frequently due to the larger measurable signal.

\begin{figure}[h]
    \centering
    \hspace*{-0.5cm}
    \includegraphics[scale=0.5]{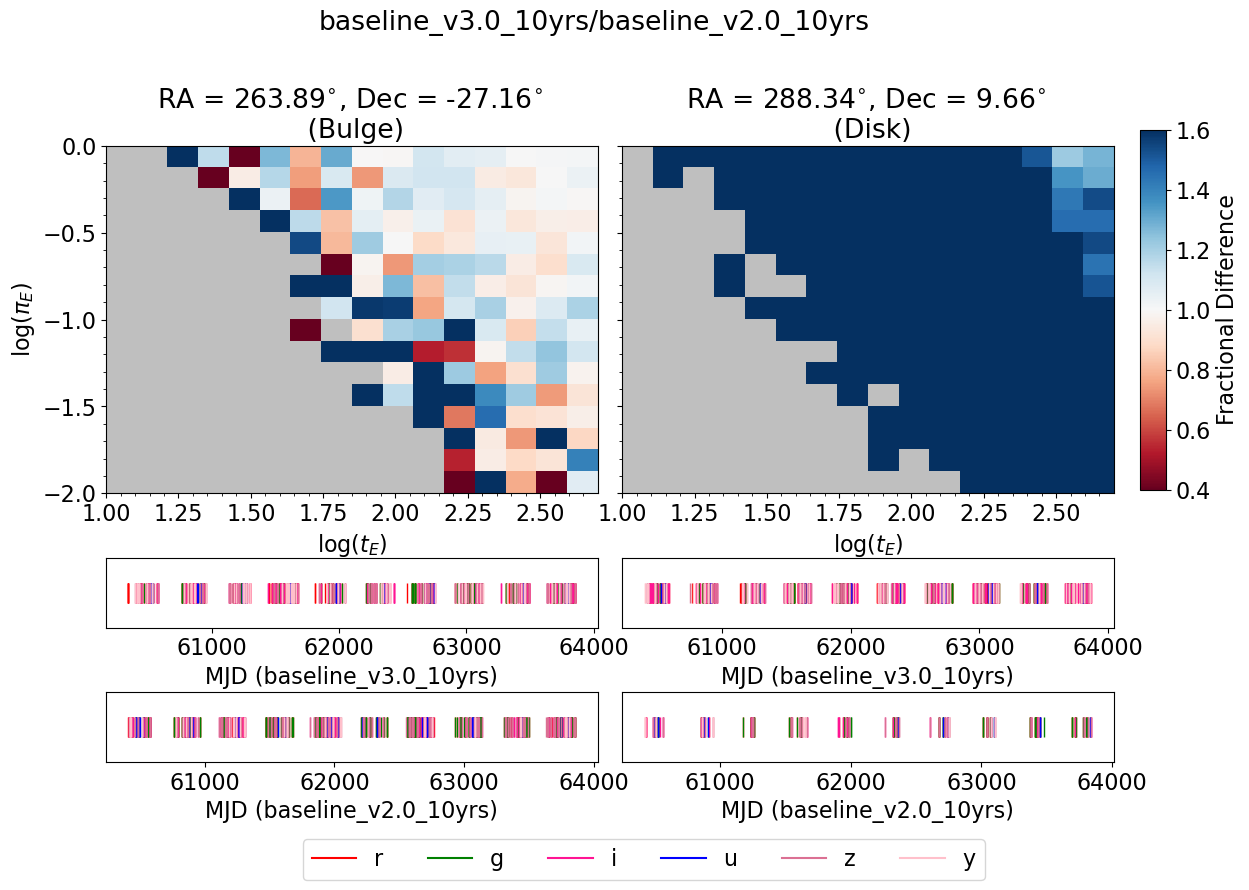}
    \caption{\reedit{2D-histogram comparing the fraction of events that are characterized ($\pi_{\rm E} > 2\sigma_{\pi_{\rm E}}$ \& $t_{\rm E} > 2\sigma_{t_{\rm E}}$) in \texttt{baseline\_v3.0\_10yrs} and \texttt{baseline\_v2.0\_10yrs} (\texttt{baseline\_v3.0\_10yrs}/\texttt{baseline\_v2.0\_10yrs}) for any given $t_{\rm E}$ and $\pi_{\rm E}$. Bluer squares mean that more events were characterized in \texttt{baseline\_v3.0\_10yrs} and redder squares mean more were characterized in \texttt{baseline\_v2.0\_10yrs}. Grey squares means that no events were characterized in either OpSim, in large part due to $\pi_{\rm E}$ being intrinsically hard to measure with a single telescope for short duration events (see Section \ref{sec: parallax characterization}).} The bottom plots show the cadence of observations where different colors represent different filters, as indicated in the legend, for the given OpSim in a $3.5^{\circ} \times 3.5^{\circ}$ square centered on the RA and Dec indicated in the title of the left and right plots. The middle row show\cedit{s} the cadence of the OpSim in the numerator (\texttt{baseline\_v3.0\_10yrs}) and the bottom row shows the cadence of the OpSim in the denominator (\texttt{baseline\_v2.0\_10yrs}). Left is a representative bulge field and right is a representative disk field in a pencil beam. See Figure \ref{fig:popsycle_piE_tE} for a realistic distribution of simulated events in $\pi_{\rm E} - t_{\rm E}$ space. \reedit{Since the \texttt{baseline\_v2.0\_10yrs} footprint included Galactic bulge and neglected the plane, the fraction of events characterized in the disk field event increased by $\sim 40-50$\% in \texttt{baseline\_v3.0\_10yrs} and the fraction of events characterized in the Galactic bulge stayed about the same.}}
    \label{fig:parallax_char_baseline_v3.0}
\end{figure}

\begin{figure}[h]
    \centering
    \hspace*{-0.5cm}
    \includegraphics[scale=0.5]{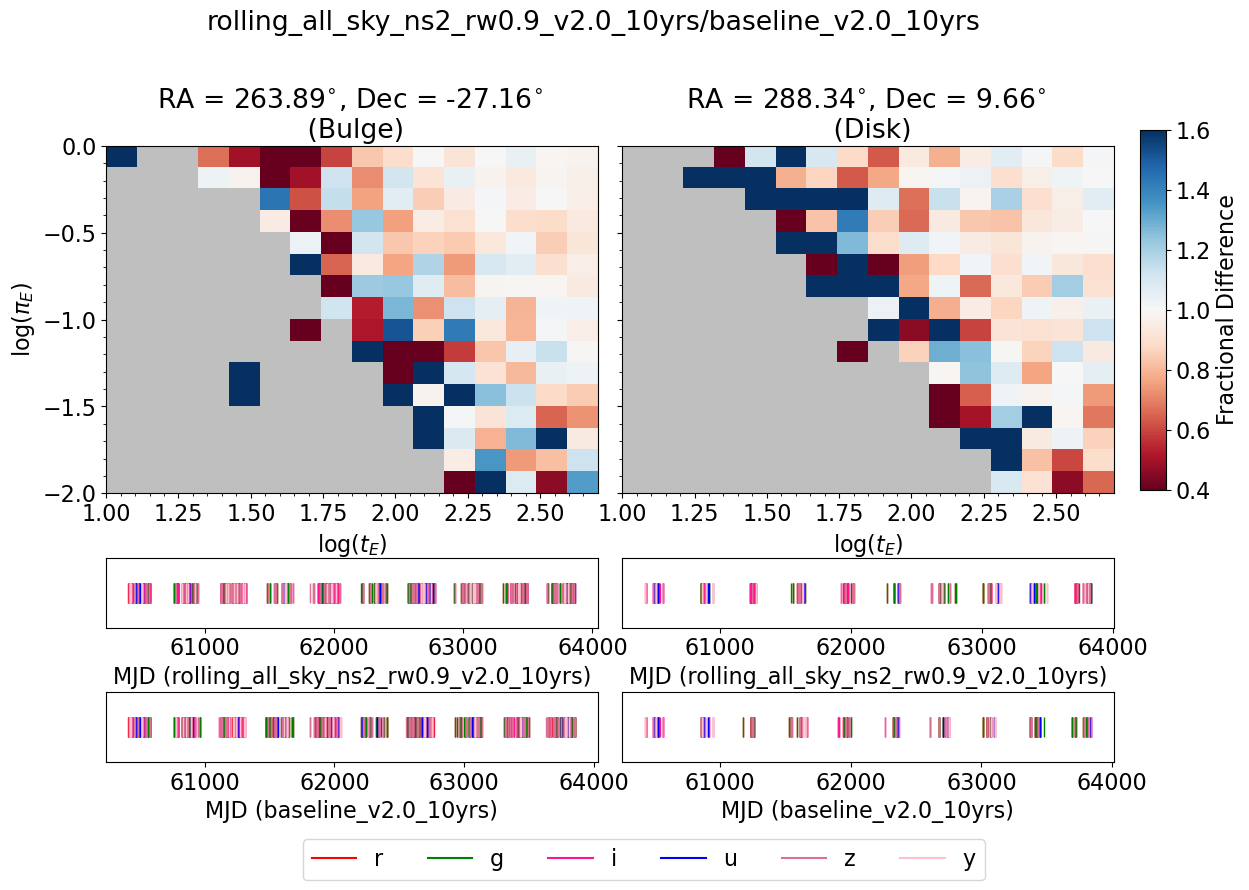}
    \caption{Same as Figure \ref{fig:parallax_char_baseline_v3.0} but \reedit{comparing \texttt{rolling\_all\_sky\_ns2\_rw0.9\_v2.0\_10yrs} to \texttt{baseline\_v2.0\_10yrs}}. This OpSim has the same footprint as \texttt{baseline\_v2.0\_10yrs} but with \reedit{a rolling cadence}. While this does not explicitly \reedit{include the Galactic plane as part of the footprint with a rolling cadence,} \reedit{the fraction of events characterized in the bulge} is $\sim$10\% less than \texttt{baseline\_v2.0\_10yrs}.}
    \label{fig:parallax_char_rolling_all_sky_v2.0}
\end{figure}

\begin{figure}[h]
    \centering
    \hspace*{-0.5cm}
    \includegraphics[scale=0.5]{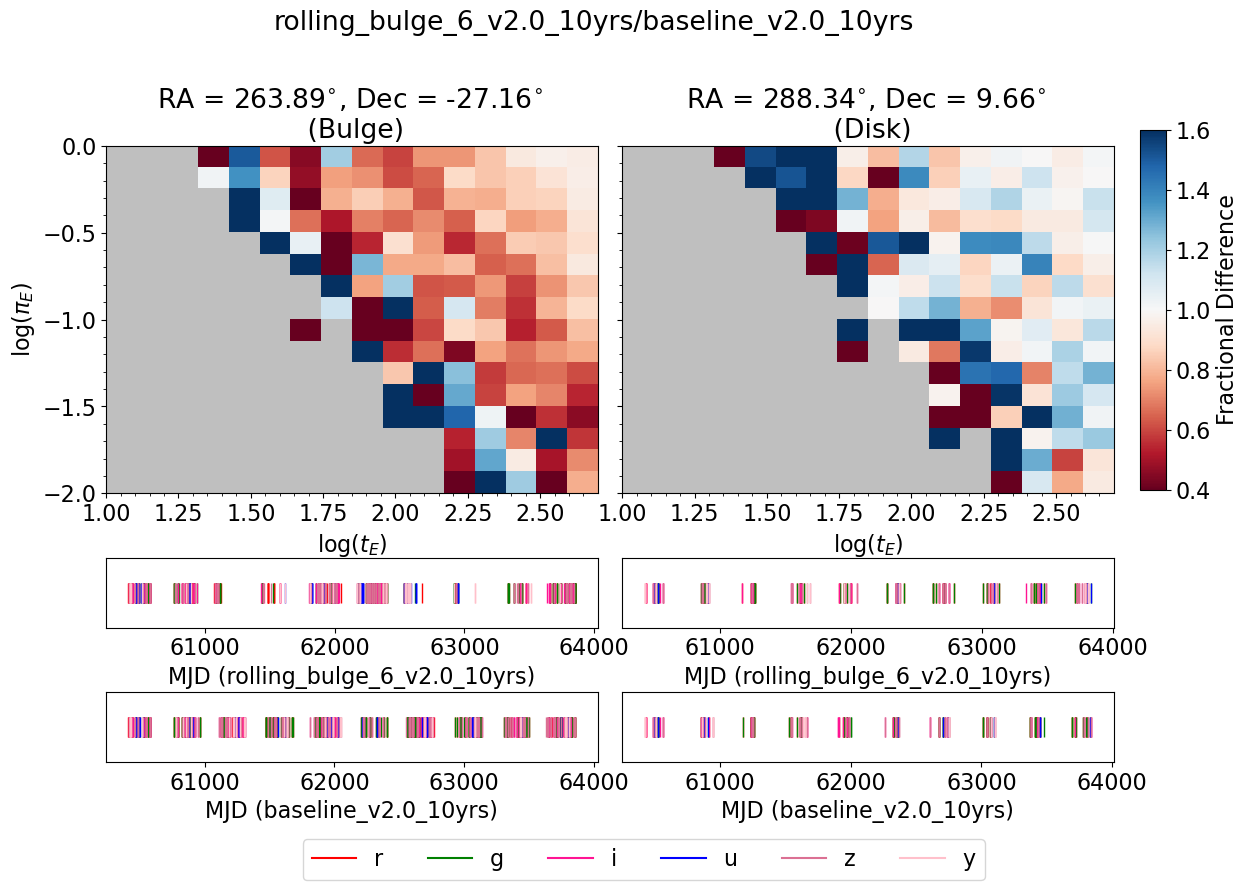}
    \caption{Same as Figure \ref{fig:parallax_char_baseline_v3.0} but \reedit{comparing \texttt{rolling\_bulge\_6\_v2.0\_10yrs} to \texttt{baseline\_v2.0\_10yrs}}. This has the same footprint as \texttt{baseline\_v2.0\_10yrs}, but splits the bulge into six sections on which it alternates focus, as can be seen in the \reedit{middle} left panel. This leads to \reedit{$\sim 20$\% fewer events being characterized} \reedit{in the Galactic bulge.}}
    \label{fig:parallax_char_rolling_bulge_v2.0}
\end{figure}

\begin{figure}[h]
    \centering
    \hspace*{-0.5cm}
    \includegraphics[scale=0.5]{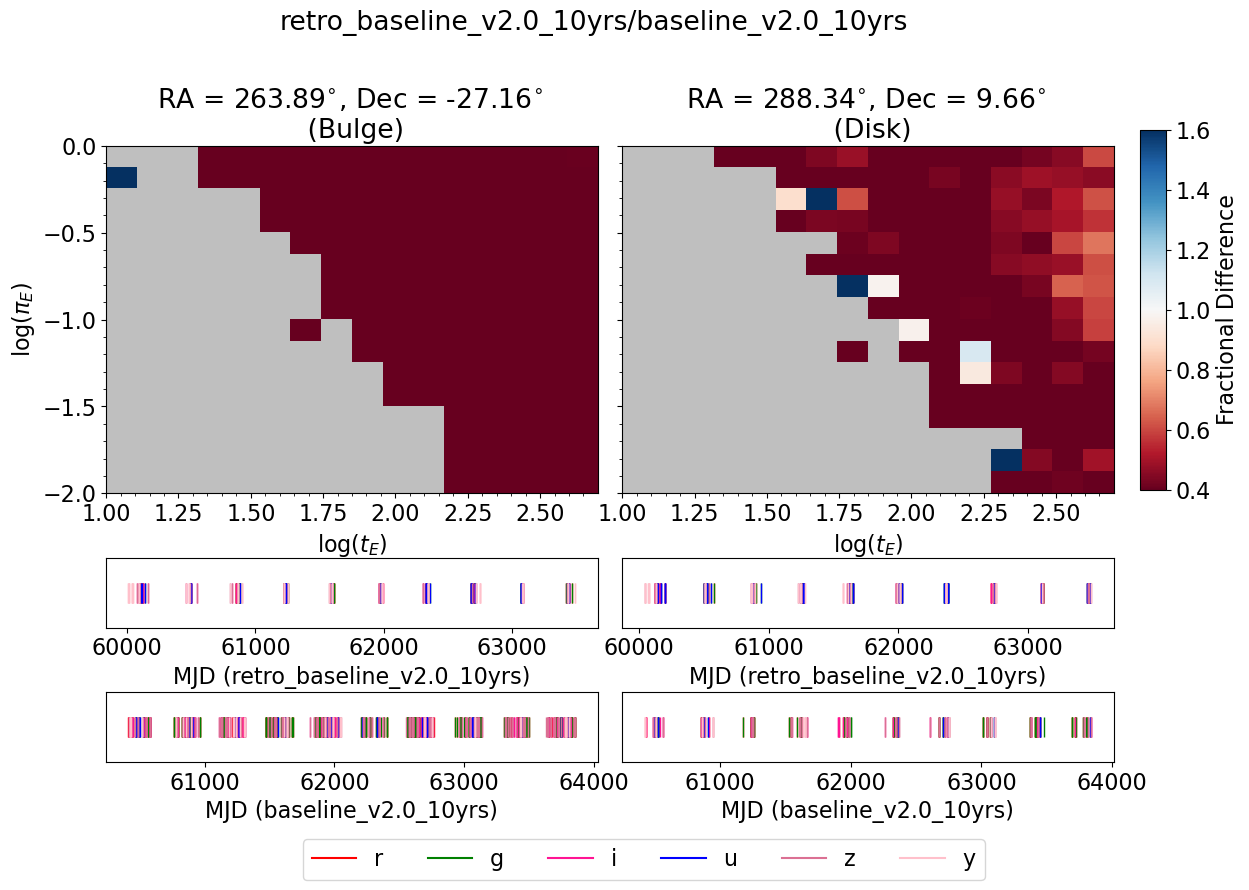}
    \caption{Same as Figure \ref{fig:parallax_char_baseline_v3.0} but \reedit{comparing \texttt{retro\_baseline\_v2.0\_10yrs} to \texttt{baseline\_v2.0\_10yrs}}. Given the extremely sparse cadence in the Galactic bulge and plane in the retro baseline, \reedit{we would characterize} 60-80\% \reedit{fewer} microlensing events.}
    \label{fig:parallax_char_retro_baseline}
\end{figure}

In Figures \ref{fig:parallax_char_baseline_v3.0}-\ref{fig:parallax_char_retro_baseline}, we plot 2D-histograms of $\log(\pi_{\rm E}$) vs log($t_{\rm E}$) dividing the fraction of events characterized by the OpSim in the numerator by the fraction of events characterized \cedit{by \texttt{baseline\_v2.0\_10yrs}}. This compares each OpSim relative to \texttt{baseline\_v2.0\_10yrs} such that the redder indicates \texttt{baseline\_v2.0\_10yrs} characterized a higher fraction of events and bluer indicates the compared OpSim characterized a higher fraction of events. In the \reedit{middle and} lower panels, we plot the cadence of observations for that OpSim and field, where lines of different colors represent observations taken with different filters, as indicated by the legend. The left plot in each figure is for a representative bulge field and the right plot in each figure is for a representative disk field in a pencil beam field. Most events fall to the left side of these plots, see Figure \ref{fig:popsycle_piE_tE} for a realistic distribution of simulated events in $\pi_{\rm E} - t_{\rm E}$ space. \reedit{The grey squares indicate that neither OpSim characterized any events there. In the low $t_{\rm E}$ parameter space, there are some events with characterizable $t_{\rm E}$ values, in particular bright events. However, these are not characterizable here because we demand that both $t_{\rm E}$ and $\pi_{\rm E}$ are characterizable and it can be intrinsically difficult to have a measurable $\pi_{\rm E}$ signature in short duration events. When an event is short compared to the duration of a year, it is not usually possible to measure the $\pi_E$ signal from a single telescope \citep[e.g.][]{Poindexter:2005, Gaudi:2012}. Instead, one can use satellite parallax, in which a single event is measured simultaneously from an Earth based observatory and far-away space-based satellite \citep{Gould1994, Refsdal1966}. This may be a possible synergy between Roman and Rubin, \cedit{although most likely only for short $t_{\rm E}$ events due to the relatively short baseline between the telescopes} \citep[e.g.][]{Yee:2013, Street:2023Roman}}

In Figure~\ref{fig:parallax_char_baseline_v3.0}, 
\reedit{we compare} \texttt{baseline\_v2.0\_10yrs} and \texttt{baseline\_v3.0\_10yrs}. \reedit{We can see in the disk there is an improvement in characterization by 40-50\%.} This is because in \texttt{baseline\_v3.0\_10yrs} there are more observations spread out across the plane, maximizing Rubin's capability to do a Galaxy-wide study. 

We can compare \texttt{baseline\_v2.0\_10yrs} to the rolling cadences since they share the same footprint. In \texttt{rolling\_all\_sky\_ns2\_rw0.9\_v2.0\_10yrs} (Figure~\ref{fig:parallax_char_rolling_all_sky_v2.0}), there is a drop in characterization efficiency on the 5-10\% level due to longer periods with long gaps between observations. Whereas in \texttt{rolling\_bulge\_6\_v2.0\_10yrs} (Figure~\ref{fig:parallax_char_rolling_bulge_v2.0}) there is a drop by $\sim 15-20\%$, particularly for high parallax events due to seasons with very sparse coverage and long gaps. In both, there is little change in the disk field as it has such sparse cadence and \reedit{has a relatively constant cadence between years}.

For reference, we can compare these cadences to \texttt{retro\_baseline\_v2.0\_10yrs} (Figure~\ref{fig:parallax_char_retro_baseline}). This covers the Galactic bulge and plane very sparsely and causes events to be 60-80\% more difficult to characterize than the current baseline. This is indicative that the strongest determiner of characterization is the footprint. If time is not dedicated to the Galactic plane and bulge, most events will not be characterizable.

\section{Discussion and Conclusion}
\label{sec: discussion and conclusion}
\cedit{Given the simplifying assumptions made in the paper,} the main survey cadence optimizations that have a major effect on microlensing discovery and characterization can be summarized as follow\reedit{s:}
\begin{enumerate}
    \item Footprint, \cedit{to} the first order, makes the most significant impact on microlensing detection and characterization. This can be seen by comparing the current baselines to the \texttt{retro\_baseline}, \cedit{which largely avoids the Galactic bulge and plane} (see Figure~\ref{fig:baseline_metrics} and \edit{Table \ref{tab:opsim_summary} for summaries of OpSims}).
    \item Rubin will be able to use its uniquely deep and wide survey area to detect and characterize microlensing events across the Galactic plane. However, if areas of low stellar density and high extinction are included, this can lead to a decrease in the fraction of characterized events (see Figures~\ref{fig:vary_gp and plane_priority metrics}-\ref{fig:pp_pbt line plot}).
    \item \edit{A rolling cadence} in the Galactic bulge and plane has the potential to improve synergies with Roman, but should be approached with caution. The survey should avoid long gaps, since many current rolling strategies decrease detection and characterization of most microlensing events.
\end{enumerate}
The survey cadences besides the \texttt{retro} footprints have also incorporated the LMC and SMC. Their inclusion will allow us to probe microlensing events cause\reedit{d} by objects in the halo.

This paper has mostly discussed what Rubin can do alone with cadence optimization. Rubin will also send out nightly alerts \reedit{that} could be used to do follow-up on candidates and is of particular interest to exoplanet science.\footnote{See the cadence note \textit{LSST Cadence Note - Alerting transient phenomena in the Galactic Plane in time to coordinate follow-up} by Hundertmark et al. \url{https://docushare.lsst.org/docushare/dsweb/Get/Document-37638/Galactic_Plane_Transients.pdf}.}
Follow-up observations could fill in gaps in Rubin's coverage, but Rubin would still require adequate cadence in the wings of the event to reliably alert on events in progress. It may also be difficult to follow-up most faint events from most ground based facilities.

\subsection{End-to-End Pipeline}
\label{sec: end-to-end pipeline}
The above assessment of cadence strategy has all been relative between cadences. We have simulated ranges of microlensing parameters such as $t_{\rm E}$ and $\pi_{\rm E}$ or plotted results as a function of those values. We have not incorporated \reedit{either} a Galactic model and simulated a microlensing survey \reedit{or a distribution of magnitudes and colors}, which \reedit{are both} necessary for predicting a realistic number and distribution of detected and characterized microlensing events \reedit{and understanding the effect of filter balance and exposure length on the survey}. 

Beyond the cadence, the pipeline, both reduction and analysis, will also play a large part in detection and characterization of microlensing events. We have seen with the first 3 years of Zwicky Transient Facility data, another all-visible-sky survey, covering one hemisphere, that $\sim 100$ events were discovered \citep{Rodriguez:2022, Medford:2023}. In order to find a pure sample, many real microlensing events were likely excluded due to the $> 10^9$ initial lightcurves requiring strict cuts to fit all of the events (see \cite{Medford:2023}, Section 6). The analysis pipeline will have a significant effect on the microlensing yield. Given that microlensing is more likely to occur in crowded fields, carefully deblending and extracting photometry will be imperative to maximizing the number of detected and well characterized events. In a true end-to-end Rubin microlensing simulation, everything from initial physical Galactic parameters to reduction pipeline would be incorporated.

Since the Rubin Survey Optimization Committee plans to assess the cadence multiple times throughout Rubin's operation, it is important to continue to develop our ability to assess cadences, including folding in real data.

\begin{acknowledgments}
\reedit{We thank Federica Bianco for helpful conversations and guidance.} This research used resources of the National Energy Research Scientific Computing Center (NERSC), a U.S. Department of Energy Office of Science User Facility located at Lawrence Berkeley National Laboratory, operated under Contract No. DE-AC02-05CH11231 using NERSC award HEP-ERCAP0023758 \reedit{and HEP-ERCAP0026816}. N.S.A. and J.R.L. acknowledge support from the National Science Foundation under grant No. 1909641 and the Heising-Simons Foundation under grant No. 2022-3542. R.A.S. gratefully acknowledges support from NSF grant number 2206828. Y.T. acknowledges the support of DFG priority program SPP 1992 “Exploring the Diversity of Extrasolar Planets” (TS 356/3-1). Support for M.R. is provided by the Direcci{\'o}n de Investigaci{\'o}n of the Universidad Cat{\'o}lica de la Sant{\'i}sima Concepci{\'o}n with the project DIREG 10/2023.

\end{acknowledgments}

%

\vspace{5mm}


\software{LSST Metric Analysis Framework \citep{rubin_maf}, \texttt{Numpy} \citep{numpy}, 
\texttt{Matplotlib} \citep{matplotlib}}, \texttt{BAGLE}, \texttt{SymPy} \citep{SymPy}, \texttt{PopSyCLE} \citep{Lam:2020}




\bibliography{microlensing_discovery_alerts_characterization}{}

\begin{thebibliography}{}
\expandafter\ifx\csname natexlab\endcsname\relax\def\natexlab#1{#1}\fi
\providecommand{\url}[1]{\href{#1}{#1}}
\providecommand{\dodoi}[1]{doi:~\href{http://doi.org/#1}{\nolinkurl{#1}}}
\providecommand{\doeprint}[1]{\href{http://ascl.net/#1}{\nolinkurl{http://ascl.net/#1}}}
\providecommand{\doarXiv}[1]{\href{https://arxiv.org/abs/#1}{\nolinkurl{https://arxiv.org/abs/#1}}}

\bibitem[{{Albrecht} {et~al.}(2006){Albrecht}, {Bernstein}, {Cahn}, {Freedman}, {Hewitt}, {Hu}, {Huth}, {Kamionkowski}, {Kolb}, {Knox}, {Mather}, {Staggs}, \& {Suntzeff}}]{albrecht2006}
{Albrecht}, A., {Bernstein}, G., {Cahn}, R., {et~al.} 2006, arXiv e-prints, astro, \dodoi{10.48550/arXiv.astro-ph/0609591}

\bibitem[{{Alcock} {et~al.}(2000){Alcock}, {Allsman}, {Alves}, {Axelrod}, {Becker}, {Bennett}, {Cook}, {Dalal}, {Drake}, {Freeman}, {Geha}, {Griest}, {Lehner}, {Marshall}, {Minniti}, {Nelson}, {Peterson}, {Popowski}, {Pratt}, {Quinn}, {Stubbs}, {Sutherland}, {Tomaney}, {Vandehei}, \& {Welch}}]{macho:alcock:2000}
{Alcock}, C., {Allsman}, R.~A., {Alves}, D.~R., {et~al.} 2000, \apj, 542, 281, \dodoi{10.1086/309512}

\bibitem[{{Beaulieu} {et~al.}(2006){Beaulieu}, {Bennett}, {Fouqu{\'e}}, {Williams}, {Dominik}, {J{\o}rgensen}, {Kubas}, {Cassan}, {Coutures}, {Greenhill}, {Hill}, {Menzies}, {Sackett}, {Albrow}, {Brillant}, {Caldwell}, {Calitz}, {Cook}, {Corrales}, {Desort}, {Dieters}, {Dominis}, {Donatowicz}, {Hoffman}, {Kane}, {Marquette}, {Martin}, {Meintjes}, {Pollard}, {Sahu}, {Vinter}, {Wambsganss}, {Woller}, {Horne}, {Steele}, {Bramich}, {Burgdorf}, {Snodgrass}, {Bode}, {Udalski}, {Szyma{\'n}ski}, {Kubiak}, {Wi{\c{e}}ckowski}, {Pietrzy{\'n}ski}, {Soszy{\'n}ski}, {Szewczyk}, {Wyrzykowski}, {Paczy{\'n}ski}, {Abe}, {Bond}, {Britton}, {Gilmore}, {Hearnshaw}, {Itow}, {Kamiya}, {Kilmartin}, {Korpela}, {Masuda}, {Matsubara}, {Motomura}, {Muraki}, {Nakamura}, {Okada}, {Ohnishi}, {Rattenbury}, {Sako}, {Sato}, {Sasaki}, {Sekiguchi}, {Sullivan}, {Tristram}, {Yock}, \& {Yoshioka}}]{Beaulieu:2006}
{Beaulieu}, J.~P., {Bennett}, D.~P., {Fouqu{\'e}}, P., {et~al.} 2006, \nat, 439, 437, \dodoi{10.1038/nature04441}

\bibitem[{{Bennett} {et~al.}(2014){Bennett}, {Batista}, {Bond}, {Bennett}, {Suzuki}, {Beaulieu}, {Udalski}, {Donatowicz}, {Bozza}, {Abe}, {Botzler}, {Freeman}, {Fukunaga}, {Fukui}, {Itow}, {Koshimoto}, {Ling}, {Masuda}, {Matsubara}, {Muraki}, {Namba}, {Ohnishi}, {Rattenbury}, {Saito}, {Sullivan}, {Sumi}, {Sweatman}, {Tristram}, {Tsurumi}, {Wada}, {Yock}, {MOA Collaboration}, {Albrow}, {Bachelet}, {Brillant}, {Caldwell}, {Cassan}, {Cole}, {Corrales}, {Coutures}, {Dieters}, {Dominis Prester}, {Fouqu{\'e}}, {Greenhill}, {Horne}, {Koo}, {Kubas}, {Marquette}, {Martin}, {Menzies}, {Sahu}, {Wambsganss}, {Williams}, {Zub}, {PLANET Collaboration}, {Choi}, {DePoy}, {Dong}, {Gaudi}, {Gould}, {Han}, {Henderson}, {McGregor}, {Lee}, {Pogge}, {Shin}, {Yee}, {{\ensuremath{\mu}}FUN Collaboration}, {Szyma{\'n}ski}, {Skowron}, {Poleski}, {Koz{\l}owski}, {Wyrzykowski}, {Kubiak}, {Pietrukowicz}, {Pietrzy{\'n}ski}, {Soszy{\'n}ski}, {Ulaczyk}, {OGLE Collaboration}, {Tsapras}, {Street}, {Dominik}, {Bramich}, {Browne}, {Hundertmark},
  {Kains}, {Snodgrass}, {Steele}, {RoboNet Collaboration}, {Dekany}, {Gonzalez}, {Heyrovsk{\'y}}, {Kandori}, {Kerins}, {Lucas}, {Minniti}, {Nagayama}, {Rejkuba}, {Robin}, \& {Saito}}]{Bennett:2014}
{Bennett}, D.~P., {Batista}, V., {Bond}, I.~A., {et~al.} 2014, \apj, 785, 155, \dodoi{10.1088/0004-637X/785/2/155}

\bibitem[{{Bianco} {et~al.}(2019){Bianco}, {Drout}, {Graham}, {Pritchard}, {Biswas}, {Narayan}, {Andreoni}, {Cowperthwaite}, {Ribeiro}, {LSST Transient}, \& {Variable Stars Collaboration}}]{Bianco:2019}
{Bianco}, F.~B., {Drout}, M.~R., {Graham}, M.~L., {et~al.} 2019, \pasp, 131, 068002, \dodoi{10.1088/1538-3873/ab121a}

\bibitem[{{Bianco} {et~al.}(2022){Bianco}, {Ivezi{\'c}}, {Jones}, {Graham}, {Marshall}, {Saha}, {Strauss}, {Yoachim}, {Ribeiro}, {Anguita}, {Bauer}, {Bauer}, {Bellm}, {Blum}, {Brandt}, {Brough}, {Catelan}, {Clarkson}, {Connolly}, {Gawiser}, {Gizis}, {Hlo{\v{z}}ek}, {Kaviraj}, {Liu}, {Lochner}, {Mahabal}, {Mandelbaum}, {McGehee}, {Neilsen}, {Olsen}, {Peiris}, {Rhodes}, {Richards}, {Ridgway}, {Schwamb}, {Scolnic}, {Shemmer}, {Slater}, {Slosar}, {Smartt}, {Strader}, {Street}, {Trilling}, {Verma}, {Vivas}, {Wechsler}, \& {Willman}}]{bianco2022}
{Bianco}, F.~B., {Ivezi{\'c}}, {\v{Z}}., {Jones}, R.~L., {et~al.} 2022, \apjs, 258, 1, \dodoi{10.3847/1538-4365/ac3e72}

\bibitem[{{Connolly} {et~al.}(2014){Connolly}, {Angeli}, {Chandrasekharan}, {Claver}, {Cook}, {Ivezic}, {Jones}, {Krughoff}, {Peng}, {Peterson}, {Petry}, {Rasmussen}, {Ridgway}, {Saha}, {Sembroski}, {vanderPlas}, \& {Yoachim}}]{Connolly:2014}
{Connolly}, A.~J., {Angeli}, G.~Z., {Chandrasekharan}, S., {et~al.} 2014, in Society of Photo-Optical Instrumentation Engineers (SPIE) Conference Series, Vol. 9150, Modeling, Systems Engineering, and Project Management for Astronomy VI, ed. G.~Z. Angeli \& P.~Dierickx, 14, \dodoi{10.1117/12.2054953}

\bibitem[{{Dal Tio} {et~al.}(2022){Dal Tio}, {Pastorelli}, {Mazzi}, {Trabucchi}, {Costa}, {Jacques}, {Pieres}, {Girardi}, {Chen}, {Olsen}, {Juric}, {Ivezi{\'c}}, {Yoachim}, {Clarkson}, {Marigo}, {Rodrigues}, {Zaggia}, {Barbieri}, {Momany}, {Bressan}, {Nikutta}, \& {da Costa}}]{DalTio:2022-LSST-TRILEGAL}
{Dal Tio}, P., {Pastorelli}, G., {Mazzi}, A., {et~al.} 2022, \apjs, 262, 22, \dodoi{10.3847/1538-4365/ac7be6}

\bibitem[{{Gaudi}(2012)}]{Gaudi:2012}
{Gaudi}, B.~S. 2012, \araa, 50, 411, \dodoi{10.1146/annurev-astro-081811-125518}

\bibitem[{{Gould}(1992)}]{Gould:1992-parallax}
{Gould}, A. 1992, \apj, 392, 442, \dodoi{10.1086/171443}

\bibitem[{{Gould}(1994)}]{Gould1994}
---. 1994, \apjl, 421, L75, \dodoi{10.1086/187191}

\bibitem[{{Gould}(2013)}]{Gould:2013}
---. 2013, arXiv e-prints, arXiv:1304.3455, \dodoi{10.48550/arXiv.1304.3455}

\bibitem[{Harris {et~al.}(2020)Harris, Millman, van~der Walt, Gommers, Virtanen, Cournapeau, Wieser, Taylor, Berg, Smith, Kern, Picus, Hoyer, van Kerkwijk, Brett, Haldane, del R{\'{i}}o, Wiebe, Peterson, G{\'{e}}rard-Marchant, Sheppard, Reddy, Weckesser, Abbasi, Gohlke, \& Oliphant}]{numpy}
Harris, C.~R., Millman, K.~J., van~der Walt, S.~J., {et~al.} 2020, Nature, 585, 357, \dodoi{10.1038/s41586-020-2649-2}

\bibitem[{Hunter(2007)}]{matplotlib}
Hunter, J.~D. 2007, Computing in Science \& Engineering, 9, 90, \dodoi{10.1109/MCSE.2007.55}

\bibitem[{{Jones} \& {Yoachim}(2022)}]{rubin_maf}
{Jones}, L., \& {Yoachim}, P. 2022, Rubin Metric Analysis Framework, https://github.com/lsst/rubin\_sim.git

\bibitem[{Jones {et~al.}(2020)Jones, Yoachim, Ivezic, Neilsen, \& Ribeiro}]{Jones_baseline_2018-2020}
Jones, R.~L., Yoachim, P., Ivezic, Z., Neilsen, E.~H., \& Ribeiro, T. 2020, {Survey Strategy and Cadence Choices for the Vera C. Rubin Observatory Legacy Survey of Space and Time (LSST)}, v1.2,  Zenodo, \dodoi{10.5281/zenodo.4048838}

\bibitem[{{Jones} {et~al.}(2014){Jones}, {Yoachim}, {Chandrasekharan}, {Connolly}, {Cook}, {Ivezic}, {Krughoff}, {Petry}, \& {Ridgway}}]{Jones:2014}
{Jones}, R.~L., {Yoachim}, P., {Chandrasekharan}, S., {et~al.} 2014, in Society of Photo-Optical Instrumentation Engineers (SPIE) Conference Series, Vol. 9149, Observatory Operations: Strategies, Processes, and Systems V, ed. A.~B. {Peck}, C.~R. {Benn}, \& R.~L. {Seaman}, 91490B, \dodoi{10.1117/12.2056835}

\bibitem[{{Jungman} {et~al.}(1996){Jungman}, {Kamionkowski}, {Kosowsky}, \& {Spergel}}]{jungman1996}
{Jungman}, G., {Kamionkowski}, M., {Kosowsky}, A., \& {Spergel}, D.~N. 1996, \prd, 54, 1332, \dodoi{10.1103/PhysRevD.54.1332}

\bibitem[{{Kim} {et~al.}(2018){Kim}, {Kim}, {Hwang}, {Albrow}, {Chung}, {Gould}, {Han}, {Jung}, {Ryu}, {Shin}, {Yee}, {Zhu}, {Cha}, {Kim}, {Lee}, {Lee}, {Lee}, {Park}, {Pogge}, \& {KMTNet Collaboration}}]{kmtnet:kim:2018}
{Kim}, D.~J., {Kim}, H.~W., {Hwang}, K.~H., {et~al.} 2018, \aj, 155, 76, \dodoi{10.3847/1538-3881/aaa47b}

\bibitem[{{Lam} \& {Lu}(2023)}]{Lam:2023-OB110462}
{Lam}, C.~Y., \& {Lu}, J.~R. 2023, arXiv e-prints, arXiv:2308.03302, \dodoi{10.48550/arXiv.2308.03302}

\bibitem[{{Lam} {et~al.}(2020){Lam}, {Lu}, {Hosek}, {Dawson}, \& {Golovich}}]{Lam:2020}
{Lam}, C.~Y., {Lu}, J.~R., {Hosek}, Matthew~W., J., {Dawson}, W.~A., \& {Golovich}, N.~R. 2020, \apj, 889, 31, \dodoi{10.3847/1538-4357/ab5fd3}

\bibitem[{{Lam} {et~al.}(2022){Lam}, {Lu}, {Udalski}, {Bond}, {Bennett}, {Skowron}, {Mr{\'o}z}, {Poleski}, {Sumi}, {Szyma{\'n}ski}, {Koz{\l}owski}, {Pietrukowicz}, {Soszy{\'n}ski}, {Ulaczyk}, {Wyrzykowski}, {Miyazaki}, {Suzuki}, {Koshimoto}, {Rattenbury}, {Hosek}, {Abe}, {Barry}, {Bhattacharya}, {Fukui}, {Fujii}, {Hirao}, {Itow}, {Kirikawa}, {Kondo}, {Matsubara}, {Matsumoto}, {Muraki}, {Olmschenk}, {Ranc}, {Okamura}, {Satoh}, {Silva}, {Toda}, {Tristram}, {Vandorou}, {Yama}, {Abrams}, {Agarwal}, {Rose}, \& {Terry}}]{Lam:2022}
{Lam}, C.~Y., {Lu}, J.~R., {Udalski}, A., {et~al.} 2022, \apjl, 933, L23, \dodoi{10.3847/2041-8213/ac7442}

\bibitem[{{Lam} {et~al.}(2023){Lam}, {Abrams}, {Andrews}, {Bachelet}, {Bahramian}, {Bennett}, {Bozza}, {Broekgaarden}, {Chakrabarti}, {Dawson}, {El-Badry}, {Fishbach}, {Fragione}, {Gaudi}, {Gautam}, {Hirai}, {Holz}, {Hosek}, {Huston}, {Jayasinghe}, {Johnson}, {Kawata}, {Koshimoto}, {Lu}, {Mandel}, {Miyazaki}, {Mr{\'o}z}, {Naoz}, {Ranc}, {Rowan}, {Sch{\"o}del}, {Shenar}, {Simon}, {Street}, {Sumi}, {Suzuki}, \& {Terry}}]{Lam2023-Roman}
{Lam}, C.~Y., {Abrams}, N., {Andrews}, J., {et~al.} 2023, arXiv e-prints, arXiv:2306.12514, \dodoi{10.48550/arXiv.2306.12514}

\bibitem[{{LSST Science Collaboration} {et~al.}(2017){LSST Science Collaboration}, {Marshall}, {Anguita}, {Bianco}, {Bellm}, {Brandt}, {Clarkson}, {Connolly}, {Gawiser}, {Ivezic}, {Jones}, {Lochner}, {Lund}, {Mahabal}, {Nidever}, {Olsen}, {Ridgway}, {Rhodes}, {Shemmer}, {Trilling}, {Vivas}, {Walkowicz}, {Willman}, {Yoachim}, {Anderson}, {Antilogus}, {Angus}, {Arcavi}, {Awan}, {Biswas}, {Bell}, {Bennett}, {Britt}, {Buzasi}, {Casetti-Dinescu}, {Chomiuk}, {Claver}, {Cook}, {Davenport}, {Debattista}, {Digel}, {Doctor}, {Firth}, {Foley}, {Fong}, {Galbany}, {Giampapa}, {Gizis}, {Graham}, {Grillmair}, {Gris}, {Haiman}, {Hartigan}, {Hawley}, {Hlozek}, {Jha}, {Johns-Krull}, {Kanbur}, {Kalogera}, {Kashyap}, {Kasliwal}, {Kessler}, {Kim}, {Kurczynski}, {Lahav}, {Liu}, {Malz}, {Margutti}, {Matheson}, {McEwen}, {McGehee}, {Meibom}, {Meyers}, {Monet}, {Neilsen}, {Newman}, {O'Dowd}, {Peiris}, {Penny}, {Peters}, {Poleski}, {Ponder}, {Richards}, {Rho}, {Rubin}, {Schmidt}, {Schuhmann}, {Shporer}, {Slater}, {Smith},
  {Soares-Santos}, {Stassun}, {Strader}, {Strauss}, {Street}, {Stubbs}, {Sullivan}, {Szkody}, {Trimble}, {Tyson}, {de Val-Borro}, {Valenti}, {Wagoner}, {Wood-Vasey}, \& {Zauderer}}]{COSEP}
{LSST Science Collaboration}, {Marshall}, P., {Anguita}, T., {et~al.} 2017, arXiv e-prints, arXiv:1708.04058, \dodoi{10.48550/arXiv.1708.04058}

\bibitem[{{Lu} {et~al.}(2016){Lu}, {Sinukoff}, {Ofek}, {Udalski}, \& {Kozlowski}}]{Lu:2016}
{Lu}, J.~R., {Sinukoff}, E., {Ofek}, E.~O., {Udalski}, A., \& {Kozlowski}, S. 2016, \apj, 830, 41, \dodoi{10.3847/0004-637X/830/1/41}

\bibitem[{{Medford} {et~al.}(2023){Medford}, {Abrams}, {Lu}, {Nugent}, \& {Lam}}]{Medford:2023}
{Medford}, M.~S., {Abrams}, N.~S., {Lu}, J.~R., {Nugent}, P., \& {Lam}, C.~Y. 2023, \apj, 947, 24, \dodoi{10.3847/1538-4357/acba8f}

\bibitem[{Meurer {et~al.}(2017)Meurer, Smith, Paprocki, \v{C}ert\'{i}k, Kirpichev, Rocklin, Kumar, Ivanov, Moore, Singh, Rathnayake, Vig, Granger, Muller, Bonazzi, Gupta, Vats, Johansson, Pedregosa, Curry, Terrel, Rou\v{c}ka, Saboo, Fernando, Kulal, Cimrman, \& Scopatz}]{SymPy}
Meurer, A., Smith, C.~P., Paprocki, M., {et~al.} 2017, PeerJ Computer Science, 3, e103, \dodoi{10.7717/peerj-cs.103}

\bibitem[{{Moniez}(2010)}]{Moniez:2010}
{Moniez}, M. 2010, General Relativity and Gravitation, 42, 2047, \dodoi{10.1007/s10714-009-0925-4}

\bibitem[{{Moniez} {et~al.}(2017){Moniez}, {Sajadian}, {Karami}, {Rahvar}, \& {Ansari}}]{Moniez:2017}
{Moniez}, M., {Sajadian}, S., {Karami}, M., {Rahvar}, S., \& {Ansari}, R. 2017, \aap, 604, A124, \dodoi{10.1051/0004-6361/201730488}

\bibitem[{{Mr{\'o}z} {et~al.}(2022){Mr{\'o}z}, {Udalski}, \& {Gould}}]{Mroz:2022}
{Mr{\'o}z}, P., {Udalski}, A., \& {Gould}, A. 2022, \apjl, 937, L24, \dodoi{10.3847/2041-8213/ac90bb}

\bibitem[{{Mr{\'o}z} {et~al.}(2017){Mr{\'o}z}, {Udalski}, {Skowron}, {Poleski}, {Koz{\l}owski}, {Szyma{\'n}ski}, {Soszy{\'n}ski}, {Wyrzykowski}, {Pietrukowicz}, {Ulaczyk}, {Skowron}, \& {Pawlak}}]{Mroz:2017}
{Mr{\'o}z}, P., {Udalski}, A., {Skowron}, J., {et~al.} 2017, \nat, 548, 183, \dodoi{10.1038/nature23276}

\bibitem[{{Mr{\'o}z} {et~al.}(2020){Mr{\'o}z}, {Udalski}, {Szyma{\'n}ski}, {Soszy{\'n}ski}, {Pietrukowicz}, {Koz{\l}owski}, {Skowron}, {Poleski}, {Ulaczyk}, {Gromadzki}, {Rybicki}, {Iwanek}, \& {Wrona}}]{Mroz:2020-galplane}
{Mr{\'o}z}, P., {Udalski}, A., {Szyma{\'n}ski}, M.~K., {et~al.} 2020, \apjs, 249, 16, \dodoi{10.3847/1538-4365/ab9366}

\bibitem[{{Naghib} {et~al.}(2019){Naghib}, {Yoachim}, {Vanderbei}, {Connolly}, \& {Jones}}]{Naghib:2019}
{Naghib}, E., {Yoachim}, P., {Vanderbei}, R.~J., {Connolly}, A.~J., \& {Jones}, R.~L. 2019, \aj, 157, 151, \dodoi{10.3847/1538-3881/aafece}

\bibitem[{{Paczynski}(1986)}]{pacynski1986halo}
{Paczynski}, B. 1986, \apj, 304, 1, \dodoi{10.1086/164140}

\bibitem[{{Penny} {et~al.}(2019){Penny}, {Gaudi}, {Kerins}, {Rattenbury}, {Mao}, {Robin}, \& {Calchi Novati}}]{roman-Penny:2019}
{Penny}, M.~T., {Gaudi}, B.~S., {Kerins}, E., {et~al.} 2019, \apjs, 241, 3, \dodoi{10.3847/1538-4365/aafb69}

\bibitem[{{Poindexter} {et~al.}(2005){Poindexter}, {Afonso}, {Bennett}, {Glicenstein}, {Gould}, {Szyma{\'n}ski}, \& {Udalski}}]{Poindexter:2005}
{Poindexter}, S., {Afonso}, C., {Bennett}, D.~P., {et~al.} 2005, \apj, 633, 914, \dodoi{10.1086/468182}

\bibitem[{{Poleski}(2016)}]{Poleski:2016}
{Poleski}, R. 2016, \mnras, 455, 3656, \dodoi{10.1093/mnras/stv2569}

\bibitem[{{Refsdal}(1966)}]{Refsdal1966}
{Refsdal}, S. 1966, \mnras, 134, 315, \dodoi{10.1093/mnras/134.3.315}

\bibitem[{{Rodriguez} {et~al.}(2022){Rodriguez}, {Mr{\'o}z}, {Kulkarni}, {Andreoni}, {Bellm}, {Dekany}, {Drake}, {Duev}, {Graham}, {Masci}, {Prince}, {Riddle}, \& {Shupe}}]{Rodriguez:2022}
{Rodriguez}, A.~C., {Mr{\'o}z}, P., {Kulkarni}, S.~R., {et~al.} 2022, \apj, 927, 150, \dodoi{10.3847/1538-4357/ac51cc}

\bibitem[{{Sahu} {et~al.}(2022){Sahu}, {Anderson}, {Casertano}, {Bond}, {Udalski}, {Dominik}, {Calamida}, {Bellini}, {Brown}, {Rejkuba}, {Bajaj}, {Kains}, {Ferguson}, {Fryer}, {Yock}, {Mr{\'o}z}, {Koz{\l}owski}, {Pietrukowicz}, {Poleski}, {Skowron}, {Soszy{\'n}ski}, {Szyma{\'n}ski}, {Ulaczyk}, {Wyrzykowski}, {Barry}, {Bennett}, {Bond}, {Hirao}, {Silva}, {Kondo}, {Koshimoto}, {Ranc}, {Rattenbury}, {Sumi}, {Suzuki}, {Tristram}, {Vandorou}, {Beaulieu}, {Marquette}, {Cole}, {Fouqu{\'e}}, {Hill}, {Dieters}, {Coutures}, {Dominis-Prester}, {Bennett}, {Bachelet}, {Menzies}, {Albrow}, {Pollard}, {Gould}, {Yee}, {Allen}, {Almeida}, {Christie}, {Drummond}, {Gal-Yam}, {Gorbikov}, {Jablonski}, {Lee}, {Maoz}, {Manulis}, {McCormick}, {Natusch}, {Pogge}, {Shvartzvald}, {J{\o}rgensen}, {Alsubai}, {Andersen}, {Bozza}, {Novati}, {Burgdorf}, {Hinse}, {Hundertmark}, {Husser}, {Kerins}, {Longa-Pe{\~n}a}, {Mancini}, {Penny}, {Rahvar}, {Ricci}, {Sajadian}, {Skottfelt}, {Snodgrass}, {Southworth}, {Tregloan-Reed}, {Wambsganss},
  {Wertz}, {Tsapras}, {Street}, {Bramich}, {Horne}, {Steele}, \& {RoboNet Collaboration}}]{Sahu:2022}
{Sahu}, K.~C., {Anderson}, J., {Casertano}, S., {et~al.} 2022, \apj, 933, 83, \dodoi{10.3847/1538-4357/ac739e}

\bibitem[{{Sajadian} \& {Poleski}(2019)}]{sajadian2019lsst}
{Sajadian}, S., \& {Poleski}, R. 2019, \apj, 871, 205, \dodoi{10.3847/1538-4357/aafa1d}

\bibitem[{{Spergel} {et~al.}(2015){Spergel}, {Gehrels}, {Baltay}, {Bennett}, {Breckinridge}, {Donahue}, {Dressler}, {Gaudi}, {Greene}, {Guyon}, {Hirata}, {Kalirai}, {Kasdin}, {Macintosh}, {Moos}, {Perlmutter}, {Postman}, {Rauscher}, {Rhodes}, {Wang}, {Weinberg}, {Benford}, {Hudson}, {Jeong}, {Mellier}, {Traub}, {Yamada}, {Capak}, {Colbert}, {Masters}, {Penny}, {Savransky}, {Stern}, {Zimmerman}, {Barry}, {Bartusek}, {Carpenter}, {Cheng}, {Content}, {Dekens}, {Demers}, {Grady}, {Jackson}, {Kuan}, {Kruk}, {Melton}, {Nemati}, {Parvin}, {Poberezhskiy}, {Peddie}, {Ruffa}, {Wallace}, {Whipple}, {Wollack}, \& {Zhao}}]{roman-Spergel:2015}
{Spergel}, D., {Gehrels}, N., {Baltay}, C., {et~al.} 2015, arXiv e-prints, arXiv:1503.03757, \dodoi{10.48550/arXiv.1503.03757}

\bibitem[{Street {et~al.}(2018)Street, Lund, Khakpash, Donachie, Dawson, Golovich, Wyrzykowski, Szkody, Naylor, \& Penny}]{street2018diverse}
Street, R., Lund, M., Khakpash, S., {et~al.} 2018, arXiv preprint arXiv:1812.03137

\bibitem[{{Street} {et~al.}(2023{\natexlab{a}}){Street}, {Li}, {Khakpash}, {Bellm}, {Girardi}, {Jones}, {Abrams}, {Tsapras}, {Hundertmark}, {Bachelet}, {Gandhi}, {Szkody}, {Clarkson}, {Szab{\'o}}, {Prisinzano}, {Bonito}, {Buckley}, {Marais}, \& {Di Stefano}}]{Street:2023}
{Street}, R.~A., {Li}, X., {Khakpash}, S., {et~al.} 2023{\natexlab{a}}, \apjs, 267, 15, \dodoi{10.3847/1538-4365/acd6f4}

\bibitem[{{Street} {et~al.}(2023{\natexlab{b}}){Street}, {Gough-Kelly}, {Lam}, {Varela}, {Makler}, {Bachelet}, {Lu}, {Abrams}, {Pusack}, {Terry}, {Di Stefano}, {Tsapras}, {Hundertmark}, {Grand}, {Daylan}, \& {Sobeck}}]{Street:2023Roman}
{Street}, R.~A., {Gough-Kelly}, S., {Lam}, C., {et~al.} 2023{\natexlab{b}}, arXiv e-prints, arXiv:2306.13792, \dodoi{10.48550/arXiv.2306.13792}

\bibitem[{{Sumi} {et~al.}(2003){Sumi}, {Abe}, {Bond}, {Dodd}, {Hearnshaw}, {Honda}, {Honma}, {Kan-ya}, {Kilmartin}, {Masuda}, {Matsubara}, {Muraki}, {Nakamura}, {Nishi}, {Noda}, {Ohnishi}, {Petterson}, {Rattenbury}, {Reid}, {Saito}, {Saito}, {Sato}, {Sekiguchi}, {Skuljan}, {Sullivan}, {Takeuti}, {Tristram}, {Wilkinson}, {Yanagisawa}, \& {Yock}}]{moa:Sumi:2003}
{Sumi}, T., {Abe}, F., {Bond}, I.~A., {et~al.} 2003, \apj, 591, 204, \dodoi{10.1086/375212}

\bibitem[{{Tsapras}(2018)}]{Tsapras:2018}
{Tsapras}, Y. 2018, Geosciences, 8, 365, \dodoi{10.3390/geosciences8100365}

\bibitem[{{Tsapras} {et~al.}(2016){Tsapras}, {Hundertmark}, {Wyrzykowski}, {Horne}, {Udalski}, {Snodgrass}, {Street}, {Bramich}, {Dominik}, {Bozza}, {Figuera Jaimes}, {Kains}, {Skowron}, {Szyma{\'n}ski}, {Pietrzy{\'n}ski}, {Soszy{\'n}ski}, {Ulaczyk}, {Koz{\l}owski}, {Pietrukowicz}, \& {Poleski}}]{Tsapras:2016}
{Tsapras}, Y., {Hundertmark}, M., {Wyrzykowski}, {\L}., {et~al.} 2016, \mnras, 457, 1320, \dodoi{10.1093/mnras/stw023}

\bibitem[{{Udalski} {et~al.}(2015){Udalski}, {Szyma{\'n}ski}, \& {Szyma{\'n}ski}}]{ogleIV:Udalski:2015}
{Udalski}, A., {Szyma{\'n}ski}, M.~K., \& {Szyma{\'n}ski}, G. 2015, \actaa, 65, 1, \dodoi{10.48550/arXiv.1504.05966}

\bibitem[{{Wyrzykowski} {et~al.}(2023){Wyrzykowski}, {Kruszy{\'n}ska}, {Rybicki}, {Holl}, {Lec{\oe}ur-Ta{\"\i}bi}, {Mowlavi}, {Nienartowicz}, {Jevardat de Fombelle}, {Rimoldini}, {Audard}, {Garcia-Lario}, {Gavras}, {Evans}, {Hodgkin}, \& {Eyer}}]{Wyrzykowski:2023}
{Wyrzykowski}, {\L}., {Kruszy{\'n}ska}, K., {Rybicki}, K.~A., {et~al.} 2023, \aap, 674, A23, \dodoi{10.1051/0004-6361/202243756}

\bibitem[{{Yee}(2013)}]{Yee:2013}
{Yee}, J.~C. 2013, \apjl, 770, L31, \dodoi{10.1088/2041-8205/770/2/L31}

\bibitem[{{Yee} {et~al.}(2012){Yee}, {Shvartzvald}, {Gal-Yam}, {Bond}, {Udalski}, {Koz{\l}owski}, {Han}, {Gould}, {Skowron}, {Suzuki}, {Abe}, {Bennett}, {Botzler}, {Chote}, {Freeman}, {Fukui}, {Furusawa}, {Itow}, {Kobara}, {Ling}, {Masuda}, {Matsubara}, {Miyake}, {Muraki}, {Ohmori}, {Ohnishi}, {Rattenbury}, {Saito}, {Sullivan}, {Sumi}, {Suzuki}, {Sweatman}, {Takino}, {Tristram}, {Wada}, {MOA Collaboration}, {Szyma{\'n}ski}, {Kubiak}, {Pietrzy{\'n}ski}, {Soszy{\'n}ski}, {Poleski}, {Ulaczyk}, {Wyrzykowski}, {Pietrukowicz}, {OGLE Collaboration}, {Allen}, {Almeida}, {Batista}, {Bos}, {Christie}, {DePoy}, {Dong}, {Drummond}, {Finkelman}, {Gaudi}, {Gorbikov}, {Henderson}, {Higgins}, {Jablonski}, {Kaspi}, {Manulis}, {Maoz}, {McCormick}, {McGregor}, {Monard}, {Moorhouse}, {Mu{\~n}oz}, {Natusch}, {Ngan}, {Ofek}, {Pogge}, {Santallo}, {Tan}, {Thornley}, {Shin}, {Choi}, {Park}, {Lee}, {Koo}, \& {{\ensuremath{\mu}}FUN Collaboration}}]{Yee:2012}
{Yee}, J.~C., {Shvartzvald}, Y., {Gal-Yam}, A., {et~al.} 2012, \apj, 755, 102, \dodoi{10.1088/0004-637X/755/2/102}

\bibitem[{{Yee} {et~al.}(2014){Yee}, {Albrow}, {Barry}, {Bennett}, {Bryden}, {Chung}, {Gaudi}, {Gehrels}, {Gould}, {Penny}, {Rattenbury}, {Ryu}, {Skowron}, {Street}, \& {Sumi}}]{Yee:2014}
{Yee}, J.~C., {Albrow}, M., {Barry}, R.~K., {et~al.} 2014, arXiv e-prints, arXiv:1409.2759, \dodoi{10.48550/arXiv.1409.2759}

\bibitem[{{Zhai} {et~al.}(2023){Zhai}, {Rodriguez}, {Lam}, {Bellm}, {Purdum}, {Masci}, \& {Wold}}]{Zhai:2023}
{Zhai}, R., {Rodriguez}, A.~C., {Lam}, C.~Y., {et~al.} 2023, arXiv e-prints, arXiv:2311.18627, \dodoi{10.48550/arXiv.2311.18627}

\end{thebibliography}
\bibliographystyle{aasjournal}



\appendix

\vspace{-5mm}
\section{Additional OpSims}
\label{appendix: additional opsims}
\begin{figure}[h]
    \centering
    \vspace*{-0.3cm}
    \includegraphics[scale=0.63]{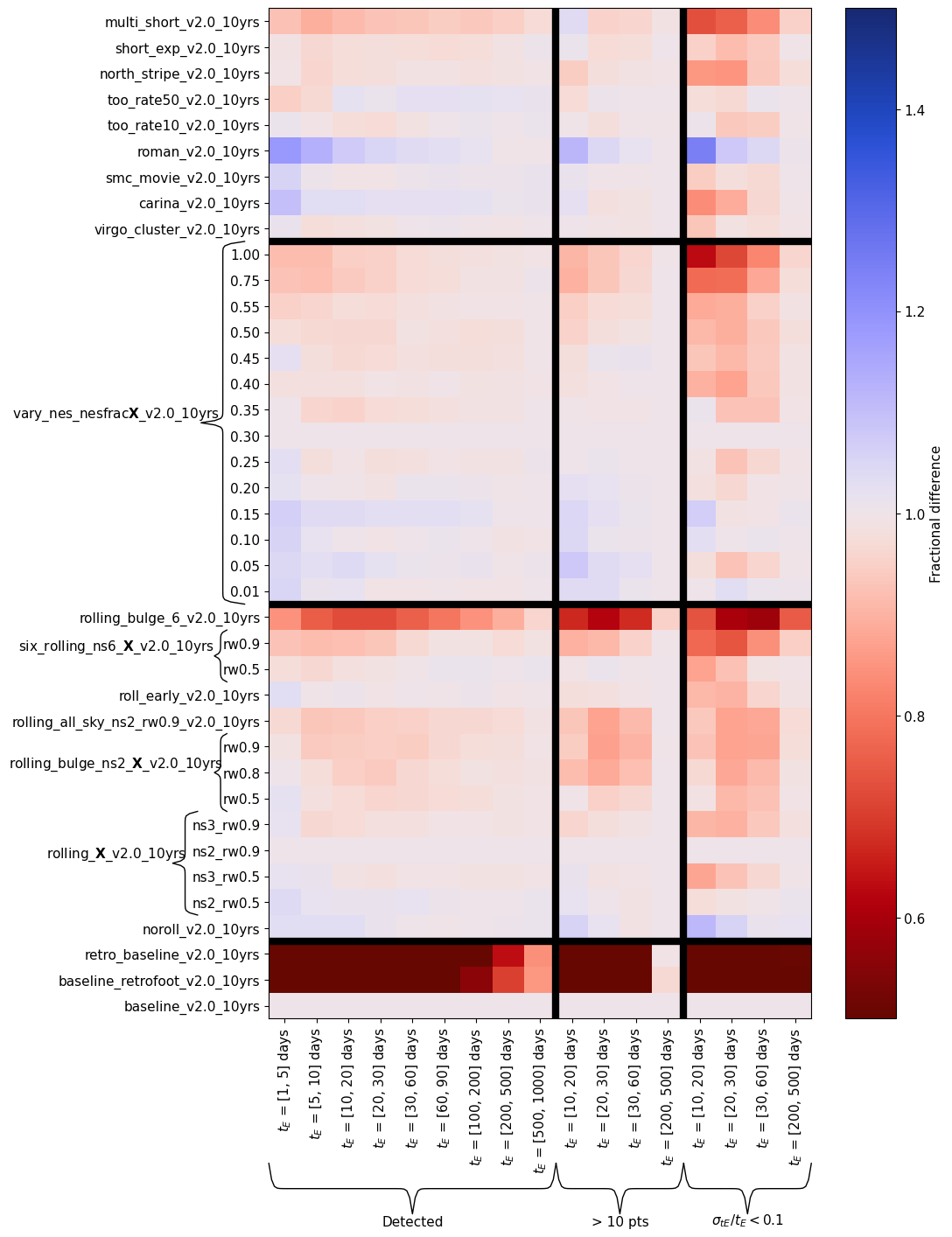}
    \caption{Same as Figure \ref{fig:baseline_metrics}, but for select other v2.0 \reedit{OpSims}, with baseline metrics for reference. This has the OpSims on the y-axis and the metrics on the x-axis for ease of plotting.}
    \label{fig:v2.0 and earlier}
    \vspace*{-0.8cm}
\end{figure}

\begin{figure}[ht]
    \centering
    \includegraphics[scale = 0.63]{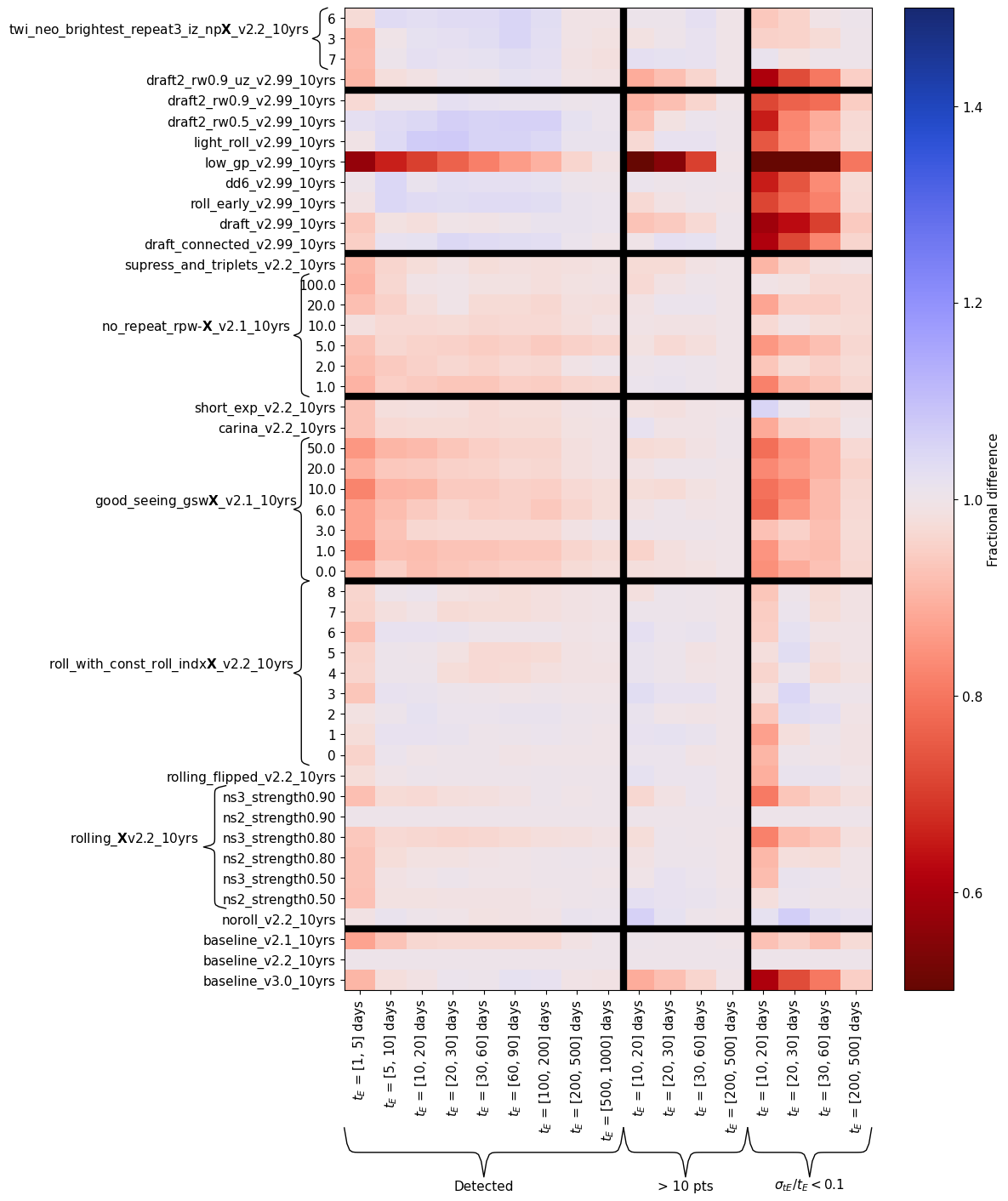}
    \caption{Same as Figure \ref{fig:baseline_metrics}, but for select other 2.1 and later \reedit{OpSims}, with baseline metrics for reference. All OpSims are plotted in reference to \texttt{baseline\_v2.2\_10yrs} in this plot. A representative 3 \texttt{twilight\_neo} runs are plotted. This has the OpSims on the y-axis and the metrics on the x-axis for ease of plotting.}
    \label{fig:v2.1 and later}
\end{figure}

Figures \ref{fig:v2.0 and earlier} and \ref{fig:v2.1 and later} summarize other select OpSims discussed in the paper, but without dedicated plots. Figure \ref{fig:v2.0 and earlier} has v2.0 OpSims and Figure \ref{fig:v2.1 and later} has v2.1-v3.0 OpSims.
In Table \ref{tab:opsim_summary}, the OpSims discussed in this paper are summarized with descriptions relevant to microlensing and other Galactic science. See \cite{Jones_baseline_2018-2020} and Rubin technical note \href{https://pstn-055.lsst.io/}{PSTN-055} for more detailed descriptions.

\edit{There are a few OpSim families that include additional surveys unrelated to microlensing to LSST which we will discuss here. The \texttt{ddf\_} OpSim family evaluates changes to the DDF strategy. Since the DDFs do not cover the Galactic plane, the more they have visits dedicated to them the fewer visits are available for regions in the Galactic plane. At the current level of the DDFs in the OpSims, it decreases short event characterization at the 10-15\% level, but besides that it does not appear to significantly affect the microlensing science case.}

\edit{The \reedit{\texttt{vary\_nes}} OpSim family varies the coverage of the North Ecliptic Spur (NES) as a percentage of the WFD survey. The more the survey strategy covers the NES, the less we are able to cover the Galactic bulge and plane which causes the microlensing metric to suffer. For an NES coverage of \reedit{\texttt{nesfrac=}}60-75\%, there is a significant drop of about 15\% in fraction of microlensing events that can be characterized (see Figure \ref{fig:v2.0 and earlier}).}

\edit{The Twilight NEO OpSim family (\texttt{twilight\_neo\_nightpatternX}) explores adding a twilight observing strategy, primarily looking for Near-Earth Objects. The SNR of observations is reduced, so the events, especially the short stellar events, suffer in characterizations by 5-10\%. Some of the long events technically have more observations that overcome the SNR downsides, but the quality loss and systematic effects would make an analysis challenging despite the technically better relative assessment.}
\edit{This is a surprising result since we did not think the twilight observations would lead to a poorer coverage of the night-time events. A representative 3 of the 84 \texttt{twilight\_neo} runs are plotted in Figure \ref{fig:v2.1 and later}.}

\newpage
\begin{table}[hb]
    \centering
    \begin{tabular}{l|l}
         OpSim (Family) Name & Description  \\
         \hline
         baseline & Baseline survey strategies. \\
         \hspace{5mm}\texttt{baseline\_v3.0\_10yrs} & Includes high priority areas across the Galactic plane and bulge. \\
         \hspace{5mm}\texttt{baseline\_v2.2\_10yrs} & Optimizations to code and DDF strategy change. \\
         \hspace{5mm}\texttt{baseline\_v2.1\_10yrs} & Includes Virgo cluster and acquisition of good seeing images in r and i bands. \\
         \hspace{5mm}\texttt{baseline\_v2.0\_10yrs} & Added in the Galactic bulge, LMC, and SMC. \\
         \hspace{5mm}\texttt{baseline\_retrofoot\_v2.0\_10yrs} & Uses 2018 footprint and v2.0 baseline strategy. \\
         \hspace{5mm}\texttt{retro\_baseline\_v2.0\_10yrs} & Uses 2018 footprint and strategy. \\
         \texttt{ddf\_\reedit{*}} & Varies survey strategy of DDFs. \\

         galactic plane & Simulations that explore the Galactic plane survey strategy. \\

         \hspace{5mm}\texttt{plane\_priority\_priorityX\_pbf\_\reedit{*}} & Includes regions of Galactic plane with priority $\geq$ X \reedit{(\texttt{priorityX})}\\ & \hspace{5mm} not including pencil beam fields \reedit{(\texttt{pbf} = pencil beams false)}. \\
         \hspace{5mm}\texttt{plane\_priority\_priorityX\_pbt\_\reedit{*}} & Includes regions of Galactic plane with priority $\geq$ X \reedit{(\texttt{priorityX})}\\ & \hspace{5mm} including pencil beam fields \reedit{(\texttt{pbt} = pencil beams true)}.  \\

         \hspace{5mm}\texttt{pencil\_fsX\_10yrs} & Varies size/number of pencil beams where X = 1 \reedit{(\texttt{fs1})} is 20 smaller fields \\ & \hspace{5mm} and X = 2 is 4 larger ones \reedit{(\texttt{fs2})}. \\
         \hspace{5mm}\texttt{vary\_gp\_gpfracX\_\reedit{*}} & Spends X\% \reedit{(\texttt{gpfracX})} of survey time on areas of the Galactic plane \\ & \hspace{5mm} not including in the WFD. \\
         
         \texttt{good\_seeing\_\reedit{*}} & Adds requirement of at least 3 good seeing images per year per pointing. \\
         microsurveys & Surveys requiring $< 3\%$ of LSST time. \\
         \hspace{5mm}\texttt{roman\_v2.0\_10yrs} & Adds microsurvey of Roman GBTDS field. \\
         \hspace{5mm}\texttt{smc\_movie\_v2.0\_10yrs} & Add two nights of observing of the SMC. \\
         rolling & Varies strategy to alternate high cadence coverage of areas of the sky. \\
         \hspace{5mm}\texttt{noroll\_v2.0\_10yrs} & No rolling \reedit{(\texttt{noroll})}. \\
         \hspace{5mm}\texttt{rolling\_nsX\_rwY\_\reedit{*}} & Splits the sky into X regions \reedit{(\texttt{nsX})} with Y\% \reedit{(\texttt{rwY})} strength of \reedit{\texttt{rolling}}.\\
         \hspace{5mm}\texttt{rolling\_nsX\_strengthY\_\reedit{*}} & Splits the sky into X regions \reedit{(\texttt{nsX})} with Y\% \reedit{(\texttt{strengthY})} strength of \reedit{\texttt{rolling}}.\\
         \hspace{5mm}\texttt{rolling\_bulge\_nsX\_rwY\_\reedit{*}} & Splits the Galactic bulge into X regions \reedit{(\texttt{nsX})} with Y\% \reedit{(\texttt{rwY})} strength of \reedit{\texttt{rolling}}. \\
         \hspace{5mm}\texttt{rolling\_early\_v2.0\_10yrs} & \reedit{Rolling cadence} beginning in year one of LSST. \\
         \hspace{5mm}\texttt{six\_rolling\_\reedit{*}} & Splits sky into six regions and \reedit{performs a rolling cadence}. \\
         \hspace{5mm}\texttt{rolling\_bulge\_6\_v2.0\_10yrs} & Splits Galactic bulge into six regions and \reedit{performs a rolling cadence}. \\
         \hspace{5mm}\texttt{rolling\_with\_const\_\reedit{*}} & Intersperses rolling \reedit{cadence} with constant years. \\

         \hspace{5mm}\texttt{rolling\_flipped\_\reedit{*}} & Flips the order of the 2-band 90\% strength rolling cadence. \\

         \hspace{5mm}\reedit{\texttt{rolling\_all\_sky\_*}} & \reedit{Rolls the whole sky with the same cadence.} \\
         
         triplets & Triplet observations in a single night strategies. \\
         \hspace{5mm}\texttt{\reedit{presto}\_gapX} & Triplets spaced X hours \reedit{(\texttt{gapX})} apart. \\
         \hspace{5mm}\texttt{\reedit{presto}\_half\_gapX} & Triplets spaced X hours \reedit{(\texttt{gapX})} apart every other night. \\

        \hspace{5mm}\texttt{long\_gaps\_nightsoffX\_\reedit{delayed-1}} & Triplets every X nights \reedit{(\texttt{nightsoffX})} . \\
         \hspace{5mm}\texttt{long\_gaps\_nightsoffX\_delayed\reedit{1827}} & Triplets every X nights \reedit{(\texttt{nightsoffX})} starting after year 5. \\

         twilight NEO & Survey added in twilight to observe Near Earth Objects. \\

         \texttt{vary\_nes\_nesfracX\_\reedit{*}} & Survey strategy spends X\% \reedit{(\texttt{nesfracX})} of survey time on the North Ecliptic Spur. 
    \end{tabular}
    \caption{Summary of OpSims alphabetical by family that are relevant to microlensing and Milky Way science with descriptions of pertinent aspects. Those that end in an underscore \reedit{and star (\_*)} indicate there are multiple OpSims related to that entry. Indented entries belong to the family listed above. See \cite{Jones_baseline_2018-2020} and Rubin technical note \href{https://pstn-055.lsst.io/}{PSTN-055} for more detailed descriptions.}
    \label{tab:opsim_summary}
\end{table}

\begin{table}[h]
    \centering
    \begin{tabular}{l|lrrr}
        \toprule
         Metric & $t_{\rm E}$ (days) & Galactic Bulge & Galactic Plane & Magellenic Clouds \\
         \hline
        Discovery & 1 - 5 & 220 & 810 & 51 \\
        & 5 - 10 & 486 & 1618 & 113 \\
        & 10 - 20 & 752 & 2497 & 162 \\
        & 20 - 30 & 926 & 3123 & 209 \\
        & 30 - 60 & 1116 & 3768 & 249 \\
        & 60 - 90 & 1267 & 4317 & 291 \\
        & 100 - 200 & 1542 & 5259 & 343 \\
        & 200 - 500 & 1862 & 6445 & 421 \\
        & 500 - 1000 & 1839 & 6490 & 430 \\
        \hline
        Npts & 10 - 20 & 707 & 2251 & 159 \\
        & 20 - 30 & 1319 & 4199 & 257 \\
        & 30 - 60 & 1761 & 5683 & 342 \\
        & 200 - 500 & 2800 & 9855 & 655 \\
        \hline
        Fisher & 10 - 20 & 224 & 792 & 26 \\
        & 20 - 30 & 512 & 1738 & 80 \\
        & 30 - 60 & 822 & 2753 & 146 \\
        & 200 - 500 & 2403 & 8252 & 529 \\
    \end{tabular}
    \caption{\cedit{Metric results for \texttt{baseline\_v2.0\_10yrs} breaking down metric results according to location since the $t_{\rm E}$ distributions are change as a function of location. Events were sorted into these locations using the Healpix maps from \cite{Street:2023}. The Discovery metric is the number of events (out of 10,000 in the entire footprint) with 2 points on the rising side of the lightcurve with $\geq 3\sigma$ difference in magnitude; the Npts metric has the number of lightcurves with at least 10 points in $t_0 \pm t_{\rm E}$ (see Section \ref{sec: microlensing metric}); and the Fisher metric has the number of lightcurves with $\frac{\sigma_{t_{\rm E}}}{t_{\rm E}} < 0.1$ (see Section \ref{sec: fisher matrix}).}}
    \label{tab:location breakdown}
\end{table}

\pagebreak

\begin{longtable}{lrrr}
\toprule
OpSim & Nvisits & Area w\/ $>$ 825 visits & Median WFD Inter-Night Gaps \\
\hline
\midrule
\endfirsthead
\toprule
OpSim & Nvisits & Area w\/ $>$ 825 visits & Median WFD Inter-Night Gaps \\
\hline
\midrule
\endhead
\midrule
\multicolumn{4}{r}{Continued on next page} \\
\midrule
\endfoot
\bottomrule
\endlastfoot
baseline\_v3.0\_10yrs & 2086079 & 2921.58 & 2.95 \\
baseline\_v2.2\_10yrs & 2074975 & 10569.22 & 2.98 \\
baseline\_v2.1\_10yrs & 2081749 & 12434.14 & 3.02 \\
baseline\_v2.0\_10yrs & 2086980 & 12893.23 & 3.00 \\
baseline\_retrofoot\_v2.0\_10yrs & 2086534 & 17510.18 & 2.98 \\
retro\_baseline\_v2.0\_10yrs & 2048566 & 14611.26 & 3.08 \\
noroll\_v2.0\_10yrs & 2083375 & 12558.35 & 3.93 \\
rolling\_ns2\_rw0.5\_v2.0\_10yrs & 2084510 & 12614.58 & 3.83 \\
rolling\_ns3\_rw0.5\_v2.0\_10yrs & 2085499 & 12729.57 & 3.08 \\
rolling\_ns2\_rw0.9\_v2.0\_10yrs & 2086980 & 12893.23 & 3.00 \\
rolling\_ns3\_rw0.9\_v2.0\_10yrs & 2090909 & 13255.80 & 2.95 \\
rolling\_bulge\_ns2\_rw0.5\_v2.0\_10yrs & 2084413 & 12600.32 & 3.83 \\
rolling\_bulge\_ns2\_rw0.8\_v2.0\_10yrs & 2086550 & 12733.76 & 3.01 \\
rolling\_bulge\_ns2\_rw0.9\_v2.0\_10yrs & 2087907 & 12940.23 & 3.00 \\
rolling\_all\_sky\_ns2\_rw0.9\_v2.0\_10yrs & 2088105 & 12900.78 & 3.00 \\
roll\_early\_v2.0\_10yrs & 2088889 & 12999.82 & 2.98 \\
six\_rolling\_ns6\_rw0.5\_v2.0\_10yrs & 2084148 & 12342.65 & 3.00 \\
six\_rolling\_ns6\_rw0.9\_v2.0\_10yrs & 2089610 & 12735.44 & 2.03 \\
rolling\_bulge\_6\_v2.0\_10yrs & 2085745 & 12492.89 & 2.99 \\
noroll\_v2.2\_10yrs & 2071905 & 10156.29 & 3.88 \\
rolling\_ns2\_strength0.50v2.2\_10yrs & 2073211 & 10356.88 & 3.07 \\
rolling\_ns3\_strength0.50v2.2\_10yrs & 2073781 & 10393.81 & 3.02 \\
rolling\_ns2\_strength0.80v2.2\_10yrs & 2074478 & 10560.83 & 2.99 \\
rolling\_ns3\_strength0.80v2.2\_10yrs & 2077231 & 10936 & 2.94 \\
rolling\_ns2\_strength0.90v2.2\_10yrs & 2074975 & 10569.22 & 2.98 \\
rolling\_flipped\_v2.2\_10yrs & 2075364 & 10481.94 & 2.98 \\
rolling\_ns3\_strength0.90v2.2\_10yrs & 2078058 & 11031.68 & 2.92 \\
roll\_with\_const\_roll\_indx0\_v2.2\_10yrs & 2074872 & 10763.10 & 2.98 \\
roll\_with\_const\_roll\_indx1\_v2.2\_10yrs & 2074885 & 10714.42 & 2.98 \\
roll\_with\_const\_roll\_indx2\_v2.2\_10yrs & 2075185 & 10791.64 & 2.98 \\
roll\_with\_const\_roll\_indx3\_v2.2\_10yrs & 2074993 & 10629.65 & 2.98 \\
roll\_with\_const\_roll\_indx4\_v2.2\_10yrs & 2075387 & 10711.07 & 2.98 \\
roll\_with\_const\_roll\_indx5\_v2.2\_10yrs & 2074878 & 10516.35 & 2.98 \\
roll\_with\_const\_roll\_indx6\_v2.2\_10yrs & 2076044 & 10046.34 & 2.97 \\
roll\_with\_const\_roll\_indx7\_v2.2\_10yrs & 2076338 & 9976.68 & 2.97 \\
roll\_with\_const\_roll\_indx8\_v2.2\_10yrs & 2075707 & 10893.19 & 2.95 \\
presto\_gap1.5\_v2.0\_10yrs & 1991926 & 2177.97 & 3.95 \\
presto\_gap2.0\_v2.0\_10yrs & 2004698 & 3255.62 & 3.91 \\
presto\_gap2.5\_v2.0\_10yrs & 1986054 & 2386.95 & 3.85 \\
presto\_gap3.0\_v2.0\_10yrs & 1986782 & 2732.74 & 3.72 \\
presto\_gap3.5\_v2.0\_10yrs & 1992713 & 3034.05 & 3.68 \\
presto\_gap4.0\_v2.0\_10yrs & 1998152 & 3530.07 & 3.22 \\
presto\_half\_gap1.5\_v2.0\_10yrs & 2039120 & 7436.14 & 3.96 \\
presto\_half\_gap2.0\_v2.0\_10yrs & 2052706 & 9016.53 & 3.93 \\
presto\_half\_gap2.5\_v2.0\_10yrs & 2044849 & 8104.22 & 3.91 \\
presto\_half\_gap3.0\_v2.0\_10yrs & 2044562 & 8098.34 & 3.89 \\
presto\_half\_gap3.5\_v2.0\_10yrs & 2046991 & 8476.87 & 3.86 \\
presto\_half\_gap4.0\_v2.0\_10yrs & 2049239 & 8649.76 & 3.81 \\
long\_gaps\_nightsoff7\_delayed1827\_v2.0\_10yrs & 2085109 & 12702.71 & 3.01 \\
long\_gaps\_nightsoff6\_delayed1827\_v2.0\_10yrs & 2084436 & 12451.76 & 3.00 \\
long\_gaps\_nightsoff5\_delayed1827\_v2.0\_10yrs & 2084570 & 12548.28 & 3.01 \\
long\_gaps\_nightsoff4\_delayed1827\_v2.0\_10yrs & 2083856 & 12463.51 & 3.01 \\
long\_gaps\_nightsoff3\_delayed1827\_v2.0\_10yrs & 2082889 & 12372.87 & 3.01 \\
long\_gaps\_nightsoff2\_delayed1827\_v2.0\_10yrs & 2081844 & 12188.22 & 3.01 \\
long\_gaps\_nightsoff1\_delayed1827\_v2.0\_10yrs & 2078434 & 11770.25 & 3.02 \\
long\_gaps\_nightsoff0\_delayed1827\_v2.0\_10yrs & 2067452 & 10288.90 & 2.99 \\
long\_gaps\_nightsoff7\_delayed-1\_v2.0\_10yrs & 2083825 & 12377.90 & 3.02 \\
long\_gaps\_nightsoff6\_delayed-1\_v2.0\_10yrs & 2081933 & 12266.28 & 3.02 \\
long\_gaps\_nightsoff5\_delayed-1\_v2.0\_10yrs & 2082106 & 12234.38 & 3.02 \\
long\_gaps\_nightsoff4\_delayed-1\_v2.0\_10yrs & 2080392 & 12098.42 & 3.02 \\
long\_gaps\_nightsoff3\_delayed-1\_v2.0\_10yrs & 2079329 & 12009.45 & 3.03 \\
long\_gaps\_nightsoff2\_delayed-1\_v2.0\_10yrs & 2076198 & 11510.91 & 3.02 \\
long\_gaps\_nightsoff1\_delayed-1\_v2.0\_10yrs & 2070980 & 10896.55 & 3.10 \\
long\_gaps\_nightsoff0\_delayed-1\_v2.0\_10yrs & 2046745 & 7792.84 & 3.00 \\
vary\_nes\_nesfrac0.01\_v2.0\_10yrs & 2087392 & 17534.52 & 2.98 \\
vary\_nes\_nesfrac0.05\_v2.0\_10yrs & 2087572 & 17223.14 & 2.99 \\
vary\_nes\_nesfrac0.10\_v2.0\_10yrs & 2087761 & 16779.16 & 2.99 \\
vary\_nes\_nesfrac0.15\_v2.0\_10yrs & 2087741 & 16146.33 & 3.00 \\
vary\_nes\_nesfrac0.20\_v2.0\_10yrs & 2087462 & 15380.06 & 2.99 \\
vary\_nes\_nesfrac0.25\_v2.0\_10yrs & 2087377 & 14228.55 & 3.00 \\
vary\_nes\_nesfrac0.30\_v2.0\_10yrs & 2086980 & 12893.23 & 3.00 \\
vary\_nes\_nesfrac0.35\_v2.0\_10yrs & 2087488 & 11059.37 & 3.01 \\
vary\_nes\_nesfrac0.40\_v2.0\_10yrs & 2087857 & 9061.01 & 3.01 \\
vary\_nes\_nesfrac0.45\_v2.0\_10yrs & 2087796 & 7089.51 & 3.02 \\
vary\_nes\_nesfrac0.50\_v2.0\_10yrs & 2087928 & 5425.19 & 3.03 \\
vary\_nes\_nesfrac0.55\_v2.0\_10yrs & 2087662 & 4007.63 & 3.03 \\
vary\_nes\_nesfrac0.75\_v2.0\_10yrs & 2088056 & 2155.31 & 3.07 \\
vary\_nes\_nesfrac1.00\_v2.0\_10yrs & 2089159 & 5678.66 & 2.99 \\
vary\_gp\_gpfrac0.01\_v2.0\_10yrs & 2088147 & 16736.35 & 2.99 \\
vary\_gp\_gpfrac0.05\_v2.0\_10yrs & 2088534 & 16665.01 & 2.99 \\
vary\_gp\_gpfrac0.10\_v2.0\_10yrs & 2088477 & 16511.42 & 2.99 \\
vary\_gp\_gpfrac0.15\_v2.0\_10yrs & 2088838 & 15621.77 & 2.99 \\
vary\_gp\_gpfrac0.20\_v2.0\_10yrs & 2088077 & 14733.80 & 3.00 \\
vary\_gp\_gpfrac0.25\_v2.0\_10yrs & 2088389 & 13427.02 & 3.00 \\
vary\_gp\_gpfrac0.30\_v2.0\_10yrs & 2087314 & 11804.67 & 3.01 \\
vary\_gp\_gpfrac0.35\_v2.0\_10yrs & 2086873 & 9751.75 & 3.01 \\
vary\_gp\_gpfrac0.40\_v2.0\_10yrs & 2086160 & 7674.50 & 3.02 \\
vary\_gp\_gpfrac0.45\_v2.0\_10yrs & 2085954 & 5786.93 & 3.02 \\
vary\_gp\_gpfrac0.50\_v2.0\_10yrs & 2085679 & 3880.05 & 3.02 \\
vary\_gp\_gpfrac0.55\_v2.0\_10yrs & 2085386 & 2317.29 & 3.03 \\
vary\_gp\_gpfrac0.75\_v2.0\_10yrs & 2084363 & 211.50 & 3.46 \\
vary\_gp\_gpfrac1.00\_v2.0\_10yrs & 2084391 & 118.34 & 3.94 \\
plane\_priority\_priority1.2\_pbf\_v2.1\_10yrs & 2019424 & 6545.65 & 3.00 \\
plane\_priority\_priority0.9\_pbf\_v2.1\_10yrs & 2019086 & 5655.16 & 3.01 \\
plane\_priority\_priority0.6\_pbf\_v2.1\_10yrs & 2018906 & 2586.70 & 3.02 \\
plane\_priority\_priority0.4\_pbf\_v2.1\_10yrs & 2019352 & 604.29 & 3.09 \\
plane\_priority\_priority0.3\_pbf\_v2.1\_10yrs & 2019186 & 156.11 & 3.87 \\
plane\_priority\_priority0.2\_pbf\_v2.1\_10yrs & 2018674 & 367.61 & 3.15 \\
plane\_priority\_priority0.1\_pbf\_v2.1\_10yrs & 2013032 & 119.18 & 2.97 \\
plane\_priority\_priority1.2\_pbt\_v2.1\_10yrs & 2019424 & 6545.65 & 3.00 \\
plane\_priority\_priority0.9\_pbt\_v2.1\_10yrs & 2019098 & 4391.18 & 3.01 \\
plane\_priority\_priority0.6\_pbt\_v2.1\_10yrs & 2019067 & 2682.38 & 3.01 \\
plane\_priority\_priority0.4\_pbt\_v2.1\_10yrs & 2020233 & 612.68 & 3.08 \\
plane\_priority\_priority0.3\_pbt\_v2.1\_10yrs & 2018919 & 160.31 & 3.87 \\
plane\_priority\_priority0.2\_pbt\_v2.1\_10yrs & 2018144 & 349.99 & 3.20 \\
plane\_priority\_priority0.1\_pbt\_v2.1\_10yrs & 2013710 & 123.38 & 2.98 \\
pencil\_fs1\_v2.1\_10yrs & 1948456 & 890.49 & 2.99 \\
pencil\_fs2\_v2.1\_10yrs & 1948335 & 1117.94 & 2.99 \\
good\_seeing\_gsw0.0\_v2.1\_10yrs & 2083703 & 12814.34 & 3.03 \\
good\_seeing\_gsw1.0\_v2.1\_10yrs & 2082749 & 12507.15 & 3.03 \\
good\_seeing\_gsw3.0\_v2.1\_10yrs & 2081749 & 12434.14 & 3.02 \\
good\_seeing\_gsw6.0\_v2.1\_10yrs & 2079596 & 12217.60 & 3.02 \\
good\_seeing\_gsw10.0\_v2.1\_10yrs & 2078628 & 12151.29 & 3.02 \\
good\_seeing\_gsw20.0\_v2.1\_10yrs & 2077820 & 11874.33 & 3.02 \\
good\_seeing\_gsw50.0\_v2.1\_10yrs & 2077887 & 11885.24 & 3.02 \\
virgo\_cluster\_v2.0\_10yrs & 2087802 & 12161.36 & 3.00 \\
carina\_v2.0\_10yrs & 2087222 & 12519.74 & 2.96 \\
carina\_v2.2\_10yrs & 2062416 & 8796.64 & 2.98 \\
smc\_movie\_v2.0\_10yrs & 2089048 & 12732.08 & 3.00 \\
roman\_v2.0\_10yrs & 2083719 & 12283.06 & 3.00 \\
north\_stripe\_v2.0\_10yrs & 2089053 & 7379.07 & 3.02 \\
short\_exp\_v2.0\_10yrs & 2180235 & 10664.07 & 2.99 \\
short\_exp\_v2.2\_10yrs & 2115446 & 9363.16 & 2.97 \\
multi\_short\_v2.0\_10yrs & 3584186 & 151.07 & 3.03 \\
twilight\_neo\_nightpattern4v2.0\_10yrs & 2329938 & 5852.39 & 3.47 \\
twilight\_neo\_nightpattern7v2.0\_10yrs & 2417826 & 3772.62 & 3.10 \\
twilight\_neo\_nightpattern3v2.0\_10yrs & 2420218 & 3776.82 & 3.05 \\
twilight\_neo\_nightpattern6v2.0\_10yrs & 2524598 & 1927.86 & 3.06 \\
twilight\_neo\_nightpattern5v2.0\_10yrs & 2587302 & 1227.89 & 3.03 \\
twilight\_neo\_nightpattern2v2.0\_10yrs & 2586114 & 1309.30 & 3.05 \\
twilight\_neo\_nightpattern1v2.0\_10yrs & 3079761 & 114.98 & 2.98 \\
no\_repeat\_rpw-1.0\_v2.1\_10yrs & 2078336 & 12360.28 & 2.97 \\
no\_repeat\_rpw-2.0\_v2.1\_10yrs & 2077671 & 12258.72 & 2.95 \\
no\_repeat\_rpw-5.0\_v2.1\_10yrs & 2076775 & 12158.01 & 2.92 \\
no\_repeat\_rpw-10.0\_v2.1\_10yrs & 2077379 & 12262.08 & 2.90 \\
no\_repeat\_rpw-20.0\_v2.1\_10yrs & 2076942 & 11854.18 & 2.90 \\
no\_repeat\_rpw-100.0\_v2.1\_10yrs & 2077286 & 12038.83 & 2.91 \\
supress\_and\_triplets\_v2.2\_10yrs & 2070268 & 9985.08 & 3.02 \\
draft\_connected\_v2.99\_10yrs & 2076864 & 1953.04 & 3.07 \\
draft\_v2.99\_10yrs & 2076066 & 2112.50 & 3.04 \\
roll\_early\_v2.99\_10yrs & 2077239 & 1945.48 & 2.97 \\
dd6\_v2.99\_10yrs & 2077797 & 5309.37 & 3.03 \\
low\_gp\_v2.99\_10yrs & 2077254 & 8264.52 & 3.01 \\
light\_roll\_v2.99\_10yrs & 2082888 & 650.45 & 3.07 \\
draft2\_rw0.5\_v2.99\_10yrs & 2082209 & 1545.14 & 3.02 \\
draft2\_rw0.9\_v2.99\_10yrs & 2085283 & 2368.49 & 2.95 \\
draft2\_rw0.9\_uz\_v2.99\_10yrs & 2086079 & 2921.58 & 2.95 \\

\caption{\reedit{Summary statistics of OpSims in this paper. Nvisits corresponds to the ``Nvisits All visits" metric and is the total number of visits through the entire survey. Area w $>$ 825 visits corresponds to the ``fOArea fO All sky HealpixSlicer" metric and is the area in square degrees that received over 825 visits. Median WFD Inter-Night Gaps corresponds to the ``Median Median Inter-Night Gap WFD all bands HealpixSubsetSlicer" metric and is the median gap in days between observations in the WFD region including all filters. The rest of the metric results for these OpSims can be found in \url{https://github.com/lsst-pst/survey_strategy/blob/main/fbs_2.0/summary_2023_01_01.csv}.}}
    \label{tab:general_metrics}
\end{longtable}

\begin{longtable}{lrrrrrrrrr}
\toprule
& \multicolumn{9}{c}{Detect Metric ($t_{\rm E}$ range in days)} \\
OpSim & 1-5 & 5-10 & 10-20 & 20-30 & 30-60 & 60-90 & 100-200 & 200-500 & 500-1000 \\
\hline
\midrule
\endfirsthead
\toprule
& \multicolumn{9}{c}{Detect Metric ($t_{\rm E}$ range in days)} \\
OpSim & 1-5 & 5-10 & 10-20 & 20-30 & 30-60 & 60-90 & 100-200 & 200-500 & 500-1000 \\
\hline
\midrule
\endhead
\midrule
\multicolumn{10}{r}{Continued on next page} \\
\midrule
\endfoot
\bottomrule
\endlastfoot
baseline\_v3.0\_10yrs & 779 & 1618 & 2518 & 3239 & 3917 & 4559 & 5563 & 6674 & 6693 \\
baseline\_v2.2\_10yrs & 860 & 1656 & 2539 & 3212 & 3900 & 4462 & 5462 & 6710 & 6765 \\
baseline\_v2.1\_10yrs & 751 & 1535 & 2440 & 3111 & 3764 & 4312 & 5290 & 6639 & 6773 \\
baseline\_v2.0\_10yrs & 761 & 1580 & 2427 & 3116 & 3747 & 4284 & 5234 & 6551 & 6600 \\
baseline\_retrofoot\_v2.0\_10yrs & 225 & 472 & 794 & 1097 & 1565 & 2064 & 2922 & 4629 & 5626 \\
retro\_baseline\_v2.0\_10yrs & 195 & 385 & 630 & 883 & 1179 & 1533 & 2256 & 4147 & 5557 \\
noroll\_v2.0\_10yrs & 785 & 1633 & 2506 & 3154 & 3754 & 4274 & 5254 & 6587 & 6668 \\
rolling\_ns2\_rw0.5\_v2.0\_10yrs & 792 & 1609 & 2464 & 3163 & 3815 & 4314 & 5239 & 6560 & 6657 \\
rolling\_ns3\_rw0.5\_v2.0\_10yrs & 773 & 1599 & 2400 & 3062 & 3722 & 4253 & 5164 & 6463 & 6574 \\
rolling\_ns2\_rw0.9\_v2.0\_10yrs & 761 & 1580 & 2427 & 3116 & 3747 & 4284 & 5234 & 6551 & 6600 \\
rolling\_ns3\_rw0.9\_v2.0\_10yrs & 774 & 1522 & 2354 & 3058 & 3680 & 4259 & 5202 & 6470 & 6569 \\
rolling\_bulge\_ns2\_rw0.5\_v2.0\_10yrs & 777 & 1552 & 2358 & 2986 & 3610 & 4156 & 5097 & 6466 & 6573 \\
rolling\_bulge\_ns2\_rw0.8\_v2.0\_10yrs & 763 & 1538 & 2298 & 2929 & 3613 & 4183 & 5158 & 6446 & 6546 \\
rolling\_bulge\_ns2\_rw0.9\_v2.0\_10yrs & 752 & 1486 & 2288 & 2957 & 3534 & 4133 & 5095 & 6409 & 6572 \\
rolling\_all\_sky\_ns2\_rw0.9\_v2.0\_10yrs & 737 & 1473 & 2270 & 2950 & 3557 & 4133 & 5047 & 6365 & 6543 \\
roll\_early\_v2.0\_10yrs & 786 & 1577 & 2445 & 3100 & 3752 & 4282 & 5255 & 6505 & 6595 \\
six\_rolling\_ns6\_rw0.5\_v2.0\_10yrs & 742 & 1519 & 2387 & 3073 & 3736 & 4328 & 5291 & 6552 & 6661 \\
six\_rolling\_ns6\_rw0.9\_v2.0\_10yrs & 707 & 1449 & 2231 & 2900 & 3617 & 4230 & 5165 & 6362 & 6519 \\
rolling\_bulge\_6\_v2.0\_10yrs & 645 & 1193 & 1759 & 2256 & 2843 & 3416 & 4419 & 5859 & 6341 \\
noroll\_v2.2\_10yrs & 853 & 1679 & 2546 & 3210 & 3851 & 4419 & 5430 & 6795 & 6823 \\
rolling\_ns2\_strength0.50v2.2\_10yrs & 795 & 1636 & 2503 & 3180 & 3864 & 4424 & 5455 & 6723 & 6795 \\
rolling\_ns3\_strength0.50v2.2\_10yrs & 798 & 1638 & 2536 & 3228 & 3885 & 4446 & 5464 & 6715 & 6753 \\
rolling\_ns2\_strength0.80v2.2\_10yrs & 797 & 1617 & 2504 & 3173 & 3874 & 4453 & 5471 & 6740 & 6769 \\
rolling\_ns3\_strength0.80v2.2\_10yrs & 806 & 1599 & 2443 & 3074 & 3748 & 4327 & 5353 & 6605 & 6706 \\
rolling\_ns2\_strength0.90v2.2\_10yrs & 860 & 1656 & 2539 & 3212 & 3900 & 4462 & 5462 & 6710 & 6765 \\
rolling\_flipped\_v2.2\_10yrs & 837 & 1651 & 2551 & 3214 & 3901 & 4474 & 5487 & 6732 & 6765 \\
rolling\_ns3\_strength0.90v2.2\_10yrs & 791 & 1605 & 2454 & 3140 & 3829 & 4425 & 5494 & 6694 & 6766 \\
roll\_with\_const\_roll\_indx0\_v2.2\_10yrs & 820 & 1669 & 2531 & 3218 & 3903 & 4440 & 5458 & 6695 & 6748 \\
roll\_with\_const\_roll\_indx1\_v2.2\_10yrs & 837 & 1688 & 2580 & 3259 & 3914 & 4488 & 5510 & 6722 & 6757 \\
roll\_with\_const\_roll\_indx2\_v2.2\_10yrs & 848 & 1666 & 2598 & 3244 & 3937 & 4516 & 5545 & 6744 & 6779 \\
roll\_with\_const\_roll\_indx3\_v2.2\_10yrs & 802 & 1684 & 2574 & 3230 & 3910 & 4460 & 5471 & 6687 & 6755 \\
roll\_with\_const\_roll\_indx4\_v2.2\_10yrs & 826 & 1663 & 2552 & 3131 & 3770 & 4327 & 5358 & 6640 & 6711 \\
roll\_with\_const\_roll\_indx5\_v2.2\_10yrs & 822 & 1667 & 2548 & 3165 & 3774 & 4308 & 5301 & 6638 & 6735 \\
roll\_with\_const\_roll\_indx6\_v2.2\_10yrs & 792 & 1687 & 2588 & 3255 & 3907 & 4464 & 5465 & 6670 & 6752 \\
roll\_with\_const\_roll\_indx7\_v2.2\_10yrs & 822 & 1628 & 2528 & 3121 & 3796 & 4346 & 5367 & 6641 & 6713 \\
roll\_with\_const\_roll\_indx8\_v2.2\_10yrs & 825 & 1661 & 2563 & 3169 & 3838 & 4347 & 5370 & 6644 & 6732 \\
presto\_gap1.5\_v2.0\_10yrs & 492 & 1046 & 1627 & 2105 & 2619 & 3141 & 3879 & 5424 & 6047 \\
presto\_gap2.0\_v2.0\_10yrs & 511 & 1058 & 1659 & 2158 & 2655 & 3149 & 3935 & 5575 & 6194 \\
presto\_gap2.5\_v2.0\_10yrs & 544 & 1123 & 1724 & 2199 & 2771 & 3314 & 4194 & 5720 & 6298 \\
presto\_gap3.0\_v2.0\_10yrs & 545 & 1100 & 1688 & 2175 & 2689 & 3224 & 4064 & 5673 & 6292 \\
presto\_gap3.5\_v2.0\_10yrs & 529 & 1069 & 1704 & 2200 & 2748 & 3260 & 4179 & 5722 & 6284 \\
presto\_gap4.0\_v2.0\_10yrs & 564 & 1096 & 1650 & 2127 & 2612 & 3095 & 3977 & 5624 & 6210 \\
presto\_half\_gap1.5\_v2.0\_10yrs & 645 & 1344 & 2232 & 2952 & 3661 & 4250 & 5208 & 6436 & 6543 \\
presto\_half\_gap2.0\_v2.0\_10yrs & 638 & 1386 & 2225 & 2938 & 3644 & 4203 & 5155 & 6341 & 6486 \\
presto\_half\_gap2.5\_v2.0\_10yrs & 711 & 1446 & 2325 & 3035 & 3745 & 4323 & 5290 & 6495 & 6567 \\
presto\_half\_gap3.0\_v2.0\_10yrs & 652 & 1393 & 2230 & 2947 & 3684 & 4276 & 5232 & 6442 & 6538 \\
presto\_half\_gap3.5\_v2.0\_10yrs & 734 & 1451 & 2353 & 3067 & 3778 & 4354 & 5355 & 6528 & 6642 \\
presto\_half\_gap4.0\_v2.0\_10yrs & 694 & 1423 & 2235 & 2927 & 3599 & 4173 & 5093 & 6412 & 6500 \\
long\_gaps\_nightsoff7\_delayed1827\_v2.0\_10yrs & 782 & 1611 & 2463 & 3165 & 3798 & 4333 & 5266 & 6533 & 6584 \\
long\_gaps\_nightsoff6\_delayed1827\_v2.0\_10yrs & 760 & 1577 & 2406 & 3103 & 3743 & 4308 & 5226 & 6512 & 6569 \\
long\_gaps\_nightsoff5\_delayed1827\_v2.0\_10yrs & 774 & 1569 & 2420 & 3112 & 3786 & 4329 & 5223 & 6484 & 6562 \\
long\_gaps\_nightsoff4\_delayed1827\_v2.0\_10yrs & 753 & 1569 & 2415 & 3089 & 3747 & 4278 & 5169 & 6506 & 6565 \\
long\_gaps\_nightsoff3\_delayed1827\_v2.0\_10yrs & 760 & 1560 & 2417 & 3100 & 3728 & 4241 & 5146 & 6468 & 6543 \\
long\_gaps\_nightsoff2\_delayed1827\_v2.0\_10yrs & 761 & 1590 & 2453 & 3138 & 3765 & 4285 & 5215 & 6499 & 6557 \\
long\_gaps\_nightsoff1\_delayed1827\_v2.0\_10yrs & 763 & 1567 & 2415 & 3072 & 3720 & 4252 & 5213 & 6466 & 6541 \\
long\_gaps\_nightsoff0\_delayed1827\_v2.0\_10yrs & 724 & 1475 & 2323 & 3019 & 3661 & 4166 & 5094 & 6405 & 6518 \\
long\_gaps\_nightsoff7\_delayed-1\_v2.0\_10yrs & 774 & 1572 & 2475 & 3120 & 3776 & 4332 & 5271 & 6555 & 6625 \\
long\_gaps\_nightsoff6\_delayed-1\_v2.0\_10yrs & 753 & 1491 & 2382 & 3079 & 3728 & 4258 & 5180 & 6500 & 6608 \\
long\_gaps\_nightsoff5\_delayed-1\_v2.0\_10yrs & 736 & 1518 & 2378 & 3048 & 3747 & 4306 & 5221 & 6553 & 6659 \\
long\_gaps\_nightsoff4\_delayed-1\_v2.0\_10yrs & 745 & 1558 & 2417 & 3101 & 3744 & 4305 & 5282 & 6566 & 6658 \\
long\_gaps\_nightsoff3\_delayed-1\_v2.0\_10yrs & 752 & 1558 & 2408 & 3070 & 3700 & 4253 & 5175 & 6449 & 6575 \\
long\_gaps\_nightsoff2\_delayed-1\_v2.0\_10yrs & 779 & 1529 & 2401 & 3098 & 3743 & 4294 & 5159 & 6428 & 6495 \\
long\_gaps\_nightsoff1\_delayed-1\_v2.0\_10yrs & 713 & 1520 & 2379 & 3069 & 3724 & 4256 & 5201 & 6455 & 6589 \\
long\_gaps\_nightsoff0\_delayed-1\_v2.0\_10yrs & 680 & 1341 & 2151 & 2824 & 3502 & 4054 & 4923 & 6222 & 6437 \\
vary\_nes\_nesfrac0.01\_v2.0\_10yrs & 802 & 1602 & 2465 & 3092 & 3726 & 4270 & 5194 & 6502 & 6610 \\
vary\_nes\_nesfrac0.05\_v2.0\_10yrs & 796 & 1627 & 2526 & 3189 & 3777 & 4317 & 5308 & 6544 & 6632 \\
vary\_nes\_nesfrac0.10\_v2.0\_10yrs & 804 & 1607 & 2431 & 3098 & 3757 & 4327 & 5242 & 6487 & 6552 \\
vary\_nes\_nesfrac0.15\_v2.0\_10yrs & 811 & 1636 & 2515 & 3211 & 3859 & 4403 & 5352 & 6558 & 6614 \\
vary\_nes\_nesfrac0.20\_v2.0\_10yrs & 778 & 1583 & 2423 & 3090 & 3785 & 4323 & 5260 & 6534 & 6609 \\
vary\_nes\_nesfrac0.25\_v2.0\_10yrs & 783 & 1548 & 2411 & 3043 & 3686 & 4261 & 5189 & 6496 & 6635 \\
vary\_nes\_nesfrac0.30\_v2.0\_10yrs & 761 & 1580 & 2427 & 3116 & 3747 & 4284 & 5234 & 6551 & 6600 \\
vary\_nes\_nesfrac0.35\_v2.0\_10yrs & 761 & 1513 & 2319 & 3027 & 3654 & 4215 & 5184 & 6487 & 6607 \\
vary\_nes\_nesfrac0.40\_v2.0\_10yrs & 747 & 1551 & 2385 & 3092 & 3706 & 4270 & 5156 & 6483 & 6571 \\
vary\_nes\_nesfrac0.45\_v2.0\_10yrs & 780 & 1543 & 2345 & 3025 & 3680 & 4188 & 5113 & 6444 & 6577 \\
vary\_nes\_nesfrac0.50\_v2.0\_10yrs & 742 & 1529 & 2336 & 3002 & 3700 & 4209 & 5108 & 6418 & 6580 \\
vary\_nes\_nesfrac0.55\_v2.0\_10yrs & 724 & 1517 & 2365 & 3026 & 3680 & 4244 & 5198 & 6517 & 6590 \\
vary\_nes\_nesfrac0.75\_v2.0\_10yrs & 706 & 1451 & 2281 & 2958 & 3631 & 4181 & 5187 & 6493 & 6646 \\
vary\_nes\_nesfrac1.00\_v2.0\_10yrs & 698 & 1448 & 2302 & 2965 & 3642 & 4192 & 5143 & 6467 & 6560 \\
vary\_gp\_gpfrac0.01\_v2.0\_10yrs & 766 & 1541 & 2396 & 3046 & 3678 & 4215 & 5172 & 6446 & 6549 \\
vary\_gp\_gpfrac0.05\_v2.0\_10yrs & 811 & 1578 & 2439 & 3082 & 3708 & 4215 & 5114 & 6397 & 6508 \\
vary\_gp\_gpfrac0.10\_v2.0\_10yrs & 761 & 1577 & 2382 & 3051 & 3699 & 4250 & 5191 & 6482 & 6619 \\
vary\_gp\_gpfrac0.15\_v2.0\_10yrs & 759 & 1582 & 2419 & 3107 & 3786 & 4310 & 5245 & 6540 & 6620 \\
vary\_gp\_gpfrac0.20\_v2.0\_10yrs & 796 & 1587 & 2395 & 3074 & 3714 & 4251 & 5175 & 6467 & 6541 \\
vary\_gp\_gpfrac0.25\_v2.0\_10yrs & 748 & 1541 & 2453 & 3150 & 3795 & 4354 & 5248 & 6533 & 6699 \\
vary\_gp\_gpfrac0.30\_v2.0\_10yrs & 758 & 1499 & 2364 & 3068 & 3732 & 4298 & 5223 & 6531 & 6646 \\
vary\_gp\_gpfrac0.35\_v2.0\_10yrs & 747 & 1562 & 2439 & 3102 & 3740 & 4338 & 5220 & 6518 & 6578 \\
vary\_gp\_gpfrac0.40\_v2.0\_10yrs & 738 & 1524 & 2374 & 3061 & 3688 & 4273 & 5200 & 6520 & 6599 \\
vary\_gp\_gpfrac0.45\_v2.0\_10yrs & 704 & 1555 & 2378 & 3063 & 3741 & 4345 & 5301 & 6586 & 6657 \\
vary\_gp\_gpfrac0.50\_v2.0\_10yrs & 725 & 1515 & 2439 & 3152 & 3739 & 4343 & 5271 & 6580 & 6685 \\
vary\_gp\_gpfrac0.55\_v2.0\_10yrs & 801 & 1571 & 2462 & 3110 & 3792 & 4374 & 5353 & 6607 & 6626 \\
vary\_gp\_gpfrac0.75\_v2.0\_10yrs & 707 & 1483 & 2345 & 3039 & 3704 & 4289 & 5273 & 6589 & 6684 \\
vary\_gp\_gpfrac1.00\_v2.0\_10yrs & 707 & 1486 & 2306 & 3008 & 3689 & 4238 & 5211 & 6549 & 6672 \\
plane\_priority\_priority1.2\_pbf\_v2.1\_10yrs & 758 & 1549 & 2427 & 3087 & 3745 & 4313 & 5310 & 6649 & 6761 \\
plane\_priority\_priority0.9\_pbf\_v2.1\_10yrs & 779 & 1567 & 2454 & 3092 & 3772 & 4370 & 5377 & 6691 & 6775 \\
plane\_priority\_priority0.6\_pbf\_v2.1\_10yrs & 828 & 1671 & 2557 & 3256 & 3941 & 4544 & 5532 & 6768 & 6769 \\
plane\_priority\_priority0.4\_pbf\_v2.1\_10yrs & 761 & 1618 & 2502 & 3166 & 3915 & 4536 & 5552 & 6801 & 6849 \\
plane\_priority\_priority0.3\_pbf\_v2.1\_10yrs & 739 & 1536 & 2446 & 3174 & 3870 & 4475 & 5457 & 6738 & 6798 \\
plane\_priority\_priority0.2\_pbf\_v2.1\_10yrs & 717 & 1528 & 2412 & 3093 & 3821 & 4427 & 5389 & 6615 & 6696 \\
plane\_priority\_priority0.1\_pbf\_v2.1\_10yrs & 675 & 1428 & 2268 & 2974 & 3709 & 4327 & 5339 & 6617 & 6674 \\
plane\_priority\_priority1.2\_pbt\_v2.1\_10yrs & 758 & 1549 & 2427 & 3087 & 3745 & 4313 & 5310 & 6649 & 6761 \\
plane\_priority\_priority0.9\_pbt\_v2.1\_10yrs & 806 & 1595 & 2475 & 3140 & 3870 & 4428 & 5473 & 6766 & 6845 \\
plane\_priority\_priority0.6\_pbt\_v2.1\_10yrs & 779 & 1601 & 2493 & 3152 & 3819 & 4421 & 5354 & 6601 & 6677 \\
plane\_priority\_priority0.4\_pbt\_v2.1\_10yrs & 760 & 1631 & 2508 & 3232 & 3943 & 4568 & 5509 & 6808 & 6823 \\
plane\_priority\_priority0.3\_pbt\_v2.1\_10yrs & 727 & 1552 & 2405 & 3075 & 3783 & 4376 & 5341 & 6685 & 6744 \\
plane\_priority\_priority0.2\_pbt\_v2.1\_10yrs & 668 & 1470 & 2320 & 3038 & 3785 & 4413 & 5392 & 6694 & 6720 \\
plane\_priority\_priority0.1\_pbt\_v2.1\_10yrs & 672 & 1480 & 2338 & 3024 & 3771 & 4410 & 5387 & 6642 & 6704 \\
pencil\_fs1\_v2.1\_10yrs & 724 & 1519 & 2397 & 3081 & 3746 & 4309 & 5304 & 6648 & 6775 \\
pencil\_fs2\_v2.1\_10yrs & 768 & 1563 & 2383 & 3044 & 3741 & 4303 & 5204 & 6546 & 6595 \\
good\_seeing\_gsw0.0\_v2.1\_10yrs & 768 & 1569 & 2335 & 2995 & 3667 & 4223 & 5168 & 6502 & 6629 \\
good\_seeing\_gsw1.0\_v2.1\_10yrs & 713 & 1525 & 2326 & 2975 & 3620 & 4183 & 5109 & 6431 & 6559 \\
good\_seeing\_gsw3.0\_v2.1\_10yrs & 751 & 1535 & 2440 & 3111 & 3764 & 4312 & 5290 & 6639 & 6773 \\
good\_seeing\_gsw6.0\_v2.1\_10yrs & 752 & 1519 & 2389 & 3084 & 3701 & 4248 & 5120 & 6436 & 6583 \\
good\_seeing\_gsw10.0\_v2.1\_10yrs & 706 & 1490 & 2300 & 3013 & 3671 & 4236 & 5171 & 6478 & 6600 \\
good\_seeing\_gsw20.0\_v2.1\_10yrs & 766 & 1548 & 2384 & 3058 & 3721 & 4317 & 5269 & 6585 & 6700 \\
good\_seeing\_gsw50.0\_v2.1\_10yrs & 735 & 1495 & 2311 & 2997 & 3690 & 4281 & 5241 & 6590 & 6703 \\
virgo\_cluster\_v2.0\_10yrs & 770 & 1540 & 2380 & 3075 & 3756 & 4314 & 5225 & 6503 & 6609 \\
carina\_v2.0\_10yrs & 836 & 1634 & 2512 & 3194 & 3847 & 4394 & 5351 & 6617 & 6690 \\
carina\_v2.2\_10yrs & 797 & 1598 & 2469 & 3114 & 3763 & 4330 & 5302 & 6623 & 6698 \\
smc\_movie\_v2.0\_10yrs & 805 & 1587 & 2417 & 3098 & 3771 & 4336 & 5271 & 6599 & 6691 \\
roman\_v2.0\_10yrs & 903 & 1786 & 2612 & 3282 & 3887 & 4415 & 5332 & 6546 & 6599 \\
north\_stripe\_v2.0\_10yrs & 758 & 1517 & 2369 & 3046 & 3705 & 4250 & 5150 & 6451 & 6555 \\
short\_exp\_v2.0\_10yrs & 754 & 1524 & 2368 & 3054 & 3649 & 4165 & 5100 & 6486 & 6651 \\
short\_exp\_v2.2\_10yrs & 799 & 1622 & 2496 & 3141 & 3778 & 4354 & 5332 & 6644 & 6729 \\
multi\_short\_v2.0\_10yrs & 702 & 1409 & 2218 & 2895 & 3495 & 4034 & 4903 & 6209 & 6410 \\
twilight\_neo\_nightpattern4v2.0\_10yrs & 762 & 1634 & 2576 & 3312 & 3982 & 4545 & 5424 & 6346 & 6392 \\
twilight\_neo\_nightpattern7v2.0\_10yrs & 784 & 1611 & 2525 & 3241 & 3953 & 4502 & 5357 & 6281 & 6283 \\
twilight\_neo\_nightpattern3v2.0\_10yrs & 800 & 1618 & 2544 & 3244 & 3931 & 4504 & 5355 & 6316 & 6349 \\
twilight\_neo\_nightpattern6v2.0\_10yrs & 750 & 1630 & 2549 & 3232 & 3966 & 4494 & 5361 & 6287 & 6320 \\
twilight\_neo\_nightpattern5v2.0\_10yrs & 783 & 1617 & 2487 & 3198 & 3938 & 4509 & 5387 & 6339 & 6345 \\
twilight\_neo\_nightpattern2v2.0\_10yrs & 709 & 1516 & 2451 & 3217 & 3929 & 4470 & 5335 & 6197 & 6261 \\
twilight\_neo\_nightpattern1v2.0\_10yrs & 752 & 1590 & 2479 & 3203 & 3907 & 4422 & 5249 & 6182 & 6293 \\
no\_repeat\_rpw-1.0\_v2.1\_10yrs & 776 & 1566 & 2384 & 2997 & 3626 & 4221 & 5156 & 6446 & 6511 \\
no\_repeat\_rpw-2.0\_v2.1\_10yrs & 789 & 1554 & 2416 & 3093 & 3729 & 4318 & 5262 & 6658 & 6775 \\
no\_repeat\_rpw-5.0\_v2.1\_10yrs & 798 & 1592 & 2420 & 3057 & 3684 & 4239 & 5141 & 6392 & 6486 \\
no\_repeat\_rpw-10.0\_v2.1\_10yrs & 844 & 1604 & 2451 & 3116 & 3759 & 4312 & 5279 & 6565 & 6686 \\
no\_repeat\_rpw-20.0\_v2.1\_10yrs & 792 & 1576 & 2486 & 3203 & 3792 & 4325 & 5257 & 6580 & 6629 \\
no\_repeat\_rpw-100.0\_v2.1\_10yrs & 774 & 1597 & 2528 & 3203 & 3852 & 4406 & 5352 & 6629 & 6691 \\
supress\_and\_triplets\_v2.2\_10yrs & 781 & 1586 & 2475 & 3179 & 3802 & 4377 & 5336 & 6602 & 6666 \\
draft\_connected\_v2.99\_10yrs & 814 & 1687 & 2593 & 3356 & 4042 & 4607 & 5630 & 6775 & 6756 \\
draft\_v2.99\_10yrs & 806 & 1631 & 2489 & 3202 & 3876 & 4466 & 5529 & 6803 & 6823 \\
roll\_early\_v2.99\_10yrs & 852 & 1736 & 2633 & 3324 & 4049 & 4628 & 5635 & 6795 & 6838 \\
dd6\_v2.99\_10yrs & 862 & 1738 & 2578 & 3304 & 4001 & 4572 & 5573 & 6762 & 6772 \\
low\_gp\_v2.99\_10yrs & 493 & 1093 & 1795 & 2457 & 3178 & 3845 & 4890 & 6434 & 6711 \\
light\_roll\_v2.99\_10yrs & 854 & 1727 & 2726 & 3457 & 4120 & 4708 & 5707 & 6798 & 6851 \\
draft2\_rw0.5\_v2.99\_10yrs & 882 & 1717 & 2655 & 3413 & 4128 & 4732 & 5803 & 6867 & 6843 \\
draft2\_rw0.9\_v2.99\_10yrs & 831 & 1657 & 2540 & 3290 & 3969 & 4528 & 5546 & 6740 & 6841 \\
draft2\_rw0.9\_uz\_v2.99\_10yrs & 779 & 1618 & 2518 & 3239 & 3917 & 4559 & 5563 & 6674 & 6693 \\

\caption{\reedit{Detect Metric values of OpSims in this paper. Each entry has the number of lightcurves detected (defined in Section \ref{sec: microlensing metric}) out of 10,000 simulated lightcurves per entry. Each column has the $t_{\rm E}$ range of events simulated. The rest of the metric results for these OpSims can be found in \url{https://github.com/lsst-pst/survey_strategy/blob/main/fbs_2.0/summary_2023_01_01.csv}.}}
    \label{tab:detect_metric}
\end{longtable}

\begin{longtable}{lrrrr|rrrr}
\toprule
& \multicolumn{4}{c}{Npts Metric ($t_{\rm E}$ range in days)} & \multicolumn{4}{c}{Fisher Metric ($t_{\rm E}$ range in days)} \\
OpSim & 10-20 & 20-30 & 30-60 & 200-500 & 10-20 & 20-30 & 30-60 & 200-500 \\
\hline
\midrule
\endfirsthead
\toprule
& \multicolumn{4}{c}{Npts Metric ($t_{\rm E}$ range in days)} & \multicolumn{4}{c}{Fisher Metric ($t_{\rm E}$ range in days)} \\
OpSim & 10-20 & 20-30 & 30-60 & 200-500 & 10-20 & 20-30 & 30-60 & 200-500 \\
\hline
\midrule
\endhead
\midrule
\multicolumn{9}{r}{Continued on next page} \\
\midrule
\endfoot
\bottomrule
\endlastfoot
baseline\_v3.0\_10yrs & 0.24 & 0.41 & 0.56 & 1 & 0.05 & 0.12 & 0.22 & 0.80 \\
baseline\_v2.2\_10yrs & 0.27 & 0.44 & 0.58 & 1 & 0.08 & 0.17 & 0.28 & 0.85 \\
baseline\_v2.1\_10yrs & 0.27 & 0.44 & 0.59 & 1 & 0.07 & 0.16 & 0.26 & 0.82 \\
baseline\_v2.0\_10yrs & 0.26 & 0.44 & 0.59 & 1 & 0.07 & 0.16 & 0.27 & 0.82 \\
baseline\_retrofoot\_v2.0\_10yrs & 0.03 & 0.08 & 0.17 & 0.97 & 0.01 & 0.01 & 0.03 & 0.39 \\
retro\_baseline\_v2.0\_10yrs & 0.07 & 0.13 & 0.24 & 0.99 & 0.01 & 0.01 & 0.02 & 0.42 \\
noroll\_v2.0\_10yrs & 0.28 & 0.45 & 0.58 & 1 & 0.07 & 0.17 & 0.27 & 0.84 \\
rolling\_ns2\_rw0.5\_v2.0\_10yrs & 0.27 & 0.45 & 0.58 & 1 & 0.06 & 0.16 & 0.27 & 0.83 \\
rolling\_ns3\_rw0.5\_v2.0\_10yrs & 0.26 & 0.44 & 0.58 & 1 & 0.06 & 0.15 & 0.26 & 0.82 \\
rolling\_ns2\_rw0.9\_v2.0\_10yrs & 0.26 & 0.44 & 0.59 & 1 & 0.07 & 0.16 & 0.27 & 0.82 \\
rolling\_ns3\_rw0.9\_v2.0\_10yrs & 0.25 & 0.43 & 0.58 & 1 & 0.06 & 0.15 & 0.25 & 0.81 \\
rolling\_bulge\_ns2\_rw0.5\_v2.0\_10yrs & 0.26 & 0.42 & 0.57 & 1 & 0.07 & 0.15 & 0.25 & 0.82 \\
rolling\_bulge\_ns2\_rw0.8\_v2.0\_10yrs & 0.24 & 0.39 & 0.54 & 1 & 0.06 & 0.14 & 0.24 & 0.81 \\
rolling\_bulge\_ns2\_rw0.9\_v2.0\_10yrs & 0.25 & 0.39 & 0.53 & 1 & 0.06 & 0.14 & 0.23 & 0.80 \\
rolling\_all\_sky\_ns2\_rw0.9\_v2.0\_10yrs & 0.24 & 0.39 & 0.54 & 1 & 0.06 & 0.14 & 0.24 & 0.80 \\
roll\_early\_v2.0\_10yrs & 0.26 & 0.44 & 0.58 & 1 & 0.06 & 0.15 & 0.26 & 0.81 \\
six\_rolling\_ns6\_rw0.5\_v2.0\_10yrs & 0.26 & 0.45 & 0.59 & 1 & 0.06 & 0.15 & 0.26 & 0.82 \\
six\_rolling\_ns6\_rw0.9\_v2.0\_10yrs & 0.23 & 0.40 & 0.56 & 1 & 0.05 & 0.12 & 0.23 & 0.78 \\
rolling\_bulge\_6\_v2.0\_10yrs & 0.18 & 0.28 & 0.40 & 0.95 & 0.05 & 0.10 & 0.16 & 0.62 \\
noroll\_v2.2\_10yrs & 0.28 & 0.45 & 0.59 & 1 & 0.08 & 0.18 & 0.29 & 0.86 \\
rolling\_ns2\_strength0.50v2.2\_10yrs & 0.27 & 0.45 & 0.59 & 1 & 0.07 & 0.17 & 0.28 & 0.85 \\
rolling\_ns3\_strength0.50v2.2\_10yrs & 0.27 & 0.45 & 0.59 & 1 & 0.07 & 0.17 & 0.28 & 0.85 \\
rolling\_ns2\_strength0.80v2.2\_10yrs & 0.27 & 0.45 & 0.59 & 1 & 0.07 & 0.17 & 0.27 & 0.84 \\
rolling\_ns3\_strength0.80v2.2\_10yrs & 0.26 & 0.45 & 0.59 & 1 & 0.06 & 0.16 & 0.26 & 0.83 \\
rolling\_ns2\_strength0.90v2.2\_10yrs & 0.27 & 0.44 & 0.58 & 1 & 0.08 & 0.17 & 0.28 & 0.85 \\
rolling\_flipped\_v2.2\_10yrs & 0.27 & 0.45 & 0.59 & 1 & 0.07 & 0.17 & 0.28 & 0.85 \\
rolling\_ns3\_strength0.90v2.2\_10yrs & 0.26 & 0.44 & 0.59 & 1 & 0.06 & 0.16 & 0.27 & 0.83 \\
roll\_with\_const\_roll\_indx0\_v2.2\_10yrs & 0.27 & 0.45 & 0.58 & 1 & 0.07 & 0.17 & 0.28 & 0.84 \\
roll\_with\_const\_roll\_indx1\_v2.2\_10yrs & 0.27 & 0.45 & 0.59 & 1 & 0.07 & 0.17 & 0.28 & 0.84 \\
roll\_with\_const\_roll\_indx2\_v2.2\_10yrs & 0.27 & 0.44 & 0.58 & 1 & 0.07 & 0.17 & 0.28 & 0.84 \\
roll\_with\_const\_roll\_indx3\_v2.2\_10yrs & 0.28 & 0.45 & 0.59 & 1 & 0.07 & 0.18 & 0.28 & 0.85 \\
roll\_with\_const\_roll\_indx4\_v2.2\_10yrs & 0.27 & 0.45 & 0.58 & 1 & 0.07 & 0.17 & 0.27 & 0.84 \\
roll\_with\_const\_roll\_indx5\_v2.2\_10yrs & 0.27 & 0.45 & 0.58 & 1 & 0.07 & 0.18 & 0.27 & 0.84 \\
roll\_with\_const\_roll\_indx6\_v2.2\_10yrs & 0.27 & 0.45 & 0.59 & 1 & 0.07 & 0.17 & 0.28 & 0.84 \\
roll\_with\_const\_roll\_indx7\_v2.2\_10yrs & 0.27 & 0.45 & 0.58 & 1 & 0.07 & 0.17 & 0.27 & 0.84 \\
roll\_with\_const\_roll\_indx8\_v2.2\_10yrs & 0.26 & 0.45 & 0.58 & 1 & 0.07 & 0.17 & 0.27 & 0.84 \\
presto\_gap1.5\_v2.0\_10yrs & 0.22 & 0.33 & 0.46 & 1 & 0.04 & 0.10 & 0.18 & 0.74 \\
presto\_gap2.0\_v2.0\_10yrs & 0.22 & 0.34 & 0.46 & 1 & 0.05 & 0.11 & 0.18 & 0.76 \\
presto\_gap2.5\_v2.0\_10yrs & 0.21 & 0.33 & 0.45 & 1 & 0.05 & 0.11 & 0.18 & 0.76 \\
presto\_gap3.0\_v2.0\_10yrs & 0.21 & 0.32 & 0.44 & 0.99 & 0.05 & 0.10 & 0.17 & 0.75 \\
presto\_gap3.5\_v2.0\_10yrs & 0.21 & 0.32 & 0.43 & 1 & 0.05 & 0.12 & 0.18 & 0.76 \\
presto\_gap4.0\_v2.0\_10yrs & 0.22 & 0.33 & 0.44 & 0.99 & 0.05 & 0.11 & 0.17 & 0.75 \\
presto\_half\_gap1.5\_v2.0\_10yrs & 0.23 & 0.41 & 0.58 & 1 & 0.05 & 0.12 & 0.23 & 0.80 \\
presto\_half\_gap2.0\_v2.0\_10yrs & 0.23 & 0.41 & 0.58 & 1 & 0.05 & 0.13 & 0.24 & 0.80 \\
presto\_half\_gap2.5\_v2.0\_10yrs & 0.22 & 0.41 & 0.57 & 1 & 0.05 & 0.13 & 0.23 & 0.79 \\
presto\_half\_gap3.0\_v2.0\_10yrs & 0.23 & 0.41 & 0.56 & 1 & 0.05 & 0.13 & 0.24 & 0.80 \\
presto\_half\_gap3.5\_v2.0\_10yrs & 0.24 & 0.41 & 0.56 & 1 & 0.05 & 0.13 & 0.23 & 0.80 \\
presto\_half\_gap4.0\_v2.0\_10yrs & 0.23 & 0.41 & 0.56 & 1 & 0.05 & 0.14 & 0.25 & 0.80 \\
long\_gaps\_nightsoff7\_delayed1827\_v2.0\_10yrs & 0.26 & 0.45 & 0.59 & 1 & 0.07 & 0.16 & 0.27 & 0.82 \\
long\_gaps\_nightsoff6\_delayed1827\_v2.0\_10yrs & 0.26 & 0.44 & 0.59 & 1 & 0.06 & 0.16 & 0.27 & 0.82 \\
long\_gaps\_nightsoff5\_delayed1827\_v2.0\_10yrs & 0.25 & 0.44 & 0.58 & 1 & 0.06 & 0.16 & 0.27 & 0.81 \\
long\_gaps\_nightsoff4\_delayed1827\_v2.0\_10yrs & 0.26 & 0.45 & 0.59 & 1 & 0.06 & 0.16 & 0.27 & 0.82 \\
long\_gaps\_nightsoff3\_delayed1827\_v2.0\_10yrs & 0.26 & 0.45 & 0.58 & 1 & 0.06 & 0.16 & 0.26 & 0.82 \\
long\_gaps\_nightsoff2\_delayed1827\_v2.0\_10yrs & 0.26 & 0.45 & 0.59 & 1 & 0.06 & 0.16 & 0.26 & 0.82 \\
long\_gaps\_nightsoff1\_delayed1827\_v2.0\_10yrs & 0.26 & 0.44 & 0.58 & 1 & 0.06 & 0.15 & 0.26 & 0.81 \\
long\_gaps\_nightsoff0\_delayed1827\_v2.0\_10yrs & 0.24 & 0.42 & 0.56 & 1 & 0.06 & 0.15 & 0.26 & 0.81 \\
long\_gaps\_nightsoff7\_delayed-1\_v2.0\_10yrs & 0.27 & 0.45 & 0.60 & 1 & 0.07 & 0.16 & 0.26 & 0.82 \\
long\_gaps\_nightsoff6\_delayed-1\_v2.0\_10yrs & 0.27 & 0.45 & 0.59 & 1 & 0.06 & 0.16 & 0.26 & 0.82 \\
long\_gaps\_nightsoff5\_delayed-1\_v2.0\_10yrs & 0.26 & 0.44 & 0.58 & 1 & 0.07 & 0.15 & 0.25 & 0.82 \\
long\_gaps\_nightsoff4\_delayed-1\_v2.0\_10yrs & 0.26 & 0.44 & 0.58 & 1 & 0.06 & 0.15 & 0.26 & 0.81 \\
long\_gaps\_nightsoff3\_delayed-1\_v2.0\_10yrs & 0.26 & 0.44 & 0.58 & 1 & 0.06 & 0.15 & 0.26 & 0.80 \\
long\_gaps\_nightsoff2\_delayed-1\_v2.0\_10yrs & 0.25 & 0.44 & 0.58 & 1 & 0.06 & 0.15 & 0.26 & 0.81 \\
long\_gaps\_nightsoff1\_delayed-1\_v2.0\_10yrs & 0.25 & 0.43 & 0.57 & 1 & 0.06 & 0.15 & 0.25 & 0.80 \\
long\_gaps\_nightsoff0\_delayed-1\_v2.0\_10yrs & 0.22 & 0.38 & 0.52 & 1 & 0.05 & 0.13 & 0.24 & 0.79 \\
vary\_nes\_nesfrac0.01\_v2.0\_10yrs & 0.27 & 0.46 & 0.59 & 1 & 0.07 & 0.17 & 0.27 & 0.83 \\
vary\_nes\_nesfrac0.05\_v2.0\_10yrs & 0.28 & 0.46 & 0.60 & 1 & 0.07 & 0.15 & 0.26 & 0.82 \\
vary\_nes\_nesfrac0.10\_v2.0\_10yrs & 0.27 & 0.45 & 0.59 & 1 & 0.07 & 0.16 & 0.27 & 0.82 \\
vary\_nes\_nesfrac0.15\_v2.0\_10yrs & 0.27 & 0.46 & 0.59 & 1 & 0.07 & 0.16 & 0.27 & 0.83 \\
vary\_nes\_nesfrac0.20\_v2.0\_10yrs & 0.27 & 0.45 & 0.59 & 1 & 0.07 & 0.16 & 0.27 & 0.82 \\
vary\_nes\_nesfrac0.25\_v2.0\_10yrs & 0.26 & 0.45 & 0.59 & 1 & 0.07 & 0.15 & 0.26 & 0.82 \\
vary\_nes\_nesfrac0.30\_v2.0\_10yrs & 0.26 & 0.44 & 0.59 & 1 & 0.07 & 0.16 & 0.27 & 0.82 \\
vary\_nes\_nesfrac0.35\_v2.0\_10yrs & 0.26 & 0.44 & 0.59 & 1 & 0.07 & 0.15 & 0.25 & 0.82 \\
vary\_nes\_nesfrac0.40\_v2.0\_10yrs & 0.26 & 0.44 & 0.59 & 1 & 0.06 & 0.14 & 0.25 & 0.81 \\
vary\_nes\_nesfrac0.45\_v2.0\_10yrs & 0.26 & 0.45 & 0.60 & 1 & 0.06 & 0.15 & 0.25 & 0.81 \\
vary\_nes\_nesfrac0.50\_v2.0\_10yrs & 0.25 & 0.44 & 0.58 & 1 & 0.06 & 0.15 & 0.25 & 0.81 \\
vary\_nes\_nesfrac0.55\_v2.0\_10yrs & 0.25 & 0.43 & 0.57 & 1 & 0.06 & 0.15 & 0.25 & 0.81 \\
vary\_nes\_nesfrac0.75\_v2.0\_10yrs & 0.23 & 0.42 & 0.57 & 1 & 0.05 & 0.13 & 0.24 & 0.80 \\
vary\_nes\_nesfrac1.00\_v2.0\_10yrs & 0.24 & 0.41 & 0.56 & 1 & 0.04 & 0.12 & 0.22 & 0.79 \\
vary\_gp\_gpfrac0.01\_v2.0\_10yrs & 0.26 & 0.44 & 0.58 & 0.98 & 0.06 & 0.16 & 0.25 & 0.80 \\
vary\_gp\_gpfrac0.05\_v2.0\_10yrs & 0.27 & 0.44 & 0.58 & 0.98 & 0.07 & 0.16 & 0.26 & 0.81 \\
vary\_gp\_gpfrac0.10\_v2.0\_10yrs & 0.28 & 0.44 & 0.58 & 0.99 & 0.07 & 0.16 & 0.26 & 0.81 \\
vary\_gp\_gpfrac0.15\_v2.0\_10yrs & 0.26 & 0.44 & 0.58 & 0.99 & 0.07 & 0.16 & 0.26 & 0.82 \\
vary\_gp\_gpfrac0.20\_v2.0\_10yrs & 0.26 & 0.43 & 0.58 & 1 & 0.07 & 0.16 & 0.26 & 0.81 \\
vary\_gp\_gpfrac0.25\_v2.0\_10yrs & 0.27 & 0.45 & 0.60 & 1 & 0.06 & 0.15 & 0.27 & 0.82 \\
vary\_gp\_gpfrac0.30\_v2.0\_10yrs & 0.27 & 0.44 & 0.58 & 1 & 0.06 & 0.15 & 0.25 & 0.82 \\
vary\_gp\_gpfrac0.35\_v2.0\_10yrs & 0.26 & 0.45 & 0.59 & 1 & 0.06 & 0.14 & 0.25 & 0.82 \\
vary\_gp\_gpfrac0.40\_v2.0\_10yrs & 0.26 & 0.44 & 0.59 & 1 & 0.06 & 0.15 & 0.26 & 0.82 \\
vary\_gp\_gpfrac0.45\_v2.0\_10yrs & 0.25 & 0.44 & 0.59 & 1 & 0.06 & 0.14 & 0.25 & 0.81 \\
vary\_gp\_gpfrac0.50\_v2.0\_10yrs & 0.25 & 0.44 & 0.59 & 1 & 0.06 & 0.15 & 0.25 & 0.82 \\
vary\_gp\_gpfrac0.55\_v2.0\_10yrs & 0.25 & 0.43 & 0.59 & 1 & 0.06 & 0.14 & 0.25 & 0.82 \\
vary\_gp\_gpfrac0.75\_v2.0\_10yrs & 0.25 & 0.44 & 0.59 & 1 & 0.05 & 0.14 & 0.26 & 0.82 \\
vary\_gp\_gpfrac1.00\_v2.0\_10yrs & 0.22 & 0.42 & 0.58 & 1 & 0.05 & 0.14 & 0.24 & 0.81 \\
plane\_priority\_priority1.2\_pbf\_v2.1\_10yrs & 0.26 & 0.43 & 0.57 & 1 & 0.06 & 0.15 & 0.25 & 0.82 \\
plane\_priority\_priority0.9\_pbf\_v2.1\_10yrs & 0.26 & 0.44 & 0.58 & 1 & 0.07 & 0.16 & 0.26 & 0.82 \\
plane\_priority\_priority0.6\_pbf\_v2.1\_10yrs & 0.26 & 0.45 & 0.59 & 1 & 0.07 & 0.16 & 0.27 & 0.83 \\
plane\_priority\_priority0.4\_pbf\_v2.1\_10yrs & 0.24 & 0.43 & 0.59 & 1 & 0.06 & 0.15 & 0.26 & 0.83 \\
plane\_priority\_priority0.3\_pbf\_v2.1\_10yrs & 0.23 & 0.42 & 0.58 & 1 & 0.05 & 0.14 & 0.26 & 0.82 \\
plane\_priority\_priority0.2\_pbf\_v2.1\_10yrs & 0.21 & 0.41 & 0.58 & 1 & 0.04 & 0.12 & 0.25 & 0.82 \\
plane\_priority\_priority0.1\_pbf\_v2.1\_10yrs & 0.18 & 0.38 & 0.56 & 1 & 0.04 & 0.11 & 0.23 & 0.80 \\
plane\_priority\_priority1.2\_pbt\_v2.1\_10yrs & 0.26 & 0.43 & 0.57 & 1 & 0.06 & 0.15 & 0.25 & 0.82 \\
plane\_priority\_priority0.9\_pbt\_v2.1\_10yrs & 0.26 & 0.43 & 0.58 & 1 & 0.06 & 0.15 & 0.26 & 0.83 \\
plane\_priority\_priority0.6\_pbt\_v2.1\_10yrs & 0.25 & 0.45 & 0.59 & 1 & 0.07 & 0.16 & 0.27 & 0.83 \\
plane\_priority\_priority0.4\_pbt\_v2.1\_10yrs & 0.24 & 0.43 & 0.59 & 1 & 0.06 & 0.15 & 0.26 & 0.82 \\
plane\_priority\_priority0.3\_pbt\_v2.1\_10yrs & 0.23 & 0.43 & 0.59 & 1 & 0.05 & 0.14 & 0.25 & 0.82 \\
plane\_priority\_priority0.2\_pbt\_v2.1\_10yrs & 0.21 & 0.41 & 0.57 & 1 & 0.05 & 0.13 & 0.25 & 0.81 \\
plane\_priority\_priority0.1\_pbt\_v2.1\_10yrs & 0.19 & 0.38 & 0.55 & 1 & 0.04 & 0.12 & 0.24 & 0.80 \\
pencil\_fs1\_v2.1\_10yrs & 0.24 & 0.43 & 0.58 & 1 & 0.06 & 0.15 & 0.25 & 0.82 \\
pencil\_fs2\_v2.1\_10yrs & 0.24 & 0.43 & 0.57 & 1 & 0.06 & 0.15 & 0.25 & 0.81 \\
good\_seeing\_gsw0.0\_v2.1\_10yrs & 0.26 & 0.44 & 0.58 & 1 & 0.06 & 0.15 & 0.26 & 0.82 \\
good\_seeing\_gsw1.0\_v2.1\_10yrs & 0.26 & 0.44 & 0.58 & 1 & 0.06 & 0.16 & 0.25 & 0.82 \\
good\_seeing\_gsw3.0\_v2.1\_10yrs & 0.27 & 0.44 & 0.59 & 1 & 0.07 & 0.16 & 0.26 & 0.82 \\
good\_seeing\_gsw6.0\_v2.1\_10yrs & 0.27 & 0.45 & 0.58 & 1 & 0.06 & 0.14 & 0.25 & 0.82 \\
good\_seeing\_gsw10.0\_v2.1\_10yrs & 0.26 & 0.43 & 0.58 & 1 & 0.06 & 0.14 & 0.25 & 0.82 \\
good\_seeing\_gsw20.0\_v2.1\_10yrs & 0.27 & 0.44 & 0.59 & 1 & 0.06 & 0.15 & 0.25 & 0.81 \\
good\_seeing\_gsw50.0\_v2.1\_10yrs & 0.26 & 0.43 & 0.58 & 1 & 0.06 & 0.14 & 0.25 & 0.82 \\
virgo\_cluster\_v2.0\_10yrs & 0.26 & 0.44 & 0.58 & 1 & 0.06 & 0.16 & 0.26 & 0.82 \\
carina\_v2.0\_10yrs & 0.27 & 0.44 & 0.58 & 1 & 0.06 & 0.14 & 0.26 & 0.82 \\
carina\_v2.2\_10yrs & 0.27 & 0.44 & 0.58 & 1 & 0.07 & 0.16 & 0.27 & 0.84 \\
smc\_movie\_v2.0\_10yrs & 0.26 & 0.44 & 0.59 & 1 & 0.06 & 0.16 & 0.26 & 0.83 \\
roman\_v2.0\_10yrs & 0.29 & 0.47 & 0.60 & 1 & 0.08 & 0.18 & 0.28 & 0.83 \\
north\_stripe\_v2.0\_10yrs & 0.25 & 0.43 & 0.58 & 0.99 & 0.06 & 0.14 & 0.25 & 0.80 \\
short\_exp\_v2.0\_10yrs & 0.26 & 0.43 & 0.57 & 1 & 0.06 & 0.15 & 0.25 & 0.82 \\
short\_exp\_v2.2\_10yrs & 0.26 & 0.44 & 0.58 & 1 & 0.08 & 0.17 & 0.27 & 0.84 \\
multi\_short\_v2.0\_10yrs & 0.27 & 0.43 & 0.56 & 0.99 & 0.05 & 0.12 & 0.22 & 0.78 \\
twilight\_neo\_nightpattern4v2.0\_10yrs & 0.27 & 0.44 & 0.60 & 1 & 0.06 & 0.16 & 0.27 & 0.82 \\
twilight\_neo\_nightpattern7v2.0\_10yrs & 0.26 & 0.44 & 0.59 & 1 & 0.06 & 0.16 & 0.27 & 0.82 \\
twilight\_neo\_nightpattern3v2.0\_10yrs & 0.27 & 0.45 & 0.59 & 1 & 0.06 & 0.16 & 0.27 & 0.82 \\
twilight\_neo\_nightpattern6v2.0\_10yrs & 0.25 & 0.44 & 0.60 & 1 & 0.06 & 0.15 & 0.27 & 0.82 \\
twilight\_neo\_nightpattern5v2.0\_10yrs & 0.25 & 0.44 & 0.59 & 1 & 0.06 & 0.15 & 0.27 & 0.82 \\
twilight\_neo\_nightpattern2v2.0\_10yrs & 0.26 & 0.45 & 0.59 & 1 & 0.06 & 0.15 & 0.27 & 0.82 \\
twilight\_neo\_nightpattern1v2.0\_10yrs & 0.23 & 0.42 & 0.57 & 1 & 0.06 & 0.14 & 0.27 & 0.81 \\
no\_repeat\_rpw-1.0\_v2.1\_10yrs & 0.27 & 0.45 & 0.59 & 1 & 0.06 & 0.15 & 0.26 & 0.82 \\
no\_repeat\_rpw-2.0\_v2.1\_10yrs & 0.27 & 0.45 & 0.59 & 1 & 0.07 & 0.16 & 0.26 & 0.82 \\
no\_repeat\_rpw-5.0\_v2.1\_10yrs & 0.26 & 0.43 & 0.57 & 1 & 0.06 & 0.15 & 0.26 & 0.82 \\
no\_repeat\_rpw-10.0\_v2.1\_10yrs & 0.27 & 0.44 & 0.58 & 1 & 0.07 & 0.17 & 0.27 & 0.82 \\
no\_repeat\_rpw-20.0\_v2.1\_10yrs & 0.27 & 0.45 & 0.59 & 1 & 0.07 & 0.16 & 0.26 & 0.82 \\
no\_repeat\_rpw-100.0\_v2.1\_10yrs & 0.26 & 0.44 & 0.59 & 1 & 0.08 & 0.17 & 0.27 & 0.82 \\
supress\_and\_triplets\_v2.2\_10yrs & 0.26 & 0.43 & 0.58 & 1 & 0.07 & 0.16 & 0.27 & 0.84 \\
draft\_connected\_v2.99\_10yrs & 0.27 & 0.46 & 0.60 & 1 & 0.05 & 0.12 & 0.23 & 0.81 \\
draft\_v2.99\_10yrs & 0.25 & 0.42 & 0.57 & 1 & 0.04 & 0.11 & 0.20 & 0.79 \\
roll\_early\_v2.99\_10yrs & 0.26 & 0.44 & 0.59 & 1 & 0.05 & 0.13 & 0.23 & 0.82 \\
dd6\_v2.99\_10yrs & 0.27 & 0.45 & 0.59 & 1 & 0.05 & 0.13 & 0.23 & 0.82 \\
low\_gp\_v2.99\_10yrs & 0.11 & 0.24 & 0.41 & 1 & 0.02 & 0.05 & 0.11 & 0.67 \\
light\_roll\_v2.99\_10yrs & 0.26 & 0.45 & 0.60 & 1 & 0.06 & 0.14 & 0.25 & 0.82 \\
draft2\_rw0.5\_v2.99\_10yrs & 0.25 & 0.44 & 0.59 & 1 & 0.05 & 0.14 & 0.25 & 0.82 \\
draft2\_rw0.9\_v2.99\_10yrs & 0.24 & 0.41 & 0.56 & 1 & 0.05 & 0.13 & 0.22 & 0.80 \\
draft2\_rw0.9\_uz\_v2.99\_10yrs & 0.24 & 0.41 & 0.56 & 1 & 0.05 & 0.12 & 0.22 & 0.80 \\
\caption{\reedit{Npts Metric and Fisher Metric values of OpSims in this paper. The first 4 columns has the fraction of lightcurves with at least 10 points in $t_0 \pm t_{\rm E}$ (see Section \ref{sec: microlensing metric}). The second 4 columns has the fraction of lightcurves with $\frac{\sigma_{t_{\rm E}}}{t_{\rm E}} < 0.1$ (see Section \ref{sec: fisher matrix}). Each column has the $t_{\rm E}$ range of events simulated. The rest of the metric results for these OpSims can be found in \url{https://github.com/lsst-pst/survey_strategy/blob/main/fbs_2.0/summary_2023_01_01.csv}.}}
    \label{tab:char_metrics}
\end{longtable}

\end{document}